\documentclass[superscriptaddress,prb,preprint,amsmath,amssymb]{revtex4} %twocolumn,
\usepackage{graphicx,natbib}
\usepackage[caption=false]{subfig}
\usepackage{grffile}

\renewcommand{\vec}[1]{\mathbf{#1}} 
\newcommand{\RM }[1]{\mathrm{#1}}
\newcommand{\ave}[1]{ {\langle {#1} \rangle} }

\def\kB{ k_{\rm{B}} }

\def\Drough{D^{\rm{rough}}} %generic rough
\def\Dref{D^{\rm{ref}}}     %generic reference

\def\DTW{D^{\rm{TW}}} %thermal walls
\def\DRW{D^{\rm{RW}}} %rotational walls
\def\DSW{D^{\rm{SW}}} %smooth walls

\def\fw{f_w}

\begin{document}
  
\title{Impact of surface roughness on diffusion of confined
fluids}

\author{William P. Krekelberg}
\email{william.krekelberg@nist.gov}
\thanks{Corresponding Author} 
\author{Vincent K. Shen}
\email{vincent.shen@nist.gov}
\address{Chemical and Biochemical Reference Data Division, 
  National Institute of Standards and Technology, 
  Gaithersburg, Maryland 02899-8380, USA}

\author{Jeffrey R. Errington}
\email{jerring@buffalo.edu}
\address{Department of Chemical and Biological Engineering, 
  University at Buffalo, The State University of New York, 
  Buffalo, New York 14260-4200, USA}

\author{Thomas M. Truskett}
\email{truskett@che.utexas.edu}
\address{Department of Chemical
  Engineering, University of
  Texas at Austin, Austin, TX 78712.}  
 \address{Institute for Theoretical Chemistry, 
   University of Texas at Austin, Austin, TX 78712.}

\begin{abstract}
  Using event-driven molecular dynamics simulations, we quantify how
  the self diffusivity of confined hard-sphere fluids depends on the
  nature of the confining boundaries. We explore systems with
  featureless confining boundaries that treat particle-boundary
  collisions in different ways and also various types of physically
  (i.e., geometrically) rough boundaries.  We show that, for
  moderately dense fluids, the ratio of the self diffusivity of a
  rough wall system to that of an appropriate smooth-wall reference
  system is a linear function of the reciprocal wall separation, with
  the slope depending on the nature of the roughness. We also discuss
  some simple practical ways to use this information to predict
  confined hard-sphere fluid behavior in different rough-wall systems.
% Using molecular simulations, we show that surface roughness effects
% the self-diffusion of a moderately dense confined fluid
% systematically.  In particular,  we show
% that the ratio of the self-diffusion in a rough system to the self
% diffusion in a reference smooth system scales inversely with the degree of
% confinement.  We find that the relationship between self-diffusion in
% rough systems quantitatively depends on the form of surface roughness,
% but the qualitative relationship does not.
\end{abstract}
\maketitle

\section{Introduction}
\label{sec:Introduction}

Predicting the dynamic properties of moderate-to-high density bulk
fluids from first principles remains an outstanding scientific
challenge. This already-difficult problem becomes considerably more
formidable when the fluid is confined to small spaces, a situation
that is commonly encountered across a broad range of technologically
important settings.\cite{Bhushan1995Nanotribology:-} In recent years,
an important step toward solving this problem for confined fluids was
taken by identifying physically meaningful static (i.e.,
thermodynamic) variables (or combinations thereof) against which
dynamic fluid properties scaled independently of the degree of
confinement.\cite{Mittal2006Thermodynamics-,Mittal2007Relationships-b,Mittal2007Does-confining-,Mittal2007Confinement-ent,Goel2008Tuning-Density-,Mittal2008Layering-and-Po,Goel2009Available-state,Mittal2009Using-Compressi}
This identification suggested that dynamic properties such as the self
diffusivity of confined fluids could be predicted based on knowledge
of their static properties and the corresponding static-dynamics
scaling relationship obtained from, e.g., bulk fluid data.
% Knowledge of this type can enable one to predict the dynamic
% properties, the self-diffusion coefficient for example, of confined
% fluids, using its bulk-phase properties or properties under other
% confinement conditions. 
To date, tests of this scaling strategy for predicting how confinement
affects dynamics of model systems have been carried out by molecular
simulation of monatomic\cite{Mittal2006Thermodynamics-,Mittal2007Relationships-b,Mittal2007Does-confining-,Mittal2007Confinement-ent,Goel2009Available-state,Mittal2009Using-Compressi}
and a limited number of molecular
\cite{Chopra2010confined-Excess}  fluids.
Although these tests suggest that the scaling method can successfully
predict the dynamic behavior of a variety of fluids confined to pores
with different geometries,
% While these earlier studies have focused on
% the robustness of the dynamic scaling relationships over a diverse
% range of different intermolecular fluid potentials, comparatively
little attention has been paid to the nature of the confining walls
themselves and how it affects the relationship between static
variables and dynamic properties. In this paper, we address
this issue by investigating how surface roughness impacts the
self diffusivity of model confined fluids.

Smooth, flat structureless walls represent a mathematically convenient
and idealized theoretical construct, and thus, historically, have been
commonly used in the study of model confined fluids. However, real
solid surfaces do in fact exhibit structure, which can significantly
impact the thermodynamic and dynamic properties of the fluids they contact.
Wetting and lubrication are two obvious examples of
phenomena where the structure and shape of the fluid-exposed solid
surface have significant implications.\cite{Grzelak2010Molecular-Simul,Grzelak2010Nanoscale-Limit}  This
should not be surprising because surface roughness, which arises from
the structural arrangement of ``wall'' particles, ultimately
influences the fluid-wall interaction. Also, it is clear that surface
roughness greatly impacts dynamics.  For low density gases, where
Knudsen diffusion\cite{Knudsen1909Knudsen} dominates, self diffusion
decreases with surface roughness.\cite{Arya2003Knudsen-Diffusi,Malek2001Effects-of-Surf,Malek2002Pore-roughness-,Malek2003Knudsen-self--a}
Furthermore, experiments of confined colloids find that some particles stick
to walls, further slowing dynamics.\cite{Edmond2010Local-influence}
Simulations have shown that position-dependent relaxation processes near
rough surfaces are much slower than those near smooth surfaces.\cite{Scheidler2002Cooperative-mot,Scheidler2004The-Relaxation-}

%new
A question which has recieved comparatively less attention is how
surface roughness impact the self-diffusivity of moderate-to-dense
confined fluids.  The goal of this paper is to answer this question
systematically via molecular simulations.  The most obvious way in which roughness
influences dynamic properties is through the modification of fluid particle--wall
collisions. Here, we investigate how
different representations of this collisional modification lead to
changes in the self-diffusion of a simple, confined fluid. As a
starting point, we study the monodisperse hard-sphere (HS) fluid
confined between smooth hard walls. Despite its simplicity, we choose
the HS fluid because it captures many of the important  effects
arising from the dominant excluded volume interactions in liquids \cite{Errington2003Quantification-,Hansen2006Theory-of-Simpl}  Obviously, a
smooth hard wall does not possess any roughness whatsoever, but it
still serves as a useful reference point for studying confined fluids,
specifically the dynamic properties in this paper. In fact, the slit
pore geometry is often assumed when determining the pore size
distribution from adsorption isotherms in real porous materials.\cite{Lastoskie1993Pore-size-distr,Olivier1994Determination-o,Lastoskie1997Chapter-15.-Str,Rouquerol1999Adsorption-of-P,Klobes2006Porosity-and-sp}
In our simulations, we introduce surface roughness in two
separate ways (i) by modifying the boundary conditions for collisions between fluid
particles and a flat wall surface and (ii) by giving the surface of the
confining walls physical roughness or shape. The former crudely
represents surface roughness on a length scale much smaller than the
diameters of the fluid particles (e.g., confined colloids), while the
latter models physical roughness on a length scale comparable to the
diameter of the fluid particles (e.g., confined molecular fluids).  
 While neither of the above types of walls can be said to
  completely represent the roughness present in real physical systems,
  they represent the types of roughness that can be incorporated in molecular
  simulations, and serve as a good starting point to address the impact of
  surface roughness on self-diffusivity.
%new

We find that surface roughness reduces average particle mobility relative to 
a smooth flat wall, where the collisions between fluid particles and 
the wall surface are perfectly reflecting.  Moreover, the reduction appears to be 
systematic with increasing degree of confinement (i.e., decreasing pore width). 
In the case of physically rough walls, we show that it is necessary to make the 
distinction between the spatially homogeneous and inhomogeneous directions because
the associated self-diffusion coefficients can differ significantly. Finally, we find that the 
dynamic properties of the hard-sphere fluid confined between rough surfaces 
can be regarded as a perturbation on the dynamic properties of an appropriately 
chosen smooth-wall reference system.

This paper is organized as follows. In Section~\ref{sec:sim_methods}, we 
describe the model fluid studied in this work, as well as the simulation 
methods used. We then discuss how wall-surface roughness is implemented in 
Section~\ref{sec:boundary_conditions}. Results are presented and discussed in 
Section~\ref{sec:Results}. Finally, we present conclusions and directions for 
future work in Section~\ref{sec:conclusions}.

\section{Model fluid}
\label{sec:sim_methods}

%new
Event-driven microcanonical molecular dynamics (MD)
simulations\cite{Rapaport2004The-Art-of-Mole} were used to study a
fluid composed of hard spheres of diameter $\sigma$ and mass $m$ with
interaction potential
\begin{equation}
  \phi(r_{ij})=
  \begin{cases}
    \infty & r_{ij}\leq \sigma, \\
    0      & r_{ij} >   \sigma,
  \end{cases}
\end{equation}
where $r_{ij}$ is the distance between particles $i$ and $j$. 
%new
Event-driven MD consists of four basic steps: (1) Calculate future events (collisions)
times. (2) Sort events to determine next event to occur. (3)
 Advance system to next event using Newtonian dynamics (free
 flight). (4) Execute event (collision) to determine new particle
 velocities, and return to (1). Various methods are available
to speed up the general algorithm.  The exact method used in this
study is described in Ref.\onlinecite{Rapaport2004The-Art-of-Mole}.
%end
A rectangular simulation cell of dimensions $L_x\times L_y \times L_z$
with $N=4000$ particles was used throughout this work. Periodic
boundary conditions were applied in the $x$ and $y$ directions in all
cases, and in the $z$ direction for the bulk fluid.  For the confined
fluid, where the $z$ direction was non-periodic, appropriate
wall-boundary conditions (see below) were employed.  The dimensions of
the simulation box were chosen to correspond to a desired reduced
number density
$\rho\sigma^3$ (or packing fraction$=\pi\rho\sigma^3/6$), and such that the dimensions of the simulation box in the
periodic directions were (nearly) equal. Initial configurations for
the MD simulations were generated by randomly inserting hard spheres
at low density with no overlaps, followed by compression and
(Monte-Carlo) relaxation steps until the desired density was
obtained. 
%change
%old
% Equilibration and production times in the MD simulations
% exceeding $1000\,(\sigma^{2}/\kB T)^{1/2}$ and $10^5\, (m \sigma^ {2}/\kB T)^{1/2}$,
% respectively, were used for all the systems studied
%new
The systems were deemed equilibrated when the number of
  collisions (particle-particle and particle-wall) per unit time
  reached a constant value as a function of time.  Equilibration times
  of $1000\,(m \sigma^2/\kB T)^{1/2}$ were found to be sufficient to
  meet this requirement.  Production simulations spanned times
  exceeding $10^5\,(m \sigma^2/\kB T)^{1/2}$, which allowed the
  slowest system to display displacements on the order of $50\sigma$.
%end
The (pore-averaged) self-diffusion
coefficients were calculated by fitting the long-time behavior of the
mean-squared displacement in a periodic direction to the Einstein
relation, e.g.,  $D_x=\lim_{t\rightarrow \infty} \ave{\Delta x^2}/(2t)$.  Where
appropriate, self-diffusion coefficients in equivalent periodic
directions were averaged together.

%\clearpage{}
\section{Surface Roughness and Boundary conditions}
\label{sec:boundary_conditions}

We study the hard-sphere fluid confined between parallel hard walls in
a slit-pore geometry. To avoid any confusion, we point out that none
of the walls studied here exhibits internal structure. That is,
the walls are of uniform density and can be regarded as a continuum
solid. Let $H$ denote the average distance between the two surfaces of
the confining walls.  In this work, we study two general types of wall
surfaces. The first involves flat walls where the
boundary conditions for the particle-wall collisions have been
modified. These walls model surface roughness on a length scale much smaller than the diameters of fluid particles, and thus, because the surfaces are geometrically flat, they are referred to as featureless
surfaces (walls). The second type of surface studied here possesses
physical (i.e., geometric) roughness, i.e., the height of the wall surface
varies with lateral position, and thus they are referred to as physically rough
surfaces (walls). Schematics of each type of surface are given in
Fig.~\ref{fig:geometry_schematics}.
%\clearpage{}
% \begin{figure}[h]
%   \centering
%   \subfloat[][]{
%     \includegraphics[width=1.5in]{Figs/HS_slit/link/smooth/figs/schematic_wall_geometry/slit}
%     \label{fig:schemitic_slit_pore}
%   }
%   \subfloat[][]{
%     \includegraphics[width=1.75in]{Figs/HS_cos/link/figs/schematic_walls/schematic}
%     \label{fig:schematic_wavy}
%   }
%   \caption{Schematics of the geometries of the confined fluid system,
%     as described in the text. (a) Featureless walls and (b) physically rough walls. }
%   \label{fig:geometry_schematics}
% \end{figure}

\subsection{Featureless Walls}
\label{sec:Flat_walls}

We considered three different types of boundary 
conditions for particle collisions with featureless walls. As a useful reference state, we 
first considered flat, smooth walls (SW), where the fluid-wall 
collisions are specular. The second type of featureless 
surface studied was a so-called thermal wall (TW). This model boundary
attempts to reproduce particle collisions with a surface uneven on length scales considerably smaller than the particle diameter, the essential feature being that pre- and post-wall collision 
velocities are uncorrelated. More importantly, the distribution of post-
collision velocities is determined by the temperature of the wall.
 The third type of featureless surface gives an
alternative approach to generate uncorrelated pre- and post-wall-collision 
velocities. We refer to it as a
rotational wall (RW) for reasons that will become apparent below. 

In order to distinguish between the featureless walls with different
boundary conditions,
consider a particle with a pre-wall-collision velocity $\vec{v}=(v_x,v_y,v_z)
$, and a post-wall collision velocity $\vec{v}'=(v_x',v_y',v_z')$.
For the cases considered here, the components of $\vec{v}'$ are as follows.
\begin{itemize}

\item Smooth walls (SW): $v'_x=v_x$, $v'_y=v_y$, and $v'_z = -v_z$

\item Thermal walls (TW): $v_z'$ is chosen according to the following 
distribution\cite{Tehver1998Thermal-walls-i} depending on wall temperature $T_{\RM{w}}$ (set equal to fluid 
temperature $T$):
  \begin{equation}
  \phi_z (v_z') = \frac{m}{k_B T_w} \ \lvert v_z' 
\rvert \ \exp{ \left[ \frac{-mv_z'^2}{2k_BT_{\RM{w}}}
  \right]  }
  \end{equation}
  where the sign of $v_z'$ depends on the location of the wall
  collision (i.e., positive for lower wall collisions and negative for
  upper wall collisions). 
  $v_i'$ (where $i=x,y$) is chosen randomly from a Gaussian distribution:
  \begin{equation}
  \phi_i (v_i') = \sqrt{\frac{m}{2 \pi k_B T_{\RM{w}}}} \ \exp
{ \left[ \frac{-m v_i'^2}{2 k_B T_w} \right] }
  \end{equation}
  
\item Rotational walls (RW): $v'_z=-v_z$, and the lateral components
  $v'_x$ and $v'_y$ are determined by rotating the corresponding
  components of the pre-wall-collision velocity $v_x$ and $v_y$ by an
  angle $\theta$:
  %projected velocities $v_x$, $v_y$ by
  \begin{equation}
    \label{eq:velocity_rotation}
    \begin{split}
      v_x'&=v_x \cos \theta'-v_y \sin \theta' 
\\
      v_y'&=v_x \sin \theta' +v_y \cos 
\theta',
    \end{split}
  \end{equation}

  where $\theta'=\pm \theta$, with the sign chosen randomly.  Note that $
  \theta=0$ corresponds to smooth walls, while $\theta=\pi$ corresponds to 
  bounce-back boundary conditions.

\end{itemize}

Note that the boundary conditions described above affect the dynamics
of the system only.  Because all three of the above walls have the
same geometrical form and produce the
same velocity component distributions, they lead to identical thermodynamic and
structural properties \cite{Tehver1998Thermal-walls-i}.  This was
verified by the simulation results.

We note that  the featureless walls described above are not
intended to model any specific physical system.  Rather, they are
mathematically convenient course-grained models for surface roughness.  For example,
the motivation for the thermal walls goes back to Maxwell
\cite{Maxwell1867roughness-somet}, who considered collisions with a
highly uneven low density granular surface. Particles that strike
this type of surface undergo a series of collisions with many
different surface molecules.
The resulting outgoing velocity is expected to be randomized, with a distribution
determined by the temperature of the wall.
We also stress that the rotational walls introduced here are not
intended to mimic any type of real system.  Instead, they should be
considered a mathematical construct that allows us to
change the surface boundary condition in a continuous fashion  from perfectly smooth
($\theta=0$) to rough (bounce-back, $\theta=\pi$) walls, with the system having identical thermodynamics for all values of $\theta$.  This allows us to study how different levels
of surface roughness impact the dynamics of the model fluid without
changing thermodynamics.

\subsection{Physically rough walls}
\label{sec:physical_roughness}

\begin{figure}[!h]
  \centering
  %\subfloat[][]{
  \subfloat{
    \includegraphics[width=1.5in]{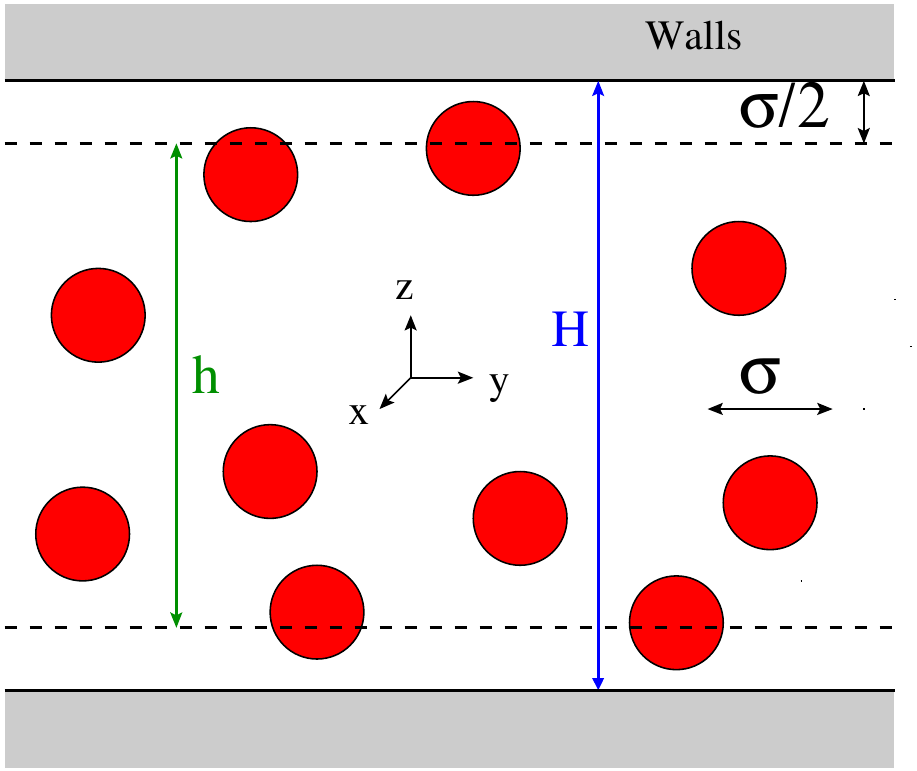}
    \label{fig:schemitic_slit_pore}
  }
  %\subfloat[][]{
  \subfloat{
    \includegraphics[width=1.75in]{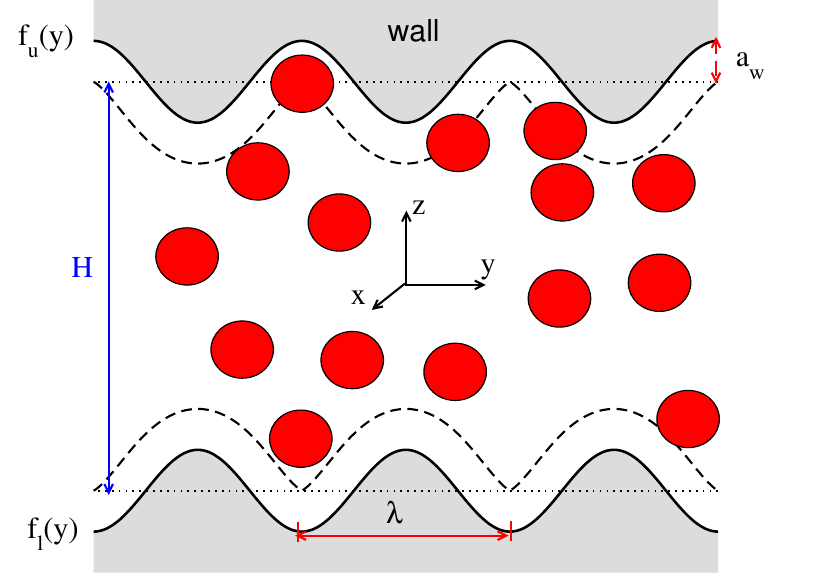}
    \label{fig:schematic_wavy}
  }
  \caption{Schematics of the geometries of the confined fluid system,
    as described in the text. (a) Featureless walls and (b) physically rough walls. }
  \label{fig:geometry_schematics}
\end{figure}

For a physically rough surface, we intuitively expect that the height of the 
surface should vary with lateral position. For simplicity, we use the 
following expressions for the upper and lower wall surfaces in a slit-
pore geometry:

\begin{equation}
  \label{eq:wavy_walls}
  \begin{split}
    f_{\rm{U}}(y)&=\frac{H}{2}+a_w \cos \left
[ \frac{2\pi
        y}{\lambda}+\pi\right],\\
    f_{\rm{L}}(y)&=-\frac{H}{2}+a_w \cos \left
[ \frac{2\pi
            y}{\lambda}\right]
  \end{split}
\end{equation}
\\
where the subscripts $\rm{L}$ and $\rm{U}$ denote the lower and upper 
surfaces, respectively, and $a_w$ and $\lambda$ are the amplitude and 
wavelength of the well-behaved surface variations, respectively. Figure~\ref
{fig:schematic_wavy} displays the geometry of the physically rough walls 
employed. As written in Eq.~\eqref{eq:wavy_walls} and depicted in 
Figure~\ref{fig:schematic_wavy}, minima in the upper wall and maxima in the 
lower wall are aligned, which imposes a maximum value upon $a_w$ if a fluid 
particle is to have access to the entire pore length. The average surface-to-surface 
distance is $H$ as long as the quantity $L_y/\lambda$ is an integer, which 
is true throughout this work. Because we have chosen to have the height of 
the surface be a function of only $y$, the system is spatially inhomogeneous 
in the $y$-direction and homogeneous in the $x$-direction. This will have 
important consequences for the self diffusivity of a fluid confined between these surfaces. 

The rigid character of the confining walls was maintained by requiring
perfectly reflecting specular particle-surface collisions. However,
instead of solving the set of nonlinear equations to determine
particle-surface collision times, which is computationally expensive,
the curved walls were discretized into a set of short, connected line
segments of length $0.015\sigma$.  Using smaller segments does not
lead to noticeable changes in the resulting properties of the systems
studied here.

We stress that the specific form of the physically rough walls studied
here is not intended to mimic a specific physical system.  Rather,
they allow us to systematically determine the impact of different
surface feature sizes on the self-diffusivity of our model fluid.

\section{Results}
\label{sec:Results}

\subsection{Featureless Flat Walls}
\label{sec:results_flat_walls}

% \begin{figure}[h]
%   \includegraphics[width=3in]{Figs/HS_slit/link/smooth/figs/D_vs_dens_isoH/D_vs_dens_isoGap_-rescale}
%   \caption{Self diffusivity $D$ versus total density $\rho=N/(AH)$ for
%     the hard-sphere fluid confined between smooth walls in slit-pore geometry.}
%   \label{fig:smooth_slit_D_vs_density}
% \end{figure}

\begin{figure}[!h]
  \includegraphics[width=3in]{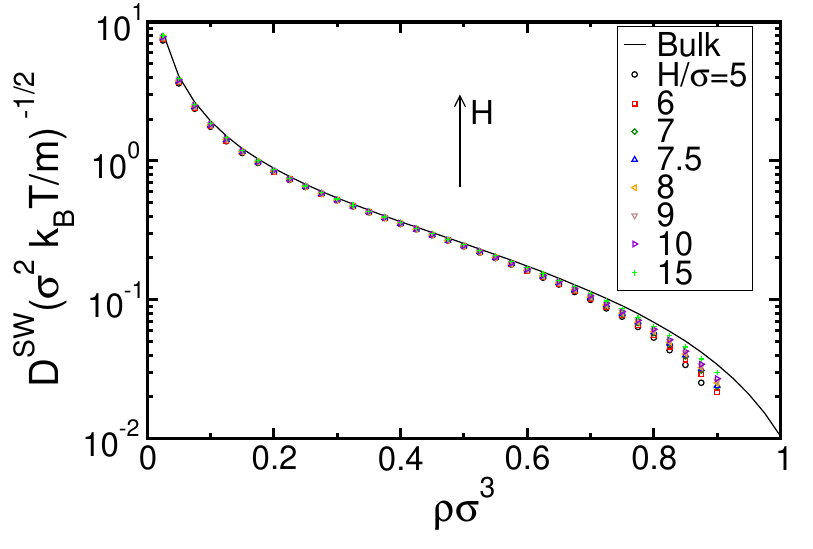}
  \caption{Self diffusivity $D$ versus total density $\rho=N/(AH)$ for
    the hard-sphere fluid confined between smooth walls in slit-pore geometry.}
  \label{fig:smooth_slit_D_vs_density}
\end{figure}

We first present results for the hard-sphere fluid confined between smooth, 
flat hard walls. In Fig.~\ref{fig:smooth_slit_D_vs_density}, we plot the 
self-diffusion coefficient, $\DSW$, where the superscript $\text{SW}
$ signifies smooth wall case, as a function of 
the total density $\rho=N/(AH)$. Data are shown for a number of pore 
widths $H$. As previously pointed out elsewhere\cite{Mittal2006Thermodynamics-,Goel2009Available-state}, all data points 
seem to fall approximately onto a single curve independently of $H$ for $\rho 
\sigma^3 \lesssim 0.75$.  Below, we investigate how modifying the nature of the confining surfaces changes the results obtained in this basic reference system.

\subsection{Thermal Walls}
\label{sec:results_thermal_walls}

\begin{figure}[!h]
  \centering
  \includegraphics[width=3in]{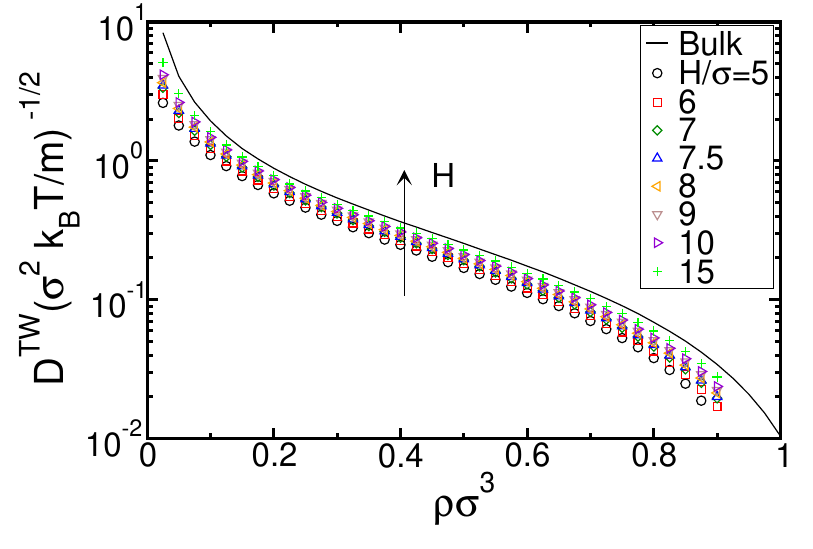}
  \caption{Self diffusivity $D$ versus total density $\rho$ for the
    hard-sphere fluid confined between thermal walls, for various wall
  separations $H$.}
  \label{fig:thermal_slit_D_vs_density}
\end{figure}

Figure~\ref{fig:thermal_slit_D_vs_density} displays the self-diffusion 
coefficient $\DTW$ for the hard-sphere fluid confined between 
featureless thermal walls as a function of $\rho$ for various values of $H$. 
Compared to the smooth, flat wall, roughness due to the thermal wall 
has a clear and noticeable influence on particle dynamics.  Thermal
walls reduce particle mobility, with the magnitude of the
reduction increasing with decreasing $H$. That average  mobility of fluid
particles confined between thermal walls decreases with $\rho$ at fixed $H$
has been previously reported.\cite{Subramanian1979Molecular-dynam}

\begin{figure}[!h]
  \centering
  \includegraphics[width=3in]{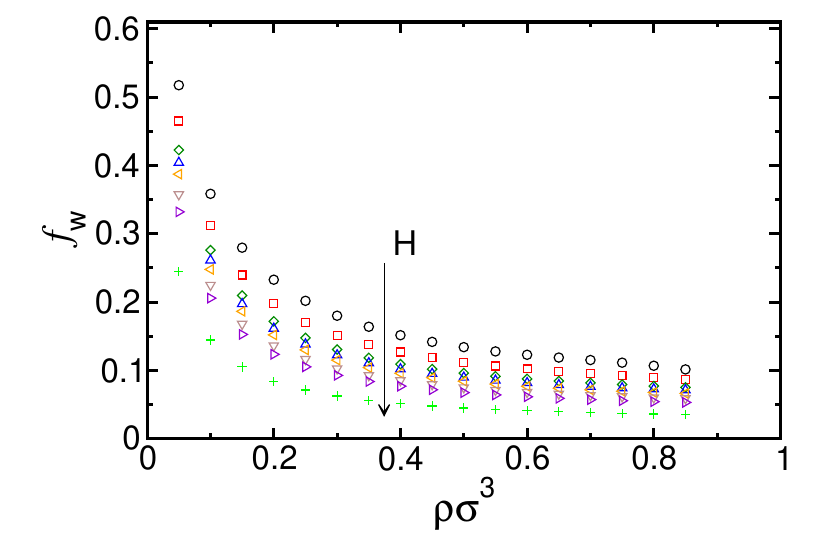}
  \caption{Fraction of collisions in the system involving the wall
    surface $\fw$ for the
    hard-sphere fluid confined to a slit-pore geometry.  Symbols have the same
    meaning as in Fig.~\ref{fig:thermal_slit_D_vs_density}}
\label{fig:slit_pore_coll_fig}
\end{figure}

The interesting trend we quantify in this study is the reduction of
$\DTW$ with increasing degree of confinement at fixed $\rho$. This
behavior can be understood by considering the quantity $\fw$, the
fraction of collisions in the system involving the surfaces of the
confining walls. Intuitively, we expect non-specular particle-surface
collisions, such as those that take place in the thermal-wall systems,
to slow dynamics in the transverse direction relative to specular
particle-surface collisions.  We base this expectation by considering
the free flight of a single particle. In this case, specular
particle-wall collisions do not change the transverse motion, while
non-specular particle-wall collisions do.
 We also expect that the
fraction of wall collisions grows with the prominence of the walls
(the fraction of the fluid particles near
the walls), 
that is, with decreasing $H$.  That is, in order to maintain a fixed
$N$ and $V$, a decrease in $H$ must be compensated by an increase in
fluid-exposed wall surface area. Figure~\ref{fig:slit_pore_coll_fig}
shows that this is indeed the case. Decreasing $H$ at constant $\rho$
systematically increases the fraction of wall collisions.  Clearly,
the reduction in $\DTW$ with decreasing $H$ observed in
Fig.~\ref{fig:thermal_slit_D_vs_density} is directly linked to the
increasing fraction of wall collisions. Also, note that the results
given in Fig.~\ref{fig:slit_pore_coll_fig} are not unique to the
thermal-wall systems.  In fact, all three kinds of featureless
surfaces studied in this work yield identical wall-collision
statistics.  As noted above, the collision boundary conditions only
affect the dynamic properties of the fluid and not the thermodynamic
properties. The thermodynamic pressure of the system is intimately
related to the wall-collision statistics.

While there are clear differences between the smooth-wall and
thermal-wall self-diffusion coefficients, $\DSW$ and $\DTW$,
respectively, these differences appear to depend systematically on
$H$. This suggests that it might be possible to develop an approach to
predict the self-diffusivity of the hard-sphere fluid confined between
thermal walls using a limited amount of information.  In particular,
note the dependence of the fraction of wall collisions on density.
Initially, increasing density leads to a pronounced decrease in the
fraction of wall collisions. This is due to the associated increase in
particle-particle collisions compared to particle-wall collisions (not
shown).  However, the fraction of wall collisions eventually becomes a
weak function of density.  Since we expect the reduction in
self diffusivity due to the presence of rough walls to scale with the
fraction of wall collisions, we should likewise expect a similar
density dependence of the ratio $\DTW/\DSW$.

\begin{figure}[!h]
  \centering
  \includegraphics[width=3.25in]{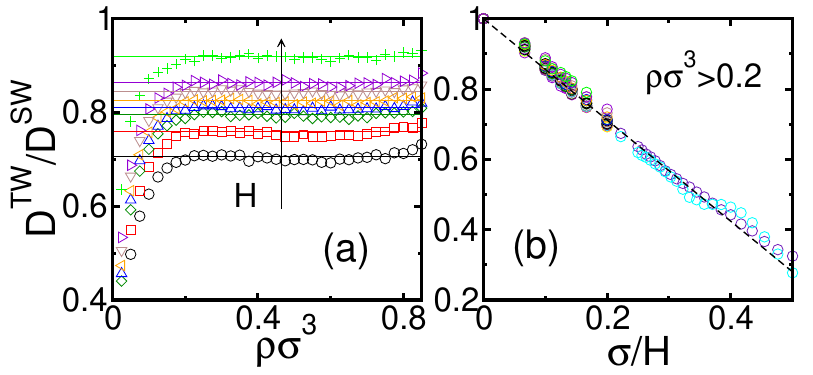}
  \subfloat{\label{fig:thermal_D_by_Dsmooth_vs_density}}
  \subfloat{\label{fig:thermal_D_by_Dsmooth_vs_Hinv}}
  \caption{Ratio of the self-diffusivity between thermal walls $\DTW$ to
  its value when confined between smooth walls $\DSW$ at the same
  $\rho$ and $H$ versus (a) $\rho$ and (b) $H^{-1}$. Symbols in (a)
  are the same as those in Fig.~\ref{fig:thermal_slit_D_vs_density}.  In (b), all data
  is for $\rho\sigma^3>0.2$, and dashed line is a fit to the form
  $\DTW/\DSW=1.0-1.44\sigma/H$.}
\label{fig:thermal_D_by_Dsmooth}
\end{figure}

In Fig.~\ref{fig:thermal_D_by_Dsmooth_vs_density}, we plot the quantity $\DTW/\DSW$ 
as a function of density. The ratio of the diffusivities is taken at the same 
thermodynamic state of the fluid, namely at the same density $\rho$ and average wall 
separation $H$. Notice that, for 
each pore width, the ratio of self-diffusion coefficients initially increases 
with density before reaching a limiting value.
Furthermore, the behavior of $\DTW/\DSW$ vs $\rho$ reflects that of
$\fw$ vs $\rho$.  Again, this points to the strong
connection between the self-diffusion in a system with rough walls to
the prominence of the walls (i.e., the fraction of the fluid near the walls).  The diffusivity ratio takes its
smallest value at low density, conditions where surface collisions are
most influential, and thus, where the greatest difference between the
two surfaces is observed.  Also, the density at which the ratio
reaches a plateau value ($\rho\sigma^3\approx0.2$) is independent of
pore width.  This is due to the dominant influence of
particle-particle collisions on the mobility of hard-spheres at
moderate-to-high densities.

The most striking feature of Fig.~\ref{fig:thermal_D_by_Dsmooth_vs_density} involves the 
$H$-dependent plateau values for $\rho \sigma^3 \gtrsim 0.2$. 
In Fig.~\ref{fig:thermal_D_by_Dsmooth_vs_Hinv}, we plot all of the data points
for which $\rho \sigma^3 \gtrsim 0.2$ versus the inverse pore width
$H^{-1}$. Fig.~\ref{fig:thermal_D_by_Dsmooth_vs_Hinv} shows that the 
ratio of diffusivities can be well described in this density range 
by a linear function of $H^{-1}$
\begin{equation}
  \label{eq:Drough_Dsmooth}
  \frac{\Drough}{\Dref}=1-C \left(\frac{\sigma}{H}\right),
\end{equation}
where the constant $C$ is a fitting parameter, $\Drough$ is the 
self-diffusivity of the fluid between the rough surface of interest, 
and  $\Dref$ is the self diffusivity of the fluid at an 
appropriately chosen reference state under the same thermodynamic
conditions.  We note that 
Eq.~\eqref{eq:Drough_Dsmooth} has, to first order, the same dependence
on $H$ as the self-diffusivity of a Brownian particle in the center of a
slit-pore.\cite{Happel1983Low-Reynolds-Nu}  This, in fact, was the
inspiration for the functional form chosen in Eq.~\eqref{eq:Drough_Dsmooth}.
 For the thermal walls, we take 
$\Drough=\DTW$ and $\Dref=\DSW$, and 
we find that $C \approx 1.44\sigma$. 

\begin{figure}[!h]
  \centering
  \includegraphics[width=3in]{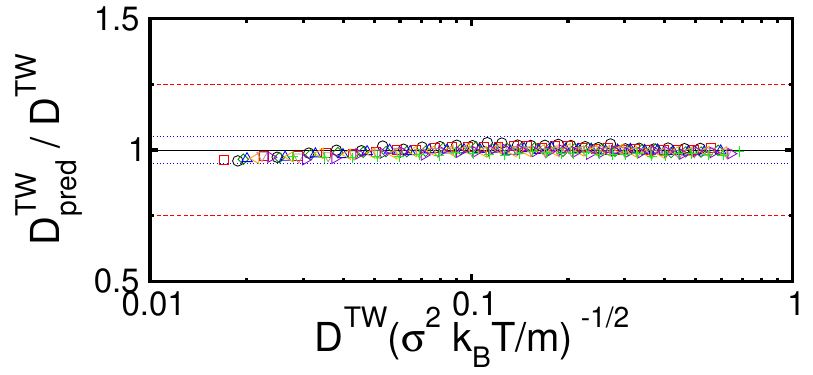}
  \caption{Ratio of predicted to observed self diffusivity (see text) for the thermal wall system.  Dotted
    blue and red dashed
    lines provide $5\%$ and $25\%$ error bounds, respectively.
    Symbols are the same as those in Fig.~\ref{fig:thermal_slit_D_vs_density}.}
  \label{fig:DTW_prediction}
\end{figure}

Assuming sufficient smooth-wall self-diffusivity data are available as a 
function of density, this provides the basis for estimating the diffusivity 
of the hard-sphere fluid confined between thermal walls for 
$\rho \sigma^{3} > 0.2$. To estimate the diffusivity of a fluid confined 
between thermal walls in a slit pore of width $H$ 
at density $\rho$, one calculates two quantities. The first quantity is the 
$H$-dependent ratio $\DTW/\DSW$ using Eq.~\ref{eq:Drough_Dsmooth}, 
and the second quantity is the self-diffusion coefficient 
of the reference system, $\DSW$, which, as discussed in
Section~\ref{sec:Introduction}, may be approximated by a scaling analysis. 
Knowledge of these two quantities then allows for an estimate of 
the self diffusivity between thermal walls. In Fig.~\ref{fig:DTW_prediction}, 
we test the predictive ability of this approach. The ratio of the predicted
to observed thermal wall self diffusivity is plotted against the
observed self diffusivity for the thermal-wall system.
If the predictions were perfect, the points would fall on the horizontal line 
$\DTW/\DSW = 1$, and this is clearly not the case. However, the predicted 
data points do fall within the indicated $5\%$ error bounds.
Moreover, we emphasize that to truly make predictions, one must still have
knowledge of $C$ and the reference self-diffusivity.

\subsection{Rotational walls}
\label{sec:results_rotation_walls}

\begin{figure}[h!]
  \centering
  \includegraphics[width=3in]{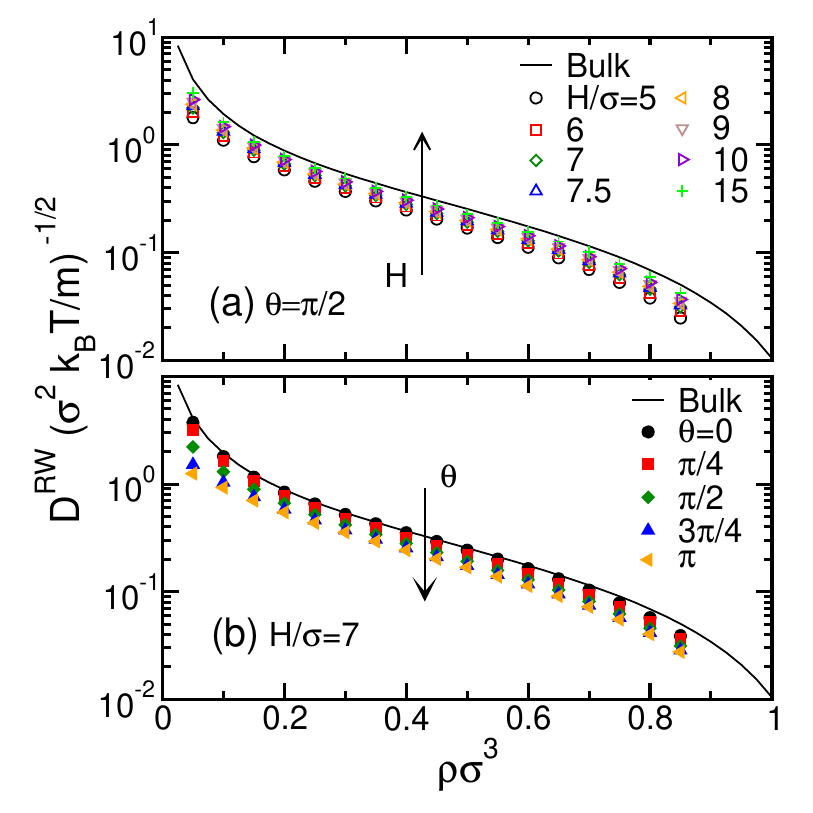}
  \subfloat{\label{fig:rotation_D_vs_density_theta_pi_2}}
  \subfloat{\label{fig:rotation_D_vs_density_Gap_7}}
  \caption{Self diffusivity $\DRW$ versus density $\rho$ for the
    hard-sphere fluid confined between rotational walls at (a)
    $\theta=\pi/2$ and various wall separations $H$ and (b)
    $H/\sigma=7$ and various values of $\theta$.}
  \label{fig:rotation_D_vs_density}
\end{figure}

In Figure~\ref{fig:rotation_D_vs_density}, we present results for the
self-diffusion coefficient $\DRW$ of the hard-sphere fluid confined
between rotational walls as a function of density $\rho$ with (a) $\theta=\pi/2$ and
various values of $H$, and (b) $H/\sigma=7.0$ and various values of
$\theta$. Qualitatively similar results were observed for other
choices of rotational-wall parameters and can be found in
Supplementary Materials.\cite{supplementary_material} Recall that $\theta$ is the angle of rotation
that the lateral velocity of a particle undergoes after it collides
with the surface.  Fig.~\ref{fig:rotation_D_vs_density_theta_pi_2}
shows that self diffusivity decreases with increasing density at fixed $H$ and
fixed $\theta$. In addition, at fixed density $\rho$, the
self-diffusivity decreases with decreasing $H$.
This latter trend, which was also observed in the
the thermal-wall systems, is not surprising considering
rotational walls, like thermal walls, inherently retard particle mobility, 
an effect that is most apparent at small wall separations (at the same fluid density).
Therefore, recalling that the collision statistics are independent
of the boundary condition for featureless walls, the same physics
explaining the reduction in mobility due to thermal walls mentioned in
Section~\ref{sec:results_thermal_walls} also applies to rotational
walls.  We also expect that the reduction in mobility should increase 
with $\theta$, since $\theta$ controls the effective roughness of the
walls. The data presented Figure~\ref{fig:rotation_D_vs_density_Gap_7} bear this out.

\begin{figure}[h!]
  \centering
  \includegraphics[width=3.in]{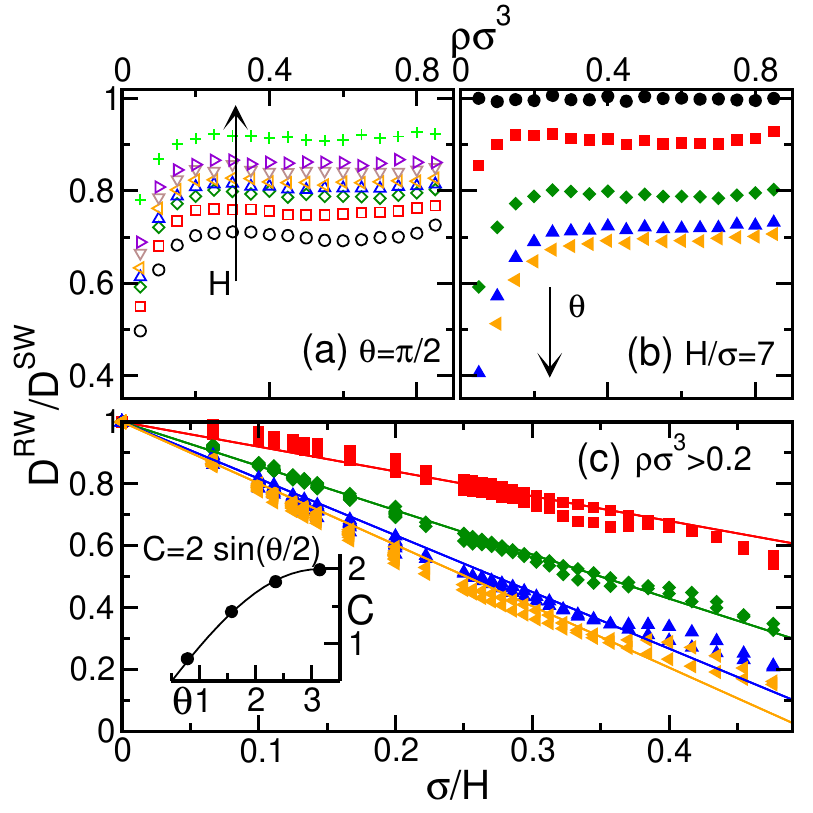}
  \subfloat{\label{fig:rotation_D_by_Dsmooth_vs_density_theta_pi_2}}
  \subfloat{\label{fig:rotation_D_by_Dsmooth_vs_density_Gap_7}}
  \subfloat{\label{fig:rotation_D_by_Dsmooth_vs_Hinv}}
  
  \caption{Ratio of self diffusivity between rotational walls $\DRW$ to
    that between smooth walls $\DSW$ at the same $\rho$ and
    $H$ versus $\rho$ at (a) $\theta=\pi/2$ and various $H$ and (b)
    $H/\sigma=7$ and various $\theta$, and versus (c) $H^{-1}$ at
    various $\theta$.  Symbols in (a) and (b-c) are the same as those
    in Figs.~\ref{fig:rotation_D_vs_density_theta_pi_2} and
    \ref{fig:rotation_D_vs_density_Gap_7}, respectively.  In (c), all
    data points are for $\rho\sigma^3>0.2$; lines are linear fits of
    the form $\DRW/\DSW=1-C (H/\sigma)^{-1}$; the inset to (c) shows
    the value of $C$ as a function of $\theta$ from linear fits in
    main panel of (c).}
  \label{fig:rotation_D_by_Dsmooth}
\end{figure}

\begin{figure}[!h]
  \centering
  \includegraphics[width=3in]{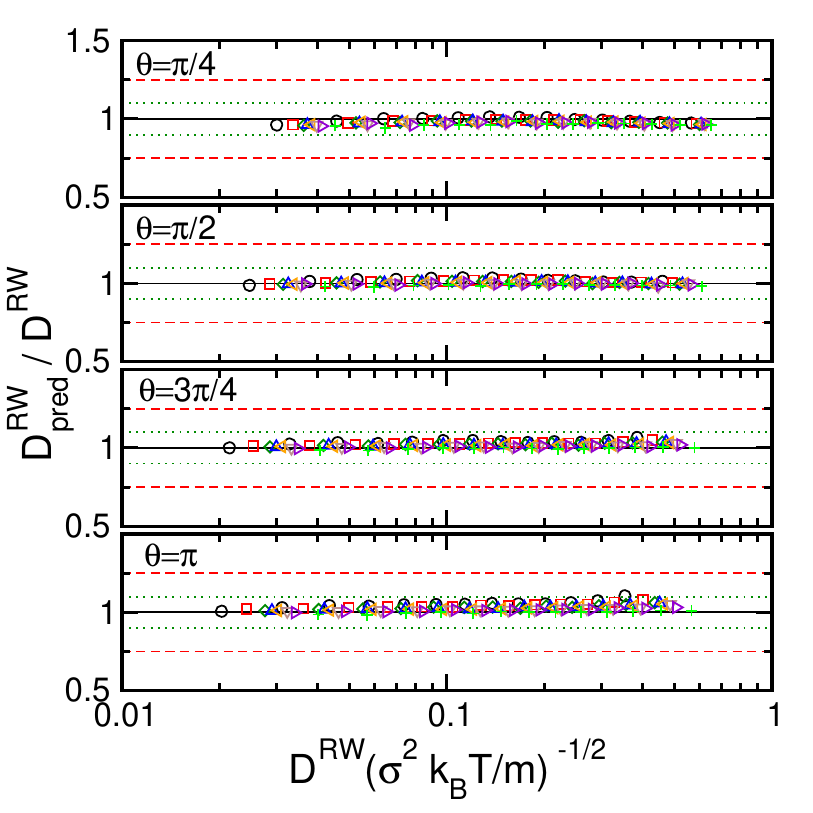}
  \caption{Ratio of predicted to observed self diffusivity (discussed
    in text) versus the observed self diffusivity for the rotational wall system with
    the value of $\theta$ denoted in each figure. Dotted blue and red dashed lines
    provide $10\%$ and $25\%$ error bounds, respectively.  Symbols have
    the same meaning as in
    Fig.~\ref{fig:rotation_D_vs_density_theta_pi_2}.}
  \label{fig:rotation_D_prediction}
\end{figure}

Figures~\ref{fig:rotation_D_by_Dsmooth_vs_density_theta_pi_2} and
\ref{fig:rotation_D_by_Dsmooth_vs_density_Gap_7} again show that
rotational walls influence self diffusivity relative to smooth walls in a
manner qualitatively similar to thermal walls.  Specifically, the
ratio of self diffusion coefficients $\DRW/\DSW$ initially increases
with density and then levels off beyond $\rho \sigma^3 \gtrsim
0.2$. In fact, for a given $\theta$, the plateau value appears to be
systematically dependent upon $H$, which suggests that an approach
similar to that adopted for the thermal wall system can be used to
estimate the diffusivity in the rotational-wall system.
Figure~\ref{fig:rotation_D_by_Dsmooth_vs_density_Gap_7} shows that the
ratio $\DRW/\DSW$ decreases with increasing $\theta$, indicating that
the constant $C$ in Eq.~\eqref{eq:Drough_Dsmooth} is dependent upon
$\theta$.  This just reflects the $\theta$ dependence noted in
Fig.~\ref{fig:rotation_D_vs_density_Gap_7}.  Accounting for this $\theta$
dependence, Fig.~\ref{fig:rotation_D_by_Dsmooth_vs_Hinv} shows 
that diffusivity ratios with densities $\rho \sigma^3 > 0.2$ are 
well described by Eq.~\eqref{eq:Drough_Dsmooth}. The inset to
Fig.~\ref{fig:rotation_D_by_Dsmooth_vs_Hinv} displays the values of
$C(\theta)$ from the fits. For simplicity, we fit $C$ to the function
$C=2.0 sin(\theta/2)$, which as shown in
Fig.~\ref{fig:rotation_D_by_Dsmooth_vs_Hinv} fits the data well. 
%new
specific form of the fit was based on the fact that $C=0$ for
$\theta=0$ (by definition), $C$ is a maximum at $\theta=\pi$, and
$C$ is periodic in $\theta$.
Using the procedure outlined above for thermal walls, 
the self-diffusion of the hard-sphere fluid between rotational walls can be 
estimated in an analogous fashion. Figure~\ref{fig:rotation_D_prediction} 
tests these estimates, plotting the ratio $\DRW_{\RM{pred}}/\DRW$ versus
$\DRW$, where $\DRW_{\RM{pred}}$ is the predicted value and $\DRW$ is
the value observed from simulation. In the case of rotational walls, the estimation
procedure yields predictions that fall within $10\%$ of the actual
values.  More accurate predictions can clearly be obtained by making
use of a more quantitative model for $C(\theta)$.

\subsection{Physically rough walls}
\label{sec:results_physical_roughness}

Here, we present results for physically rough surfaces at fixed
wavelength $\lambda / \sigma = 3$ and several densities $\rho = N /
V$, where $V$ is the surface accessible volume, while (1) fixing the amplitude of the features
$a_w/\sigma=1.0$ and varying $H$ and (2) fixing $H/\sigma=7.5$ and
varying $a_w$. Since the surface height of a physically rough wall
depends on lateral position, these parameters (see
Eq.~\ref{eq:wavy_walls} and Fig.~\ref{fig:schematic_wavy}) allow us to
study the influence of surface height variation on mobility in a
systematic way. Results for $\lambda/\sigma=1.5$, $2.0$, and $6.0$
give results qualitatively similar to those presented below, and can be
found in Supplementary Materials.\cite{supplementary_material}

\begin{figure}[!h]
  \centering
  \includegraphics[width=3.25in]{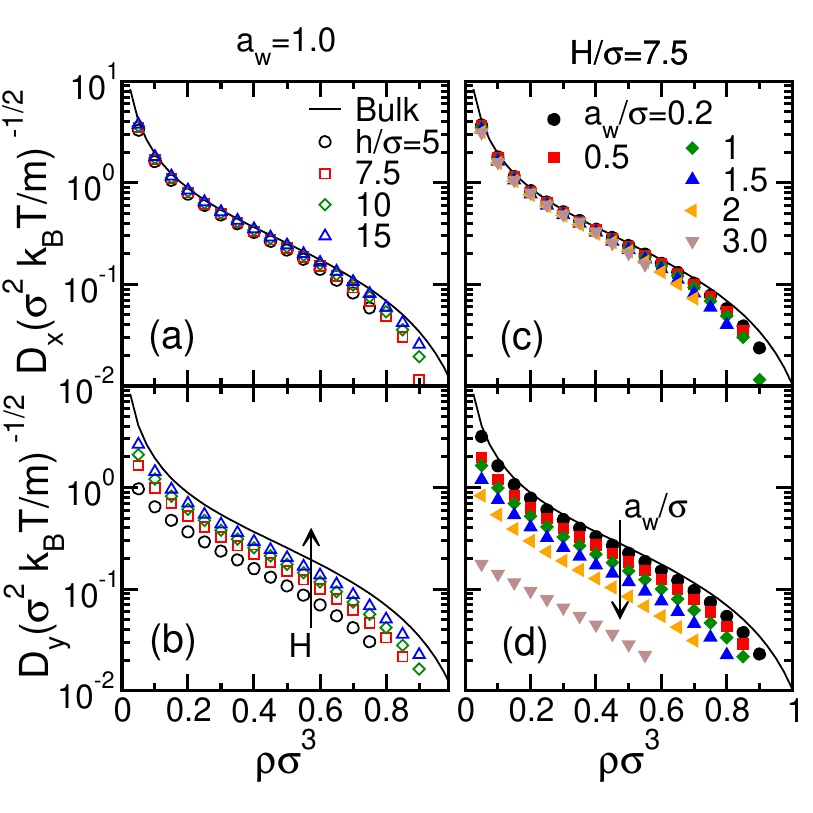}
  \subfloat{\label{fig:wave_DX_vs_dens_A_1_WL_3}}
  \subfloat{\label{fig:wave_DY_vs_dens_A_1_WL_3}}

  \subfloat{\label{fig:wave_DX_vs_dens_H_7p5_WL_3}}
  \subfloat{\label{fig:wave_DY_vs_dens_H_7p5_WL_3}}

  \caption{Self diffusivity in (a,c) smooth $x$  and (b,d) rough $y$ directions versus density for the hard-sphere fluid
    confined between physically rough walls (see
    Fig.~\ref{fig:schematic_wavy}) with $\lambda/\sigma=3.0$ and
    (a,b) $a_w/\sigma=1.0$ and various $H$ and (c,d) $H/\sigma=7.5$
    and various $a_w$.}
  \label{fig:wavy_D_vs_density}
  
\end{figure}

Notice that in the geometry of the physically rough wall system
studied (see Fig.~\ref{fig:schematic_wavy}), the $x$ and $y$
directions, though both infinite, are not equivalent. Because the
surface height is a function of $y$ only, the $y$ direction is termed
rough (inhomogeneous), while the $x$ direction is termed smooth (homogeneous).  
Also, because of this height variation, we expect the self
diffusivities in the $x$ and $y$ directions at the same state point to be significantly different, 
with $D_x > D_y$. Figure~\ref{fig:wavy_D_vs_density} shows these 
two self-diffusion coefficients separately as a function of density for
various values of $H$ and $a_w$.  Observe that the self diffusivity in
the $x$ direction has little dependence on $H$ or $a_w$. That is, 
the $D_x$--$\rho$ correlation is almost equivalent to the bulk 
correlation, much like the case for hard spheres between smooth flat walls (see
Fig.~\ref{fig:smooth_slit_D_vs_density}).  This seems consistent with
previous results for hard-spheres in cylindrical pores where the surface
is smooth and exhibits curvature.\cite{Goel2009Available-state}

In contrast, the self-diffusion coefficient in the $y$ direction
depends strongly on the degree of confinement
(Fig.~\ref{fig:wave_DY_vs_dens_A_1_WL_3}) and on the amplitude $a_w$ of
the surface variation (Fig.~\ref{fig:wave_DY_vs_dens_H_7p5_WL_3}).
Specifically, decreasing $H$  at fixed wall feature size
(constant $a_w$ and $\lambda$) systematically decreases $D_y$.  This behavior resembles 
that for the fluid confined between thermal walls
(Fig.~\ref{fig:thermal_slit_D_vs_density}) and rotational walls
(Figs.~\ref{fig:rotation_D_vs_density_theta_pi_2}).  Also, at fixed $H$,
$D_y$ systematically decreases with increasing $a_w$ (i.e., increasing
surface roughness).  This effect is analogous to increasing
surface roughness ($\theta$) in the rotational wall system (see
Fig.~\ref{fig:rotation_D_vs_density_Gap_7}).  Also, the slowing down of
dynamics due to physical roughness is consistent with previous 
studies.\cite{Sofos2010Effect-of-wall-}

\begin{figure}[!h]
  \centering
  \includegraphics[width=3in]{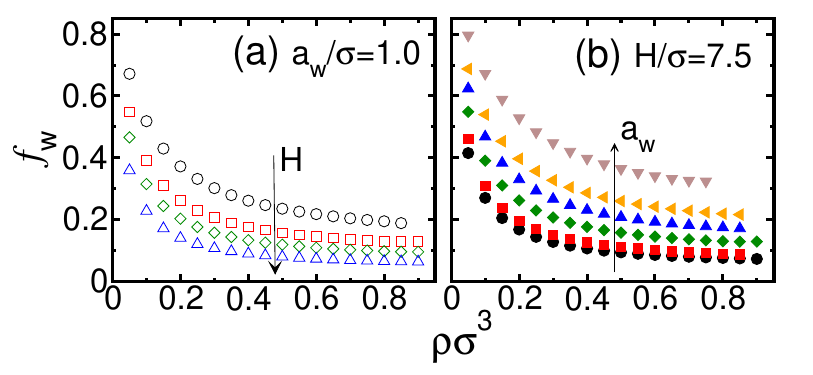}
  \subfloat{\label{fig:wavy_frac_wall_vs_den_A_1_WL_3}}
  \subfloat{\label{fig:wavy_frac_wall_vs_dens_H_7p5_WL_3}}
  \caption{Fraction of particle collisions that involve the wall $\fw$ for
    the hard-sphere fluid confined between physically rough walls versus density.  (a)
    $a_w/\sigma=1.0$ at various $H$ and (b) $H/\sigma=7.5$ and various
    $a_w$.  Symbols have the same meaning as in Fig.~\ref{fig:wavy_D_vs_density}.}
  \label{fig:wavy_frac_wall_vs_density}
\end{figure}

The disparity between $D_x$ and $D_y$ grows with decreasing $H$ at fixed $a_w$ 
and $\lambda$, and with increasing roughness ($a_w$) and fixed $H$.  Given the close
parallels between $D_y$ and the self-diffusion coefficient of the fluid 
between featureless rough walls, we expect the reduction in
$D_y$ due to surface roughness should be connected to the
fraction of collisions in the system involving the walls.
In Figure~\ref{fig:wavy_frac_wall_vs_density}, we plot $\fw$ versus density 
for the state points considered in Fig.~\ref{fig:wavy_D_vs_density}.
Fig.~\ref{fig:wavy_frac_wall_vs_den_A_1_WL_3} shows that $\fw$ grows 
with decreasing $H$ at fixed $\rho$ and $a_w$, 
while Fig.~\ref{fig:wavy_frac_wall_vs_dens_H_7p5_WL_3} shows that the 
wall-collision fraction grows with $a_w$ at fixed $\rho$ and $H$. In both 
cases, $\fw$ grows with the increasing prominence 
of the walls (caused by decreasing $H$, or increasing $a_w$)
Also, at fixed wall conditions (e.g., $a_w$ and $\lambda$), $\fw$ initially decreases with 
increasing density ($\lesssim 0.2$). However, this quantity changes only moderately with
further increases in density ($\gtrsim 0.2$).  This behavior is analogous to that 
observed in the featureless-wall systems.

\begin{figure}[!h]
  \centering
  \includegraphics[width=3.25in]{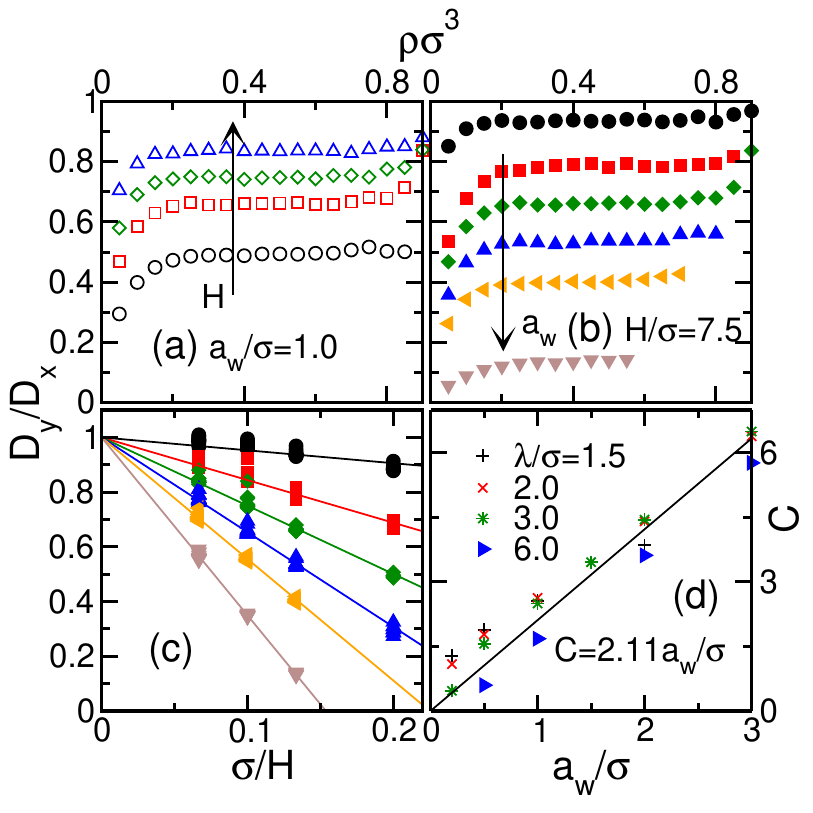}
  \subfloat{\label{fig:wavy_DY_by_DX_vs_dens_A_1}}
  \subfloat{\label{fig:wavy_DY_by_DX_vs_dens_h_7p5}}
  \subfloat{\label{fig:wavy_DY_by_DX_vs_Hinv_WL_3}}
  \subfloat{\label{fig:wavy_C_vs_A}}
  \caption{Ratio of self diffusivity in rough direction $D_y$ and smooth
    direction $D_x$ with $\lambda/\sigma=3.0$ versus (a,b) $\rho$ and
    (c) $H^{-1}$ for the hard-sphere fluid confined between physically
    rough walls.  Symbols in (a-c) are the same as in
    Fig.~\ref{fig:wavy_D_vs_density}.  In (c) all data points are for 
    $\rho\sigma^3>0.2$ and lines are fits to the form $D_y/D_x=1-C
    (H/\sigma)^{-1}$. Panel (d) displays the values of $C$ versus
    $a_w$ at all the wavelengths studied.}
  \label{fig:wavy_DY_by_DX}
\end{figure}

Because of the qualitative similarities between the properties of the
physically rough-wall system and the featureless rough-wall systems studied
above, we now examine the  diffusivity ratio $D_y/\Dref$. However, unlike the other 
surfaces encountered in this work, the choice of $\Dref$ is not obvious for 
the physically rough walls.  For the thermal and rotational walls, we chose the 
smooth (flat) wall self-diffusion coefficient at the same $\rho$ and
$H$, which corresponds to the self-diffusion of a fluid at the same thermodynamic
state without surface roughness.  Likewise, for physically rough-wall
system, the reference state should be the self-diffusion of a system
at the same thermodynamic state, but without surface roughness.  For
this, we choose $\Dref=D_x$. In Figure~\ref{fig:wavy_DY_by_DX_vs_dens_A_1},
we plot the ratio $D_y/D_x$ versus density for various values of $H$ and $a_w$.  
At fixed $H$, the ratio initially increases with $\rho$ and then levels 
off beyond $\rho\sigma\gtrsim 0.2$.  For a specified value of $\rho$, the 
ratio $D_y/D_x$ decreases with decreasing $H$ at fixed surface roughness
(Fig.~\ref{fig:wavy_DY_by_DX_vs_dens_A_1}) and with increasing
roughness at fixed $H$ (Fig.~\ref{fig:wavy_DY_by_DX_vs_dens_h_7p5}). Comparison with
Figs.~\ref{fig:thermal_D_by_Dsmooth_vs_density} and
\ref{fig:rotation_D_by_Dsmooth} shows that physical roughness alters $D_y$ 
relative to $D_x$ in a manner similar to how featureless rough walls alter the self-diffusion relative to
smooth walls. Figure~\ref{fig:wavy_DY_by_DX_vs_Hinv_WL_3} shows that, for given
surface features, $D_y/D_x$ is a linear function of $H^{-1}$.  That
is, $D_y$ follows Eq.~\eqref{eq:Drough_Dsmooth} with $\Drough=D_y$ and
$\Dref=D_x$.  Figure~\ref{fig:wavy_C_vs_A} shows the fit parameter $C$
versus $a_w$ for the different surface feature wavelengths $\lambda$
studied (see Supplementary material\cite{supplementary_material}).  Clearly, $C$ is much more
sensitive to $a_w$ than $\lambda$.  We find that a linear fit to $C$
as a function of $a_w$, although crude, describes that data reasonably
well.

\begin{figure}[!h]
  \centering
  \includegraphics[width=3.2in]{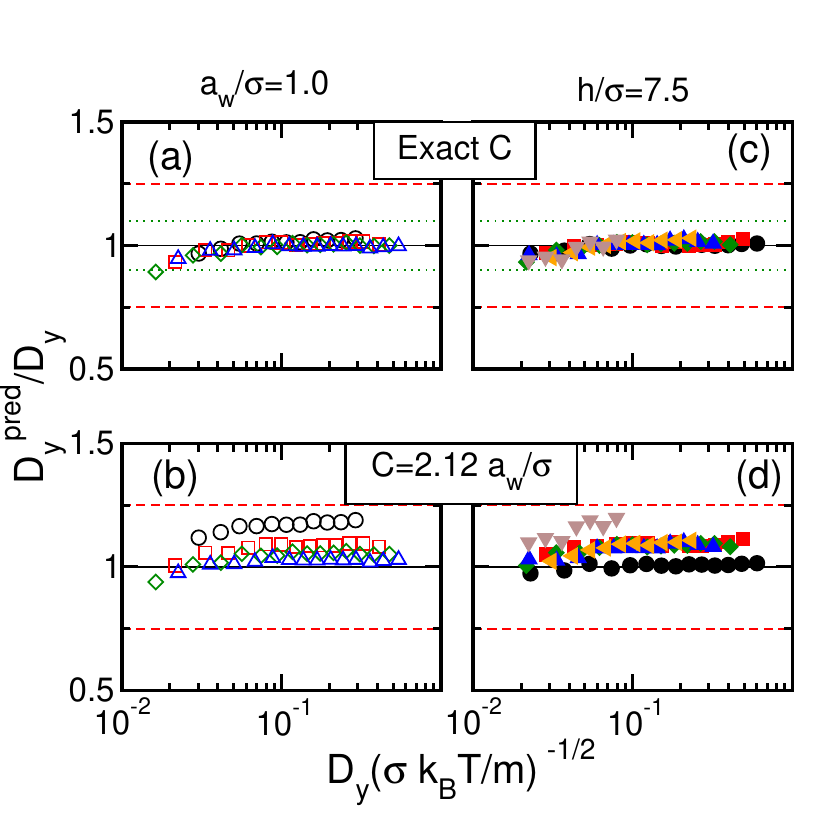}
  \subfloat{\label{fig:DY_prediction_exact_A_1}}
  \subfloat{\label{fig:DY_prediction_approx_A_1}}
  \subfloat{\label{fig:DY_prediction_exact_h_7p5}}
  \subfloat{\label{fig:DY_prediction_approx_h_7p5}}
  \caption{Ratio of predicted $D^{\RM{pred}}_y$ (see text) to observed
    value of $D_y$ plotted versus
    $D_y$ for the hard-sphere fluid confined between physically rough
    walls with    $\lambda/\sigma=3.0$ and (a,b) $a_w/\sigma=1$ and various $H$ and
    (c,d) $H/\sigma=7.5$ and various $a_w$.  Symbols in (a,b) and
    (c,d) correspond to those in
    Figs.~\ref{fig:wave_DX_vs_dens_A_1_WL_3} and
    \ref{fig:wave_DX_vs_dens_H_7p5_WL_3}, respectively. In (a,c) we use
    values of $C$ obtained from the fits in
    Fig.~\ref{fig:wavy_DY_by_DX_vs_Hinv_WL_3}, while in (b,d) we use
    $C=2.12a_w/\sigma$ (see Fig.~\ref{fig:wavy_C_vs_A}).  The dotted green and dashed red
    lines correspond to $10\%$ and $25\%$ error bounds, respectively.}
  \label{fig:wavy_DY_prediction}
\end{figure}

From the above analysis, we can now formulate a means to predict the
self-diffusion in the rough $y$ direction from the self diffusion in the
smooth $x$ direction.  Specifically, from Eq.~\eqref{eq:Drough_Dsmooth},
$D_y^{\RM{pred}}=D_x[1-C\,(H/\sigma)^{-1}]$.
Figure~\ref{fig:wavy_DY_prediction} shows the quality of the prediction
based on this formalism for the state points in
Fig.~\ref{fig:wavy_D_vs_density}.  In
Figs.~\ref{fig:DY_prediction_exact_A_1},\subref*{fig:DY_prediction_exact_h_7p5}
we use $C$ from the fits at a given $\lambda$ (i.e., data points in
Fig.~\ref{fig:wavy_C_vs_A}), while in
Figs.~\ref{fig:DY_prediction_approx_A_1},\subref*{fig:DY_prediction_approx_h_7p5}
we use the approximation $C=2.12a_w/\sigma$ (linear fit in
Fig.~\ref{fig:wavy_C_vs_A}).  We find that using the $C$ for a given
wall configuration yields good predictions, with the majority of
the data within $10\%$ of the actual data.  On the other hand, using
the approximate $C$ yields predictions within $25\%$ of the actual
value.  The information necessary to estimate $D_y$ in this
case is greatly reduced, but one would still need knowledge of $D_x$ to
truly make predictions of $D_y$.

\section{Conclusions}
\label{sec:conclusions}
We have systematically studied how surface roughness affects the 
self-diffusion of confined hard-sphere fluids. Specifically, for $\rho \sigma^3 > 0.2$, we have shown that 
the ratio of the self-diffusion coefficient in a rough wall system 
to the self-diffusion coefficient in smooth-wall reference system 
is a linear function of reciprocal wall separation [see Eq.~\eqref{eq:Drough_Dsmooth}], 
with the precise slope depending on the specific nature of the surface
roughness.  If this slope and the reference self-diffusivity
are known, accurate predictions of the self-diffusion in rough wall
systems can be obtained.

In real world applications, however, the specific nature of
surface roughness is often unknown.  This study leads to two possible ways
of making useful self-diffusivity estimates without this knowledge.  
First, our analysis shows that a only a limited number of measurements 
are needed to determine $C$. Second, if only qualitative information is sought, 
our analysis shows that $C$ is typically of $O(1)$, and that the basic physics of
self-diffusion in a rough system relative to that in a smooth system is
captured by the $H^{-1}$ correction in Eq.~\eqref{eq:Drough_Dsmooth}.
%new
That is, the specific nature of the surface roughness does not
  appear to impact radically or significantly the resulting dynamics.
  This is particularly important since modeling the specific
  details of real surfaces is a daunting task.  However, it is not our
  intention to suggest
that any type of surface roughness can be modeled in any way one
chooses, for example, modeling geometric roughness with featureless
rough walls.  Rather, the specific details of the surface roughness do
not appear to be overly important.
%end

To make predictions for self-diffusion of confined fluids between
rough surfaces using Eq.~\eqref{eq:Drough_Dsmooth}, one also requires
a prediction of the reference state self-diffusion.  As discussed in
Section~\ref{sec:Introduction}, one may use a scaling method to map
the bulk system to the confined system.  For the case of smooth
featureless walls, for example, this has been shown to be very effective.\cite{Goel2009Available-state}  However, for the case of the
physically rough system, one must be able to predict $D_x$.  We plan
to address this in a future publication.

%new
 Another question is whether the results presented in this paper
  apply to more complex pore scenarios, such as diffusion through porous
  materials.  Such systems have complex geometry and connectivity, as
  well as rough surfaces.  We believe that the simplified geometry
  studied here, slit pores, is applicable, however, when one considers
  that many of the properties in such complex systems can be considered
  as an average over a collection of simple slit pore (or cylindrical
  pore) systems. \cite{Lastoskie1993Pore-size-distr,Olivier1994Determination-o,Lastoskie1997Chapter-15.-Str,Rouquerol1999Adsorption-of-P,Klobes2006Porosity-and-sp}
%end

The lack of particle-particle or particle-wall attractions is an obvious shortcoming of this study.
Clearly, wall attractions will greatly alter the number of particles
near the wall, and therefore the fraction of wall collisions.  As
we point out, the ratio of self diffusivity in a rough-wall system to
that in a corresponding smooth-wall system appears to be closely linked with the
fraction of wall collisions.  Therefore, attractions can have a
strong effect on the behavior of the self diffusivity ratio.  However, we speculate
that the introduction of moderately strong wall attractions to the
rough systems studied here will result in an
increased retardation of self-diffusion and not a fundamental change to
the physics.  We also plan to explore this issue in detail in a future publication.

%roughness important
%\cite{Ravikovitch2006Density-Functio,Salazar2007A-Computational,Zhang0Quantitative-St}
%\cite{Gelb1999Phase-separatio}

%characterization

\section*{Acknowledgements}
\label{Acknowledgements}

W.P.K. gratefully acknowledges financial support from a National
Research Council postdoctoral research associateship at the National
Institute of Standards and Technology.  T.M.T. acknowledges support of
the Welch Foundation (F-1696) and the National Science Foundation
(CBET 1065357).  J.R.E. acknowledges financial support of the National
Science Foundation (CBET 0828979). The Biowulf Cluster at the National Institutes of
Health provided computational resources for this paper.

\newpage
\textbf{\Large{References and Notes}}

\newpage
\def\captionspace{\vspace{0.0in}}

\textbf{\Large{Figure Captions}}

\captionspace

\captionspace

\captionspace

\captionspace

\captionspace

\captionspace

\captionspace

\captionspace

\captionspace

\captionspace

\captionspace

\captionspace

\clearpage
\newpage
\appendix{}
\subsection*{Rotational walls}
\begin{figure*}[h]
  \centering
  \mbox{
    \subfloat[][$\theta=\pi/4$]{
      \includegraphics[width=0.2\textwidth]{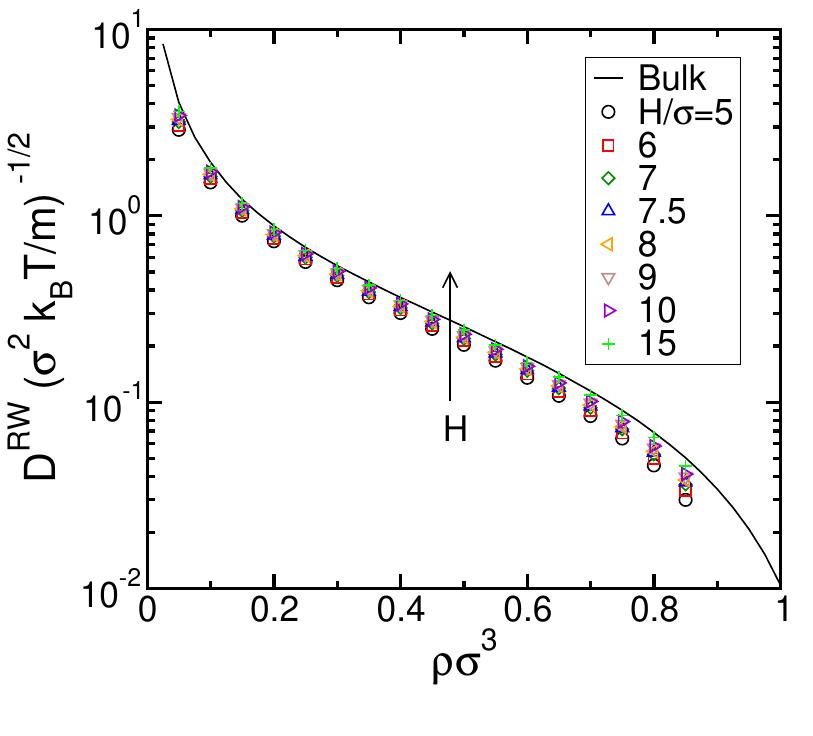}
      \label{fig:sup_rot_D_vs_dens_pi_4}
    }
    \subfloat[][$\theta=\pi/2$]{
      \includegraphics[width=0.2\textwidth]{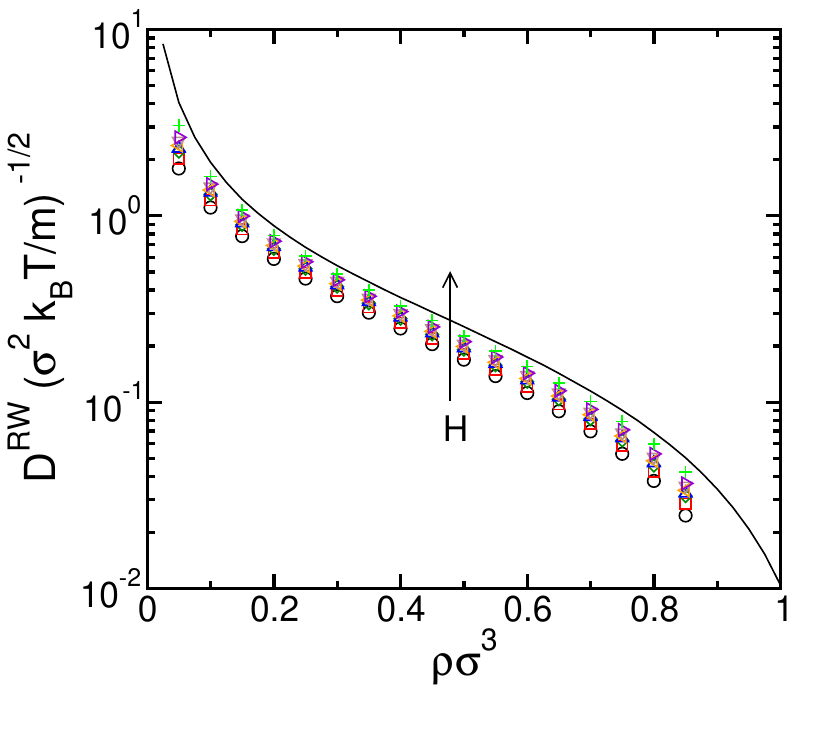}
      \label{fig:sup_rot_D_vs_dens_pi_2}
    }
    \subfloat[][$\theta=3\pi/4$]{
      \includegraphics[width=0.2\textwidth]{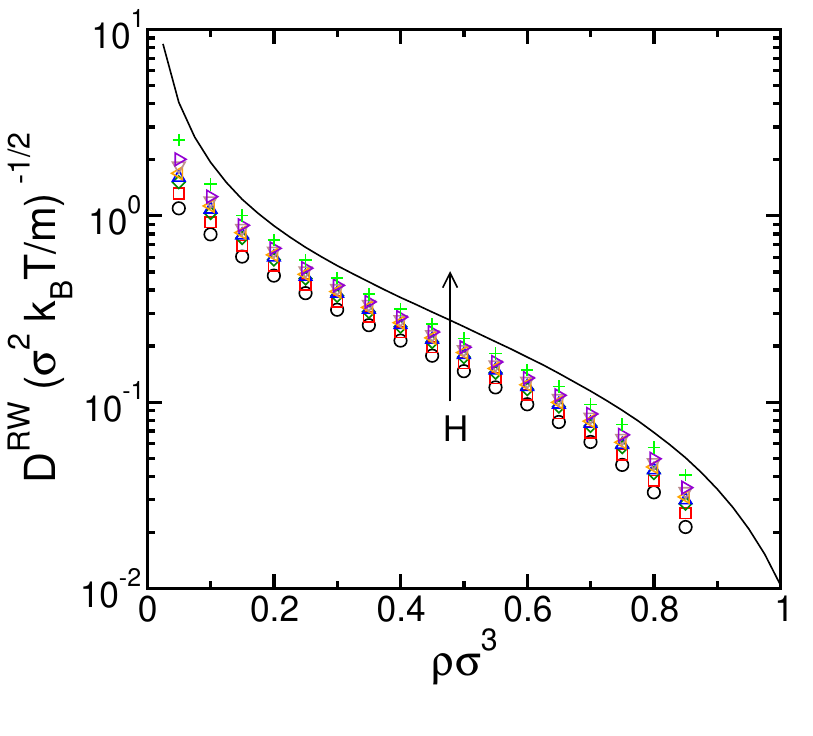}
      \label{fig:sup_rot_D_vs_dens_3pi_4}
    }
    \subfloat[][$\theta=\pi$]{
      \includegraphics[width=0.2\textwidth]{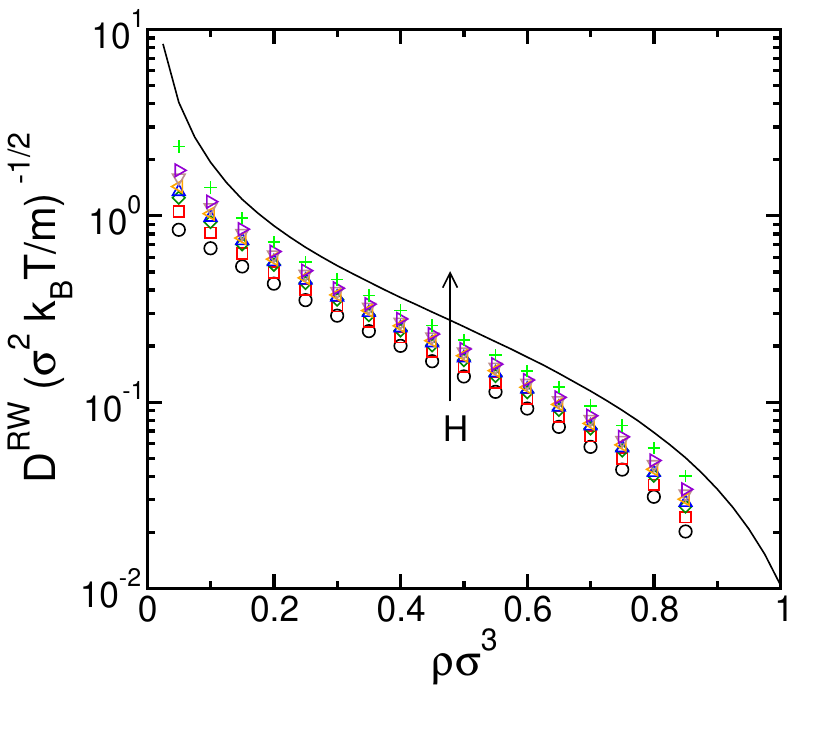}
      \label{fig:sup_rot_D_vs_dens_pi}
    }
  }

  \mbox{
    \subfloat[][$\theta=\pi/4$]{
      \includegraphics[width=0.2\textwidth]{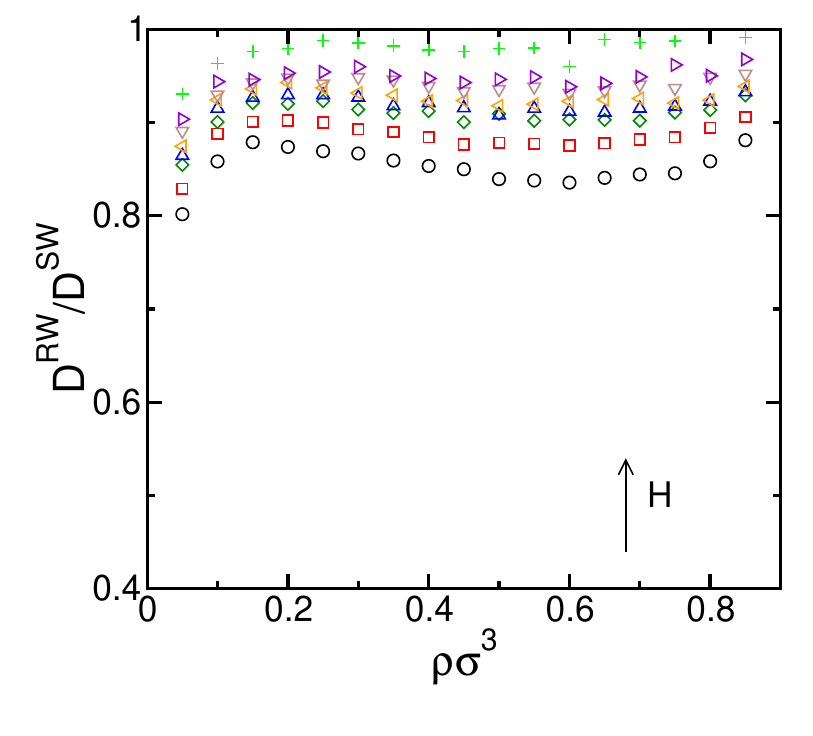}
      \label{fig:sup_rot_D_by_Dslit_vs_dens_pi_4}
    }
    \subfloat[][$\theta=\pi/2$]{
      \includegraphics[width=0.2\textwidth]{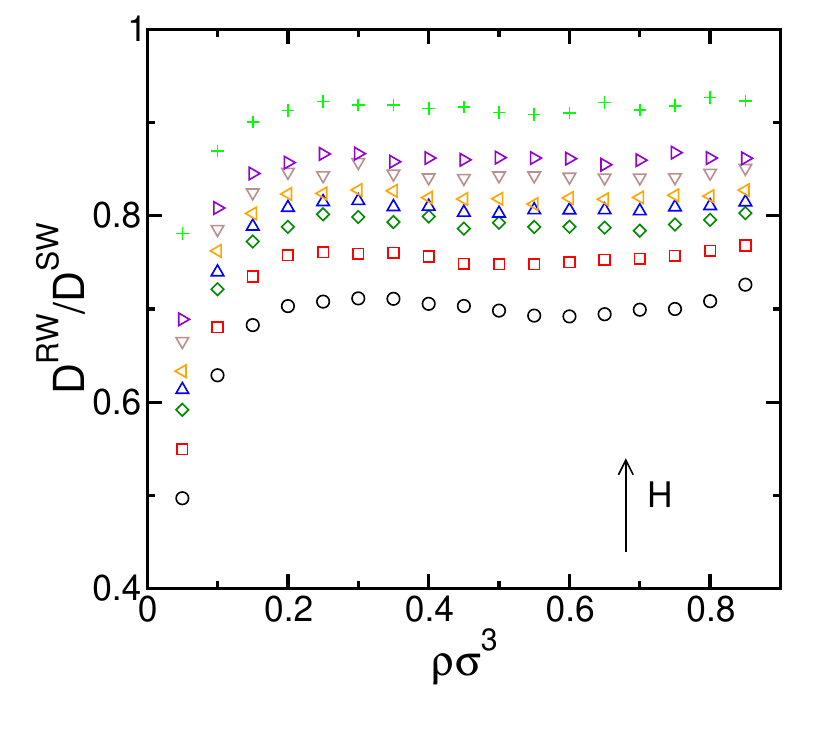}
      \label{fig:sup_rot_D_by_Dlsit_vs_dens_pi_2}
    }
    \subfloat[][$\theta=3\pi/4$]{
      \includegraphics[width=0.2\textwidth]{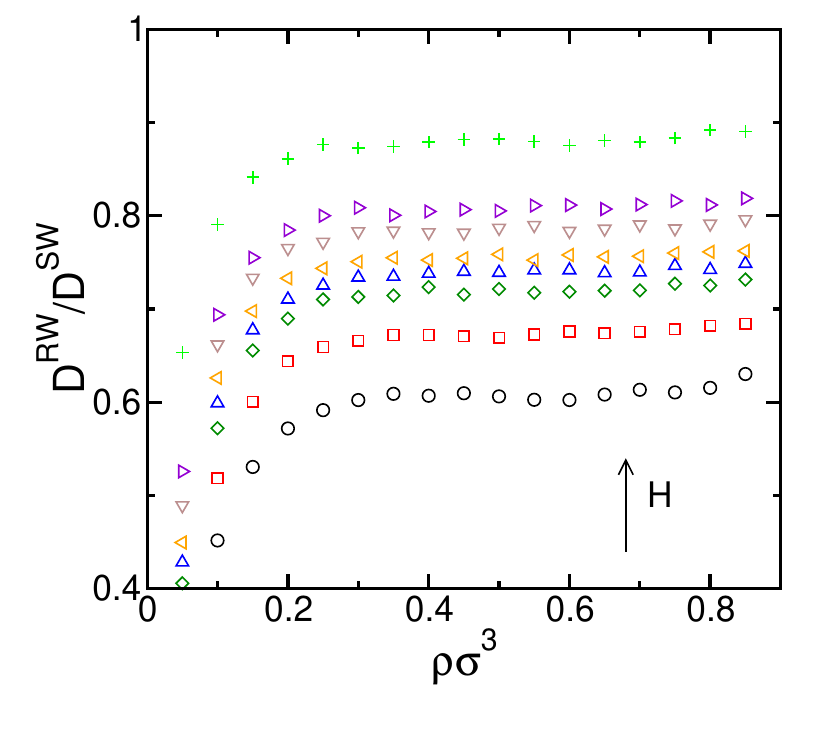}
      \label{fig:sup_rot_D_by_Dslit_vs_dens_3pi_4}
    }
    \subfloat[][$\theta=\pi$]{
      \includegraphics[width=0.2\textwidth]{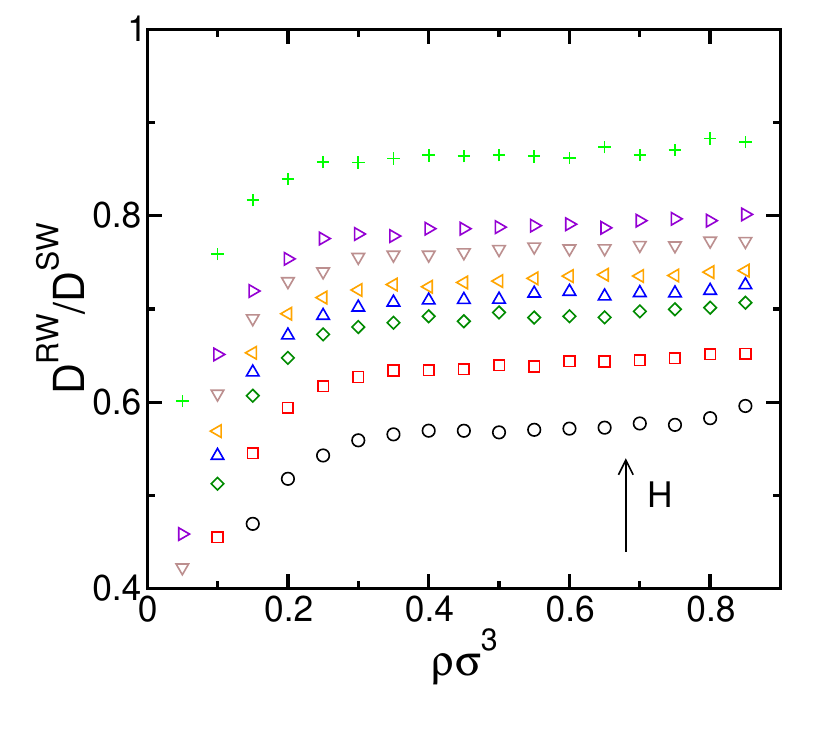}
      \label{fig:sup_rot_D_by_Dslit_vs_dens_pi}
    }
  }

  \caption{Properties for rotational walls. (a-d) Self-diffusion
    versus density. (e-h) Ratio of the self-diffusion between
    rotational walls and between smooth walls at some $H$ and $\rho$
    versus density.  Values of $\theta$ are indicted above each figure.}
  \label{fig:sup_rot_walls}
\end{figure*}

%\clearpage
\subsection*{Physically rough walls}

\begin{figure*}[t]
  \centering
  %DX
  \mbox{
    \subfloat[][$H/\sigma=5.0$]{
      \includegraphics[width=0.2\textwidth]{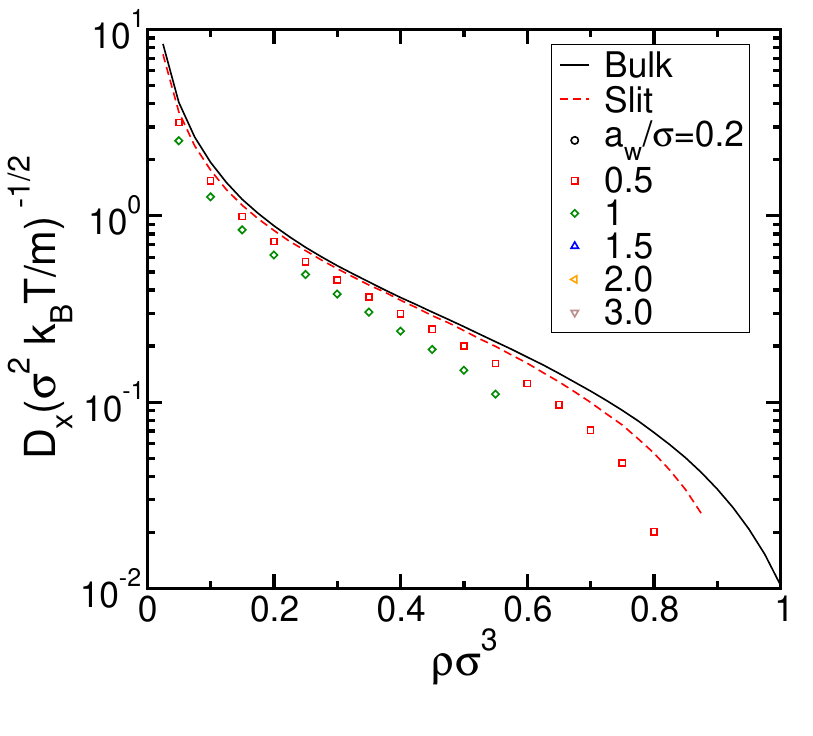}
      \label{fig:wavy_DX_vs_dens_h_5_WL_1.5}
    }
    \subfloat[][$H/\sigma=7.5$]{
      \includegraphics[width=0.2\textwidth]{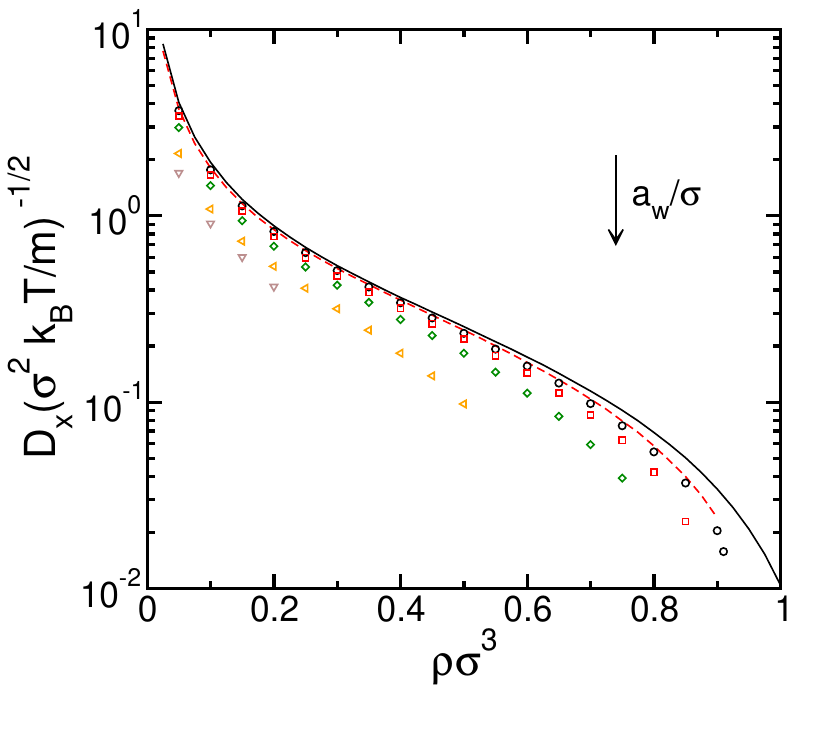}
      \label{fig:wavy_DX_vs_dens_h_7.5_WL_1.5}
    }
    \subfloat[][$H/\sigma=10.0$]{
      \includegraphics[width=0.2\textwidth]{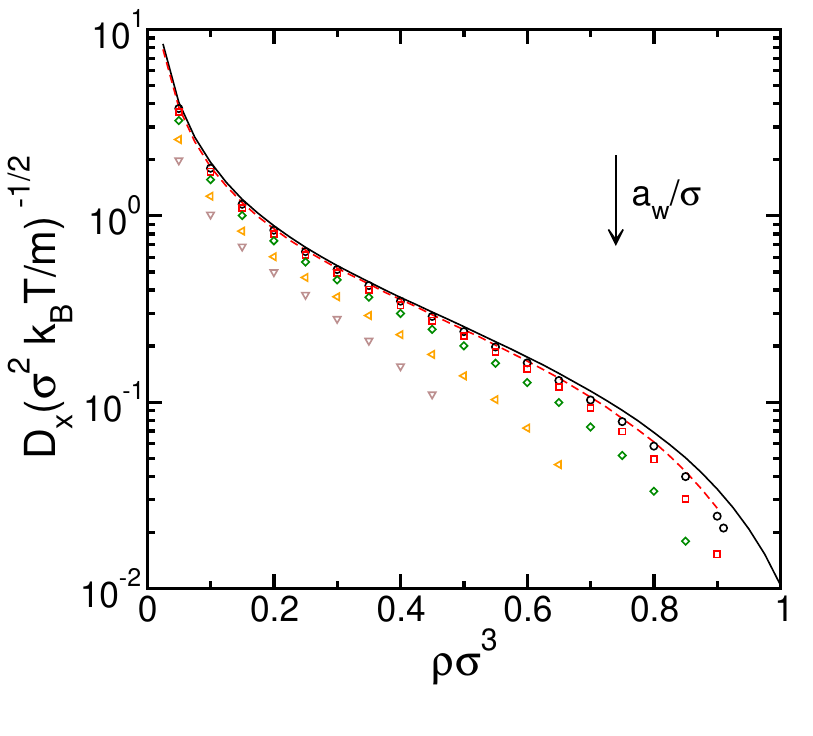}
      \label{fig:wavy_DX_vs_dens_h_10_WL_1.5}
    }
    \subfloat[][$H/\sigma=15.0$]{
      \includegraphics[width=0.2\textwidth]{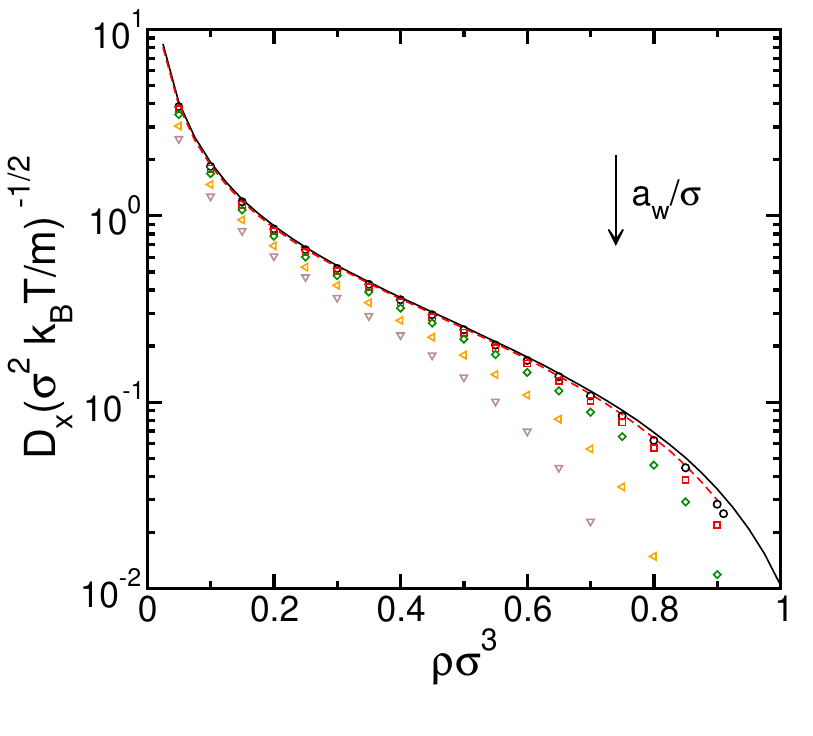}
      \label{fig:wavy_DX_vs_dens_h_15_WL_1.5}
    }

  }
  %DY
  \mbox{
    \subfloat[][$H/\sigma=5.0$]{
      \includegraphics[width=0.2\textwidth]{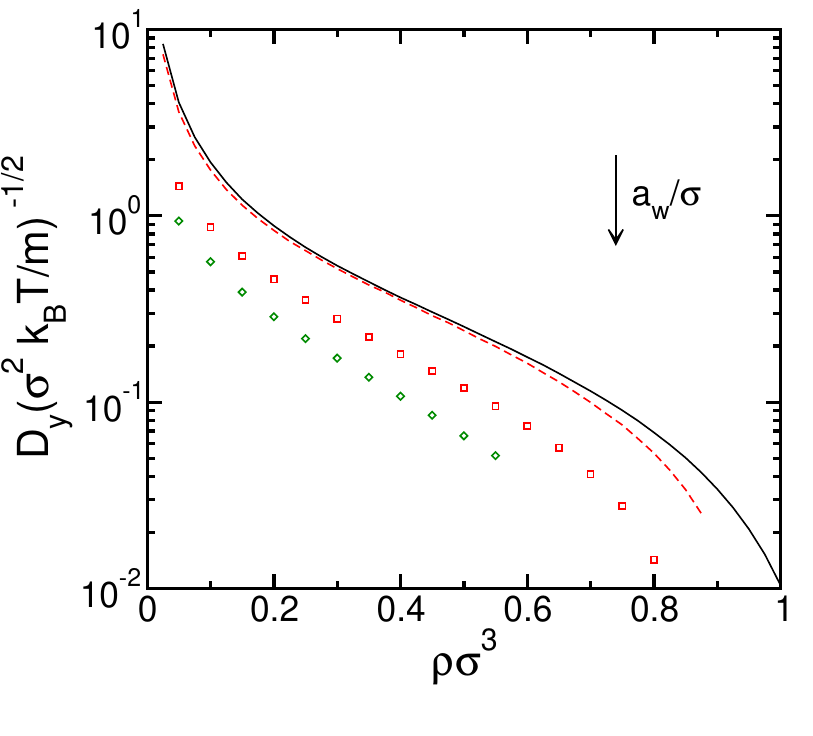}
      \label{fig:wavy_DY_vs_dens_h_5_WL_1.5}
    }
    \subfloat[][$H/\sigma=7.5$]{
      \includegraphics[width=0.2\textwidth]{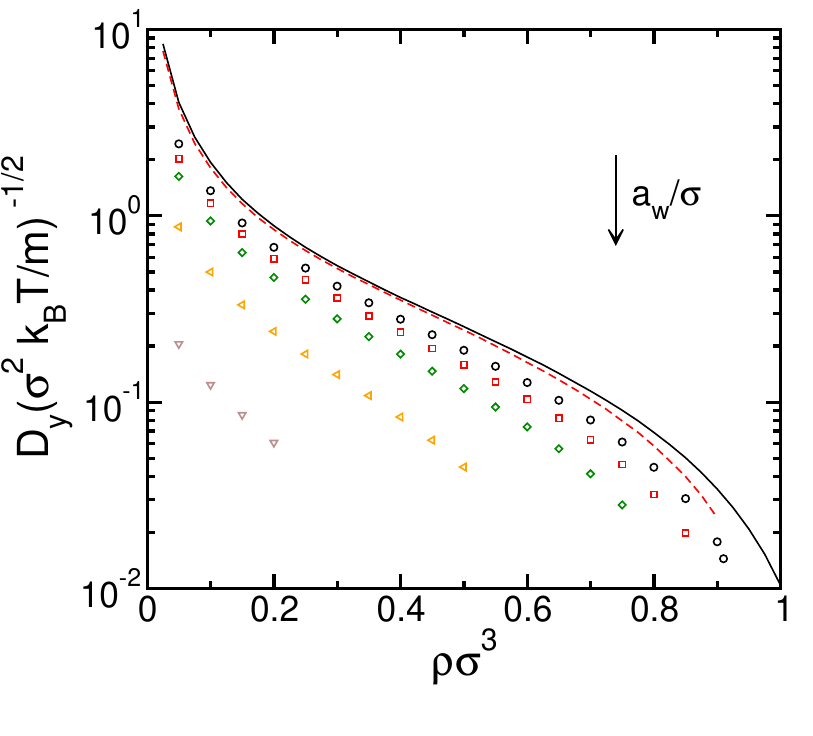}
      \label{fig:wavy_DY_vs_dens_h_7.5_WL_1.5}
    }
    \subfloat[][$H/\sigma=10.0$]{
      \includegraphics[width=0.2\textwidth]{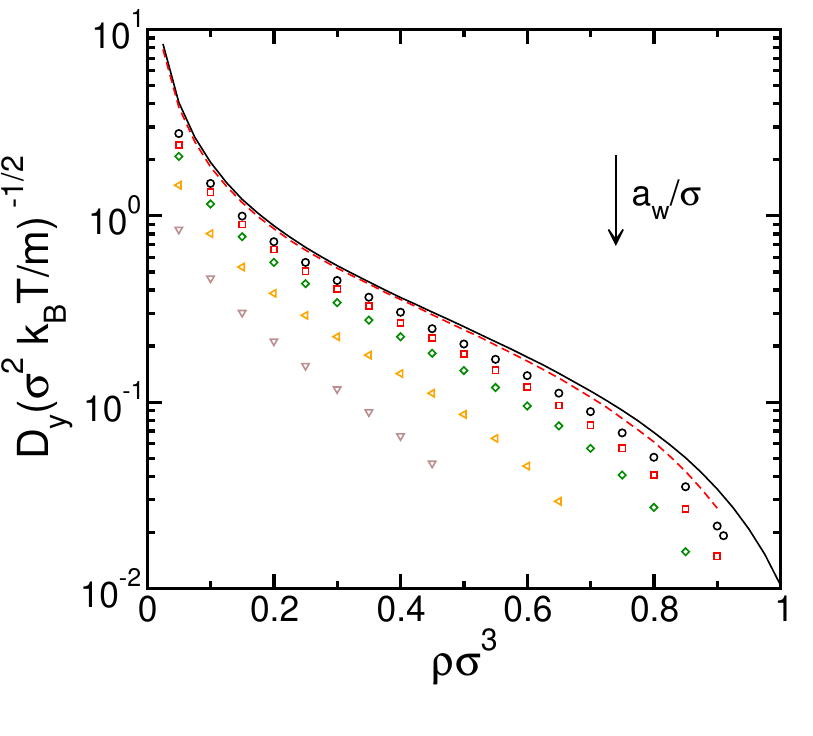}
      \label{fig:wavy_DY_vs_dens_h_10_WL_1.5}
    }
    \subfloat[][$H/\sigma=15.0$]{
      \includegraphics[width=0.2\textwidth]{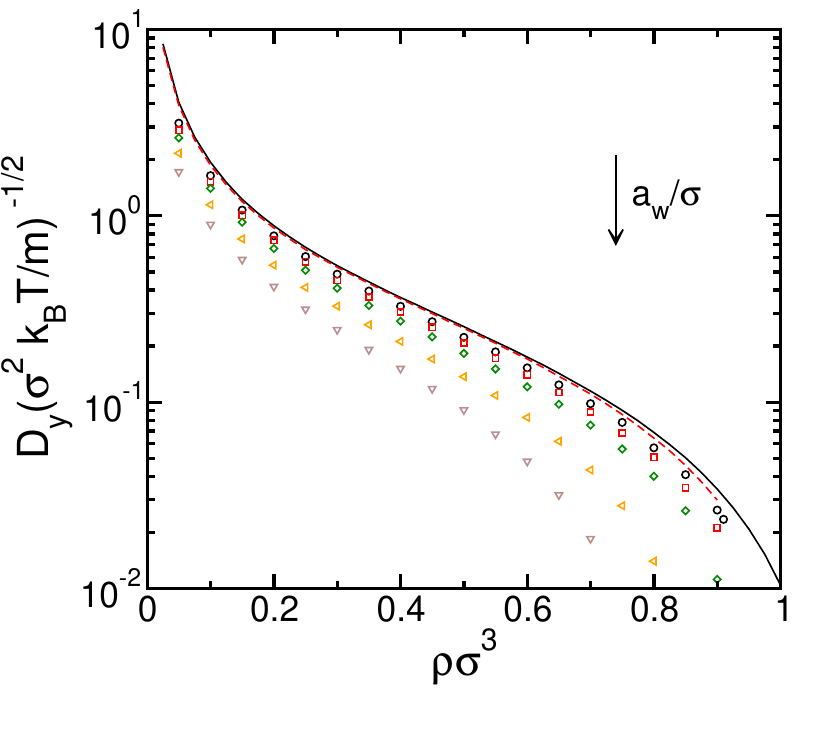}
      \label{fig:wavy_DY_vs_dens_h_15_WL_1.5}
    }
  }

  %DY/DX
  \mbox{
    \subfloat[][$H/\sigma=5.0$]{
      \includegraphics[width=0.2\textwidth]{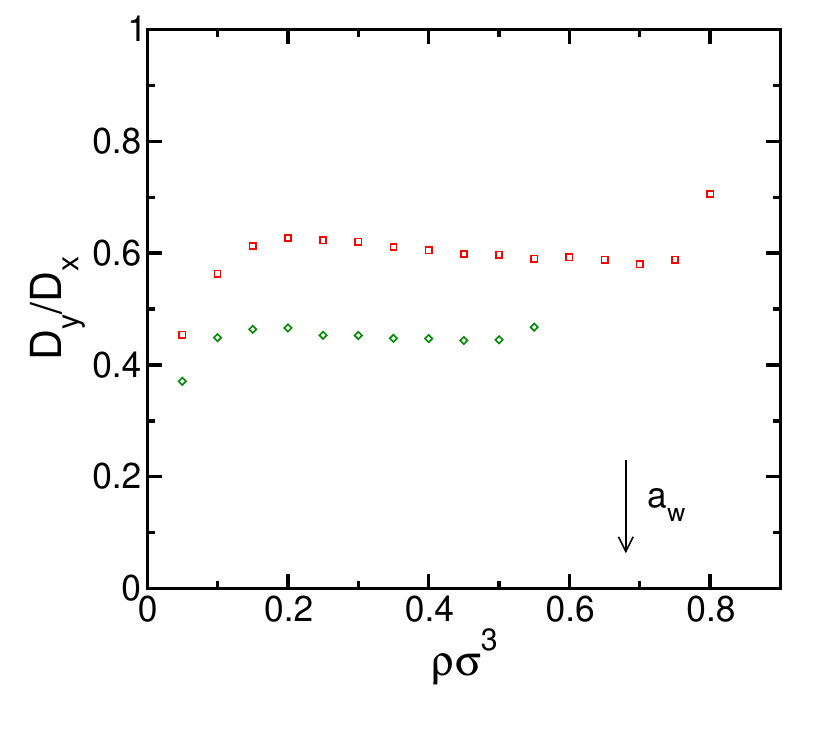}
      \label{fig:wavy_DY_by_DX_vs_dens_h_5_WL_1.5}
    }
    \subfloat[][$H/\sigma=7.5$]{
      \includegraphics[width=0.2\textwidth]{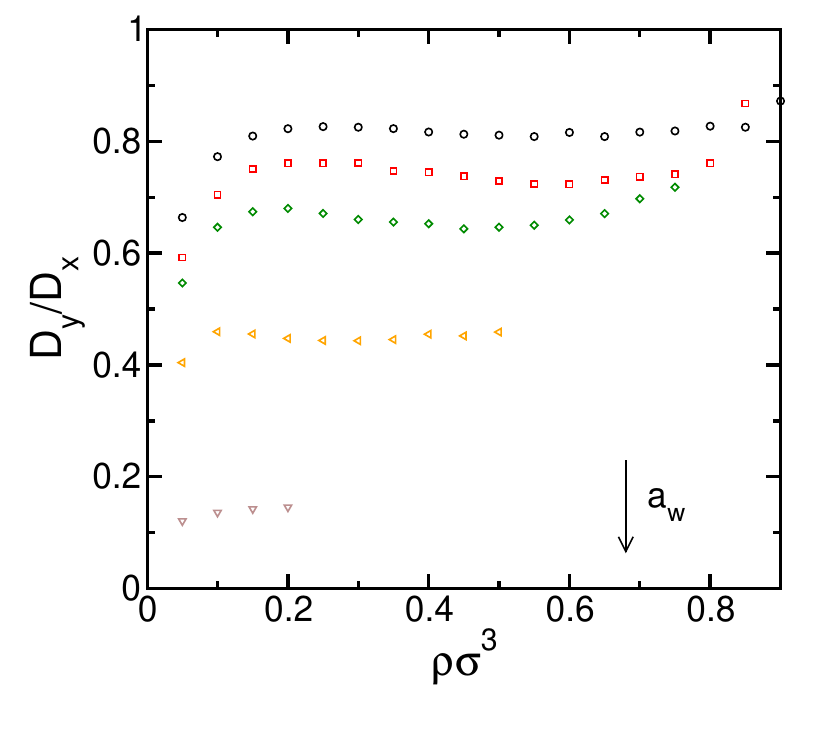}
      \label{fig:wavy_DY_by_DX_vs_dens_h_7p5_WL_1.5}
    }
    \subfloat[][$H/\sigma=10.0$]{
      \includegraphics[width=0.2\textwidth]{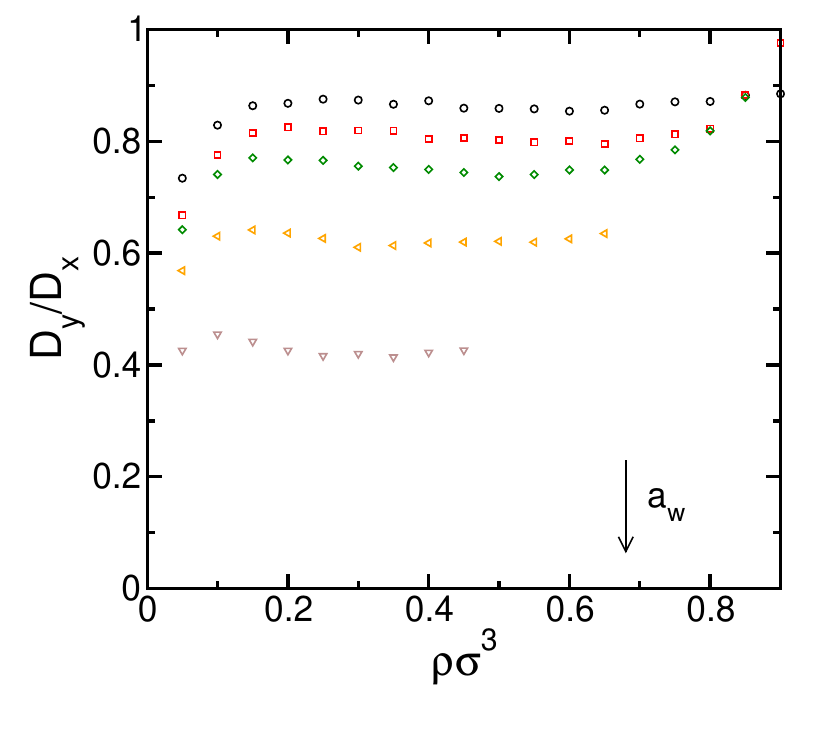}
      \label{fig:wavy_DY_by_DX_vs_dens_h_10_WL_1.5}
    }
    \subfloat[][$H/\sigma=15.0$]{
      \includegraphics[width=0.2\textwidth]{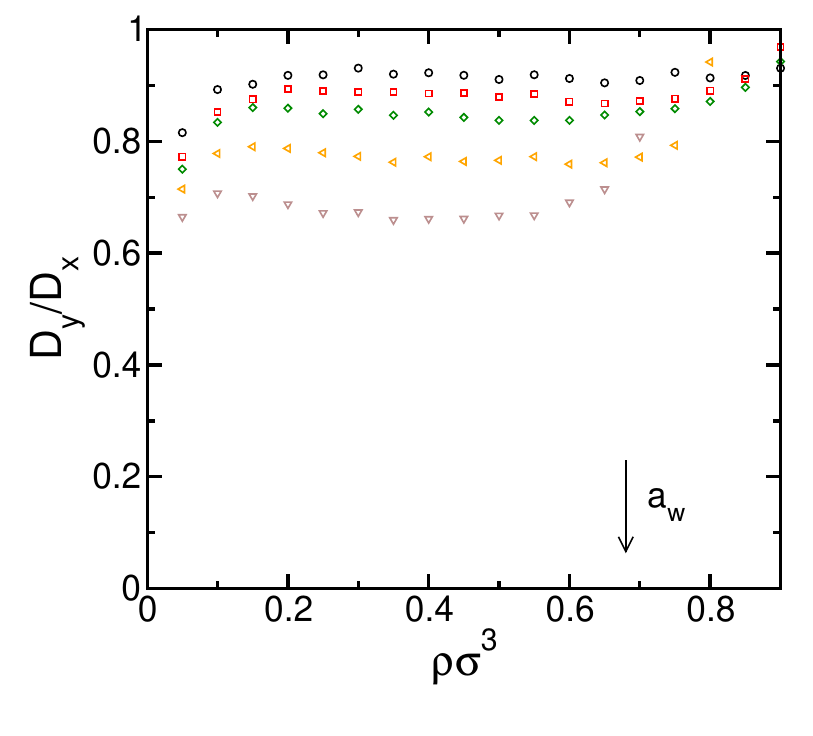}
      \label{fig:wavy_DY_by_DX_vs_dens_h_15_WL_1.5}
    }

  }

  %frac
  \mbox{
    \subfloat[][$H/\sigma=5.0$]{
      \includegraphics[width=0.2\textwidth]{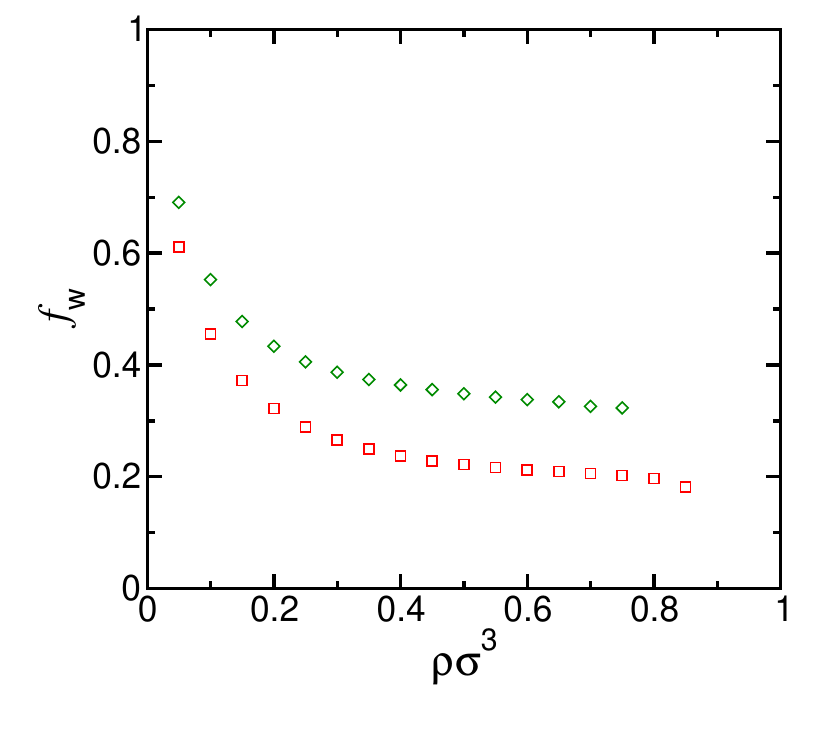}
    }
    \subfloat[][$H/\sigma=7.5$]{
      \includegraphics[width=0.2\textwidth]{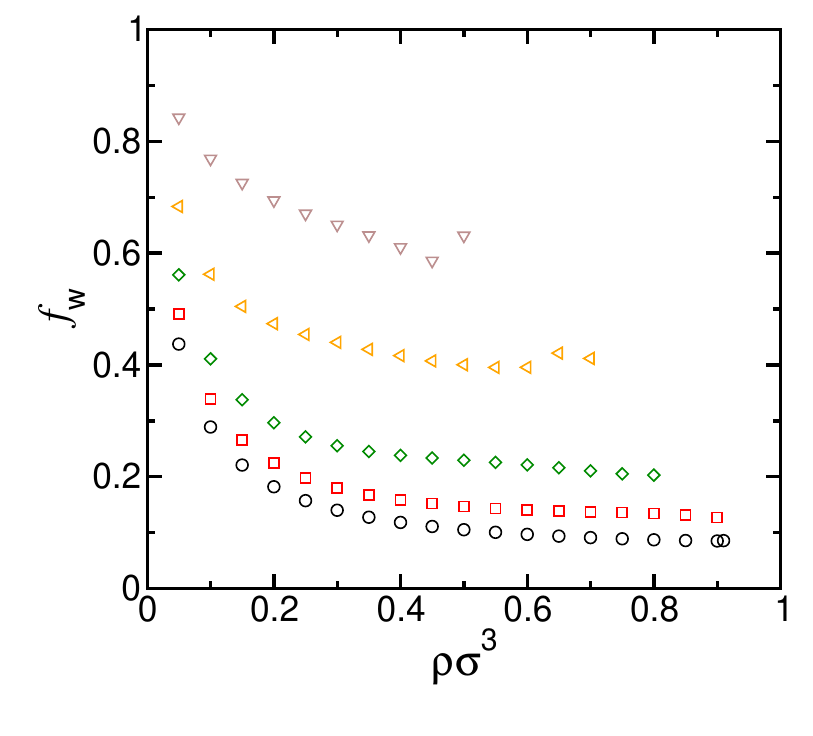}
    }
    \subfloat[][$H/\sigma=10.0$]{
      \includegraphics[width=0.2\textwidth]{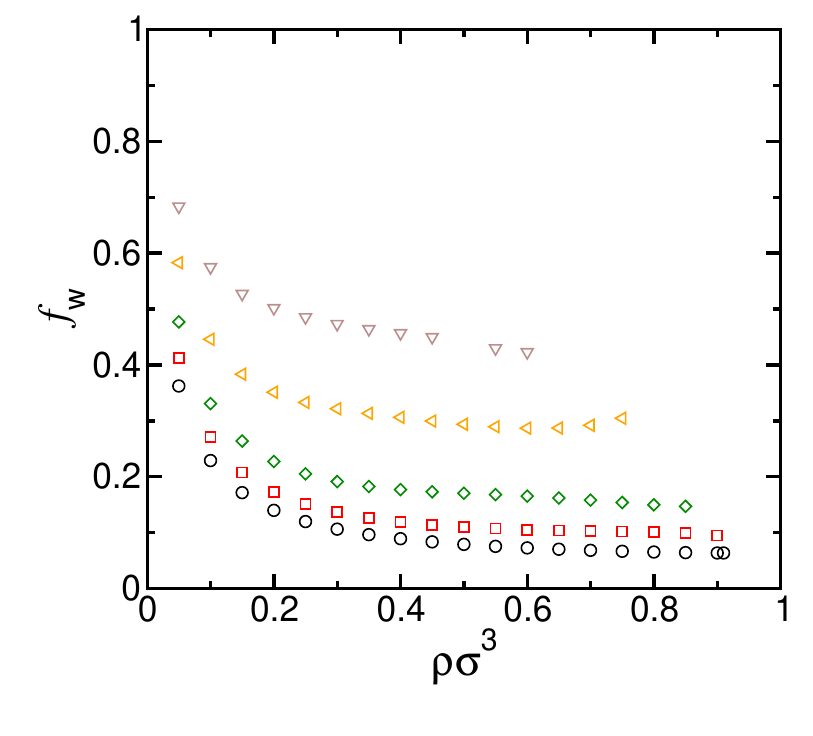}
    }
    \subfloat[][$H/\sigma=15.0$]{
      \includegraphics[width=0.2\textwidth]{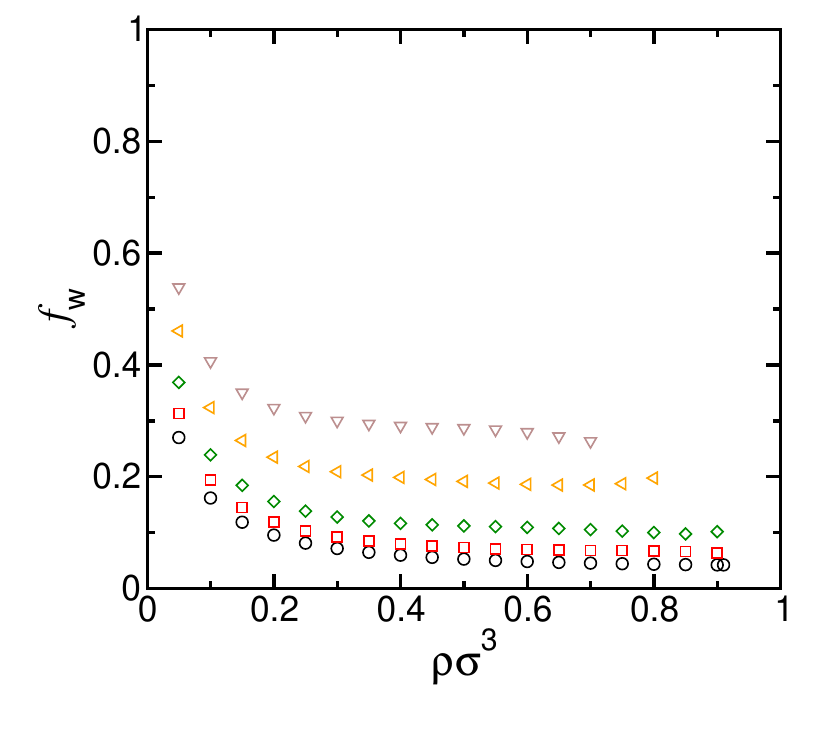}
    }
  }

  % DY/DX vs 1/H
  \subfloat[][$\lambda/\sigma=1.5$]{
      \includegraphics[width=0.2\textwidth]{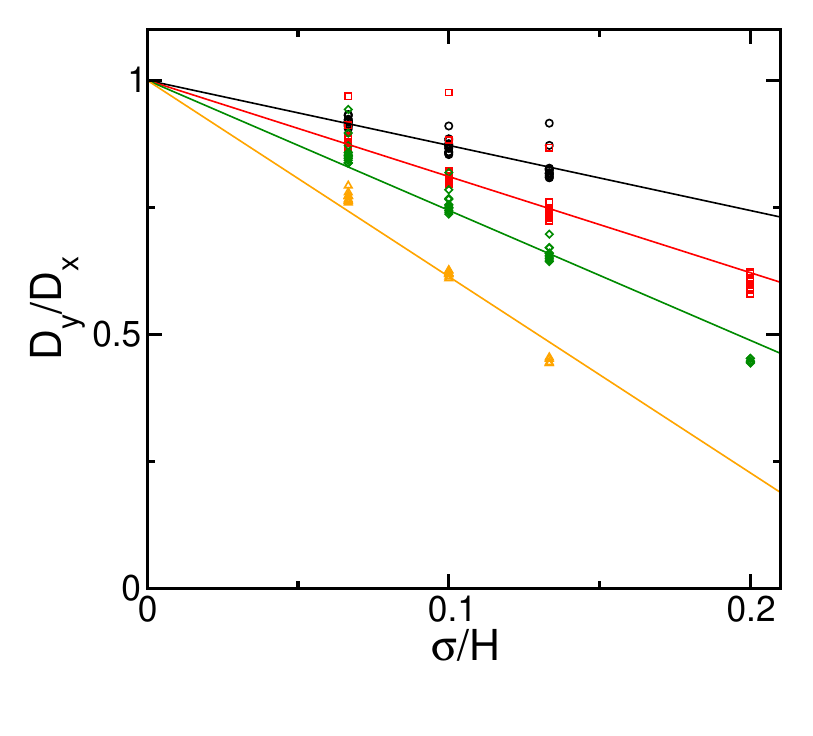}
    }\\

    \caption{Self-diffusion for physically rough system with
      $\lambda/\sigma=1.5$. (a-d) Self-diffusion in $x$ direction
      $D_x$, (e-h) self-diffusion in $y$ direction $D_y$, (i-l)
      $D_y/D_x$, and (m-p) fraction of wall collisions $f_w$ vs density. (q) $D_y/D_x$ vs $H^{-1}$ for points with
      $\rho\sigma^3>0.2$. }
    \label{fig:more_roughness_WL_1.5}
\end{figure*}

\begin{figure*}[t]
  \centering
  %DX
  \mbox{
    \subfloat[][$H/\sigma=5.0$]{
      \includegraphics[width=0.2\textwidth]{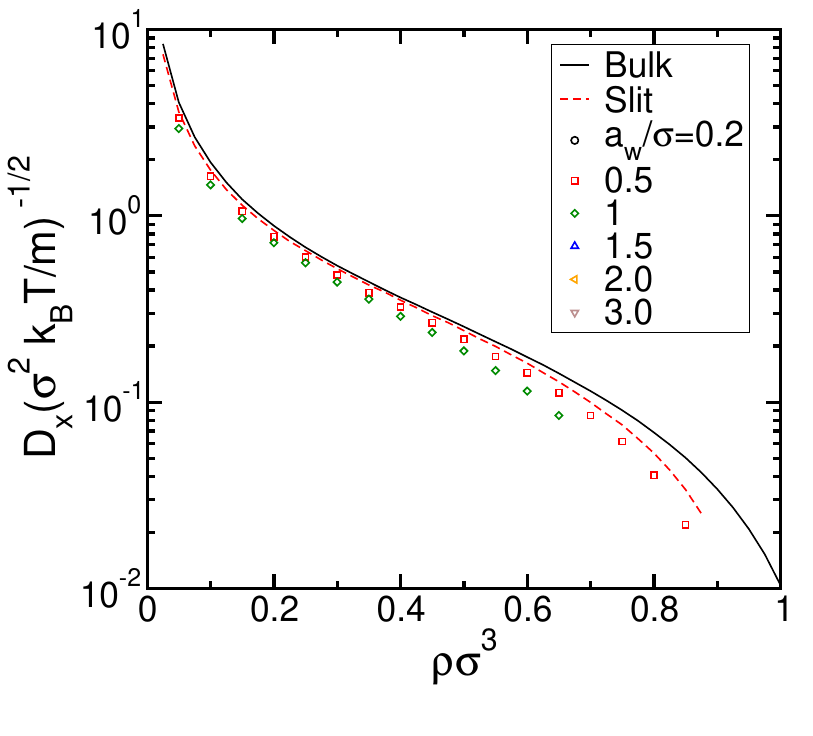}
      \label{fig:wavy_DX_vs_dens_h_5_WL_2.0}
    }
    \subfloat[][$H/\sigma=7.5$]{
      \includegraphics[width=0.2\textwidth]{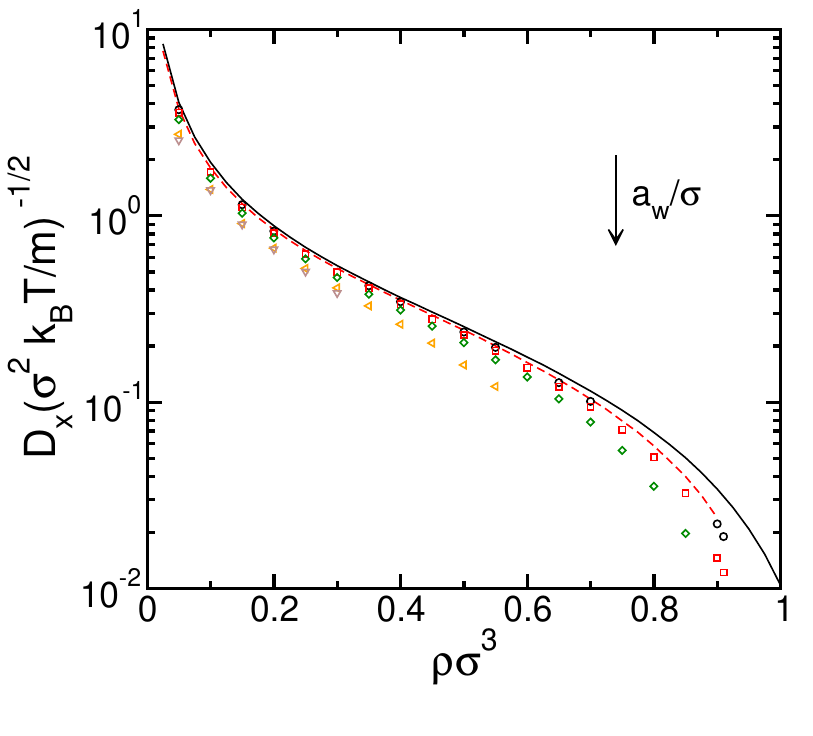}
      \label{fig:wavy_DX_vs_dens_h_7.5_WL_2.0}
    }
    \subfloat[][$H/\sigma=10.0$]{
      \includegraphics[width=0.2\textwidth]{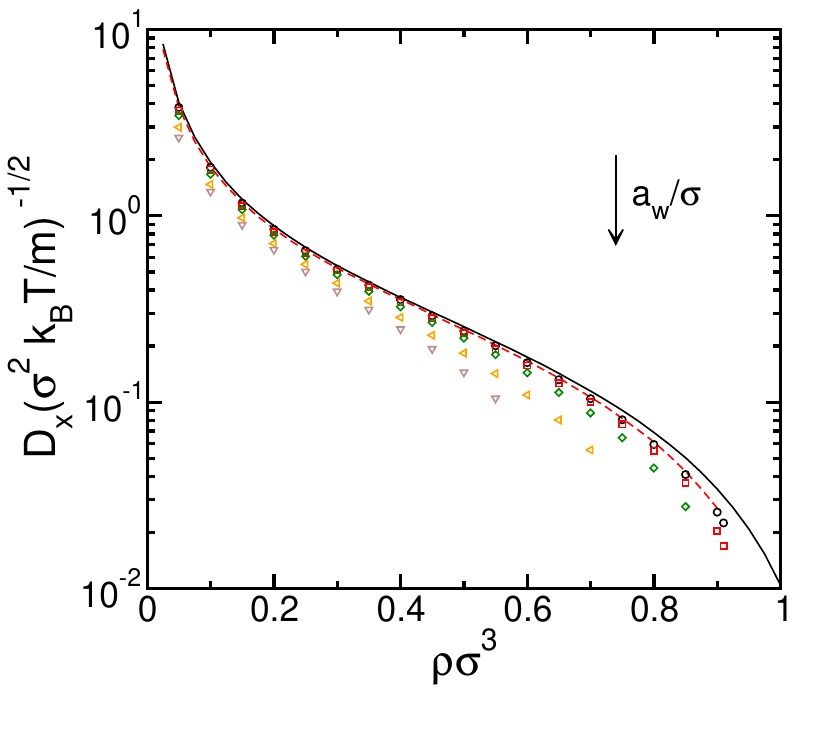}
      \label{fig:wavy_DX_vs_dens_h_10_WL_2.0}
    }
    \subfloat[][$H/\sigma=15.0$]{
      \includegraphics[width=0.2\textwidth]{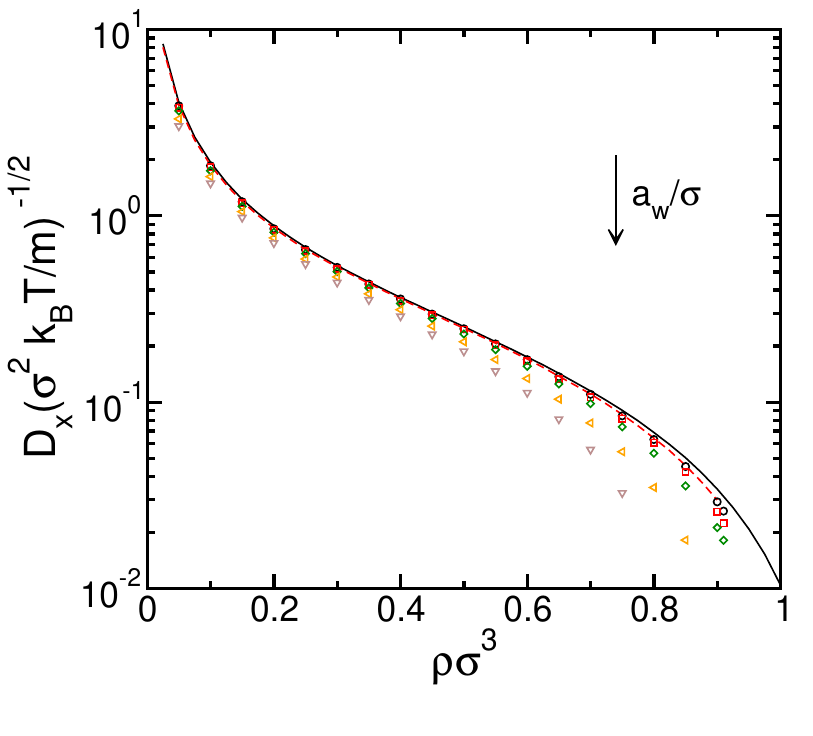}
      \label{fig:wavy_DX_vs_dens_h_15_WL_2.0}
    }

  }
  %DY
  \mbox{
    \subfloat[][$H/\sigma=5.0$]{
      \includegraphics[width=0.2\textwidth]{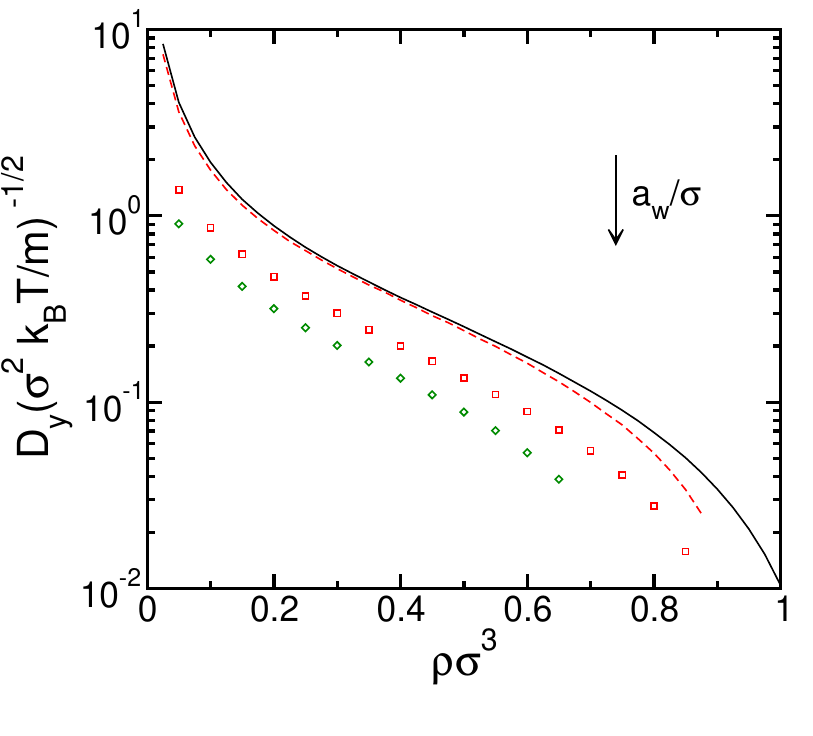}
      \label{fig:wavy_DY_vs_dens_h_5_WL_2.0}
    }
    \subfloat[][$H/\sigma=7.5$]{
      \includegraphics[width=0.2\textwidth]{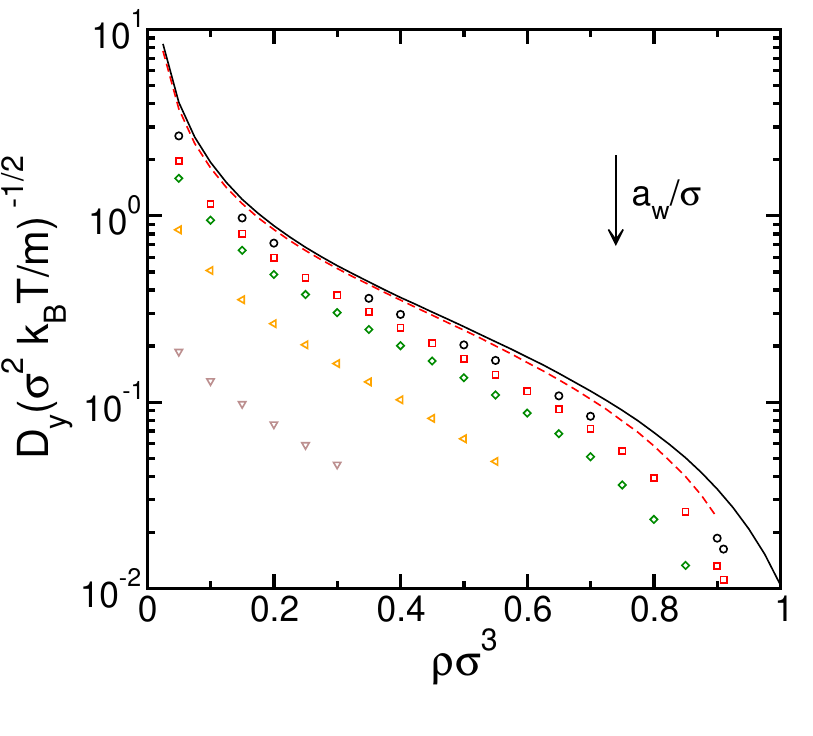}
      \label{fig:wavy_DY_vs_dens_h_7.5_WL_2.0}
    }
    \subfloat[][$H/\sigma=10.0$]{
      \includegraphics[width=0.2\textwidth]{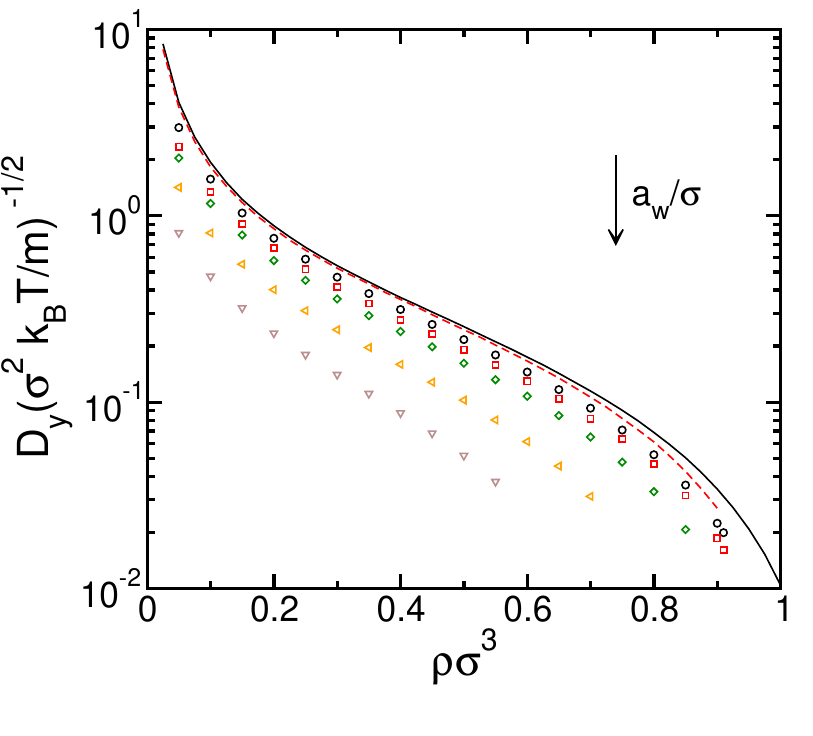}
      \label{fig:wavy_DY_vs_dens_h_10_WL_2.0}
    }
    \subfloat[][$H/\sigma=15.0$]{
      \includegraphics[width=0.2\textwidth]{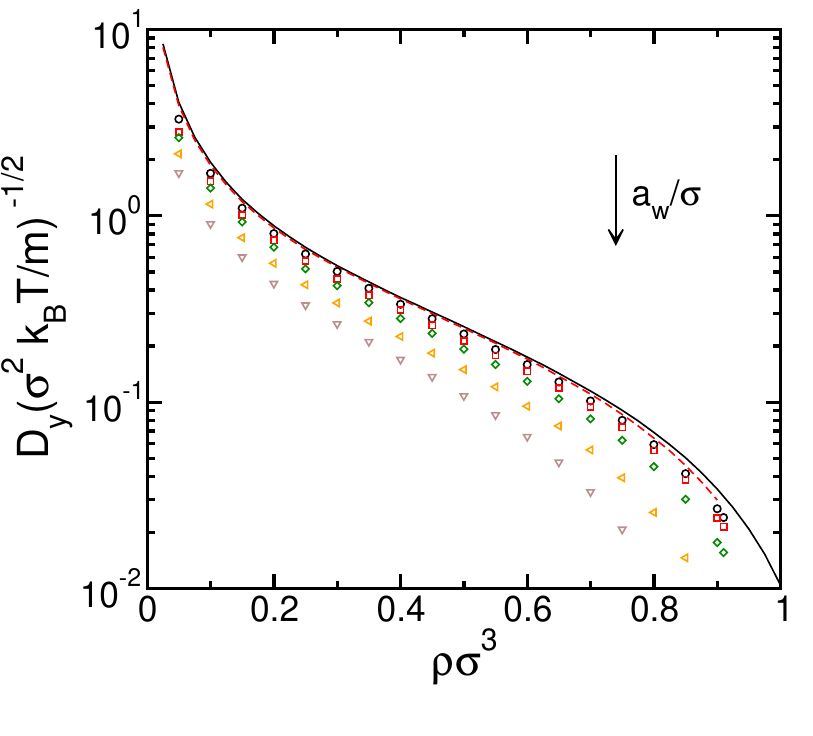}
      \label{fig:wavy_DY_vs_dens_h_15_WL_2.0}
    }
  }

  %DY/DX
  \mbox{
    \subfloat[][$H/\sigma=5.0$]{
      \includegraphics[width=0.2\textwidth]{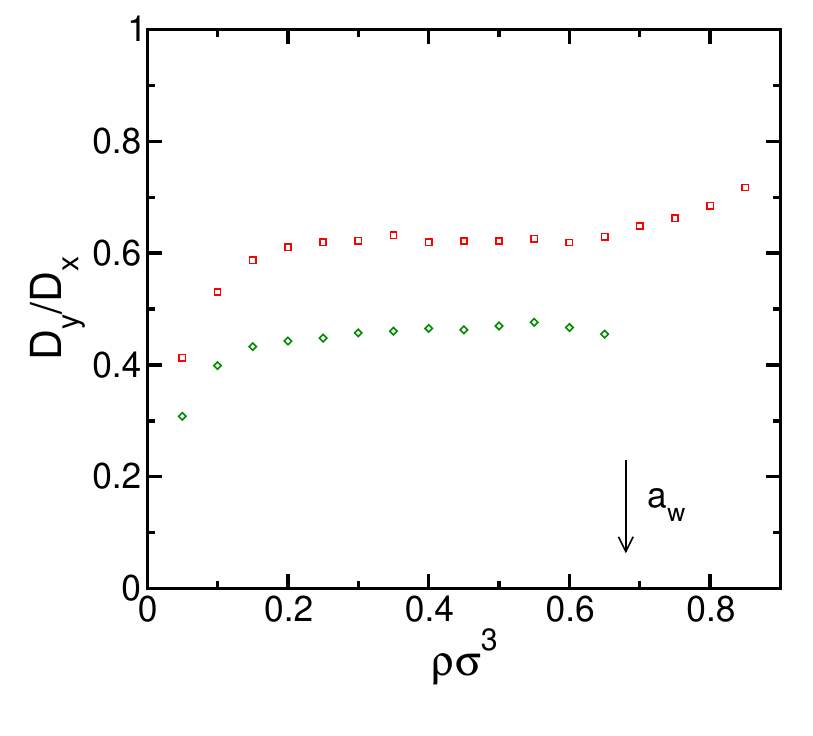}
      \label{fig:wavy_DY_by_DX_vs_dens_h_5_WL_2.0}
    }
    \subfloat[][$H/\sigma=7.5$]{
      \includegraphics[width=0.2\textwidth]{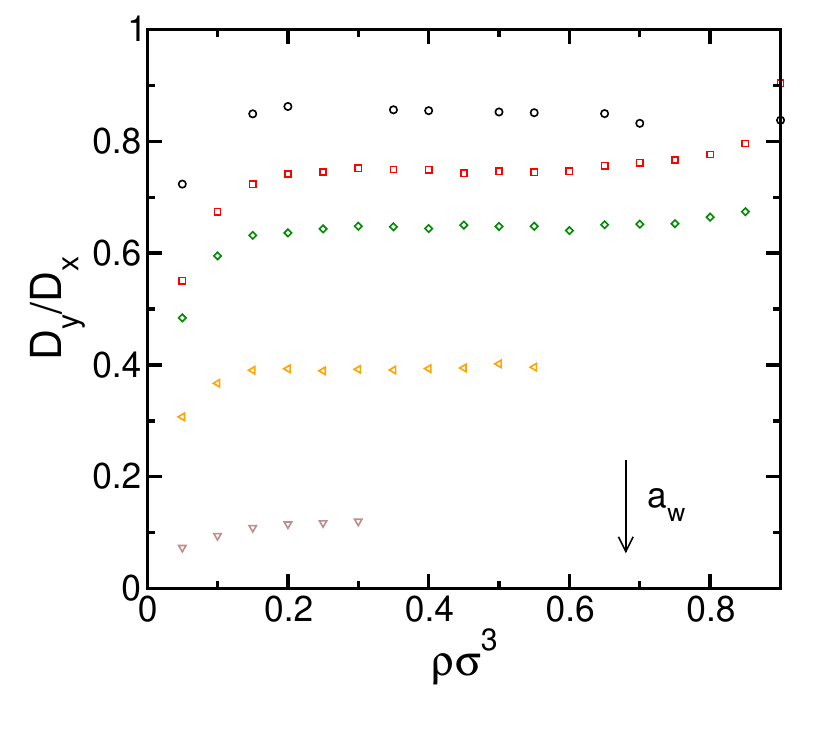}
      \label{fig:wavy_DY_by_DX_vs_dens_h_7p5_WL_2.0}
    }
    \subfloat[][$H/\sigma=10.0$]{
      \includegraphics[width=0.2\textwidth]{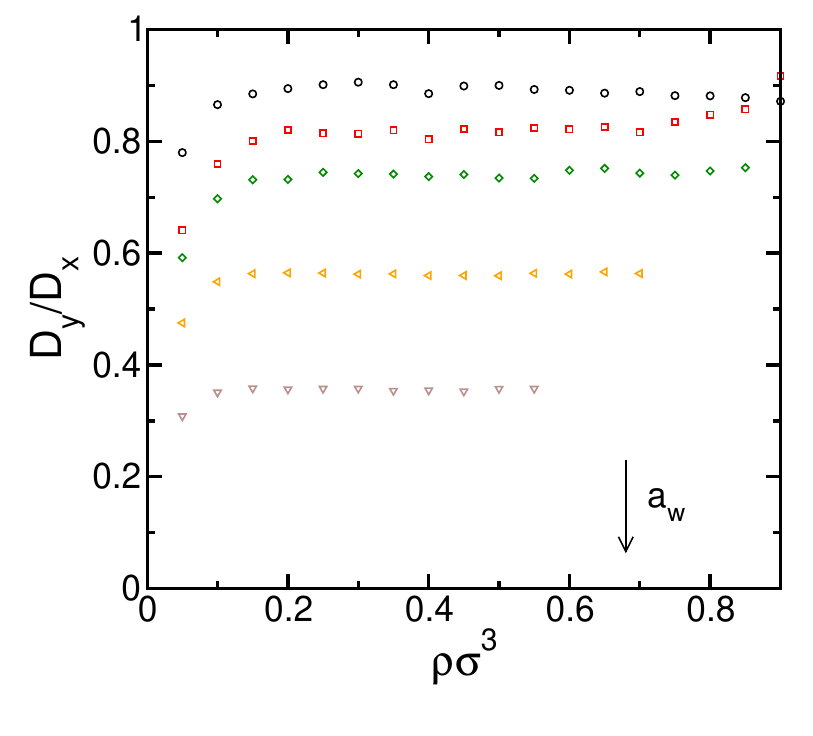}
      \label{fig:wavy_DY_by_DX_vs_dens_h_10_WL_2.0}
    }
    \subfloat[][$H/\sigma=15.0$]{
      \includegraphics[width=0.2\textwidth]{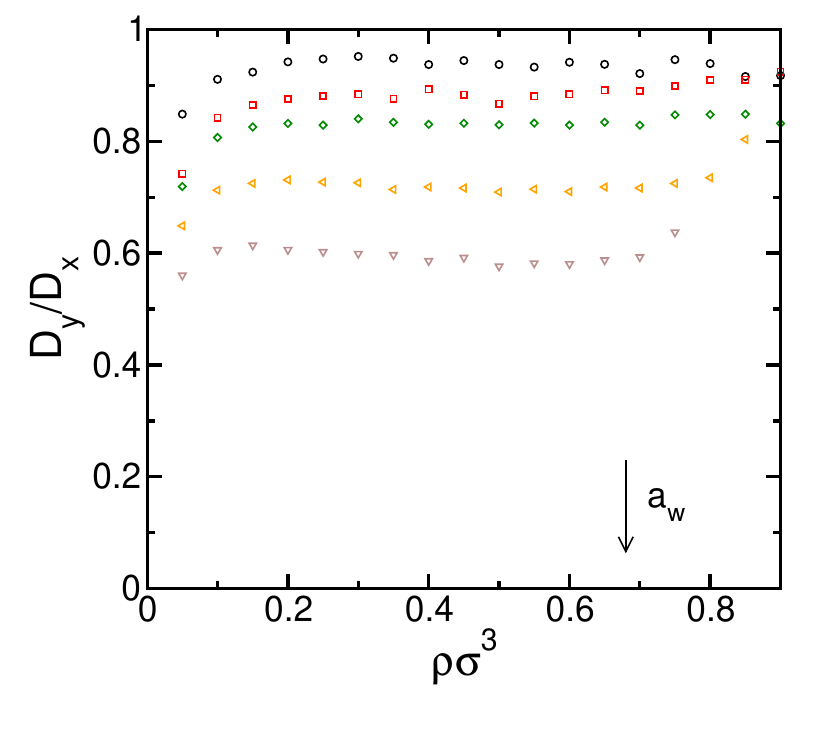}
      \label{fig:wavy_DY_by_DX_vs_dens_h_15_WL_2.0}
    }
  }

  \mbox{
    \subfloat[][$H/\sigma=5.0$]{
      \includegraphics[width=0.2\textwidth]{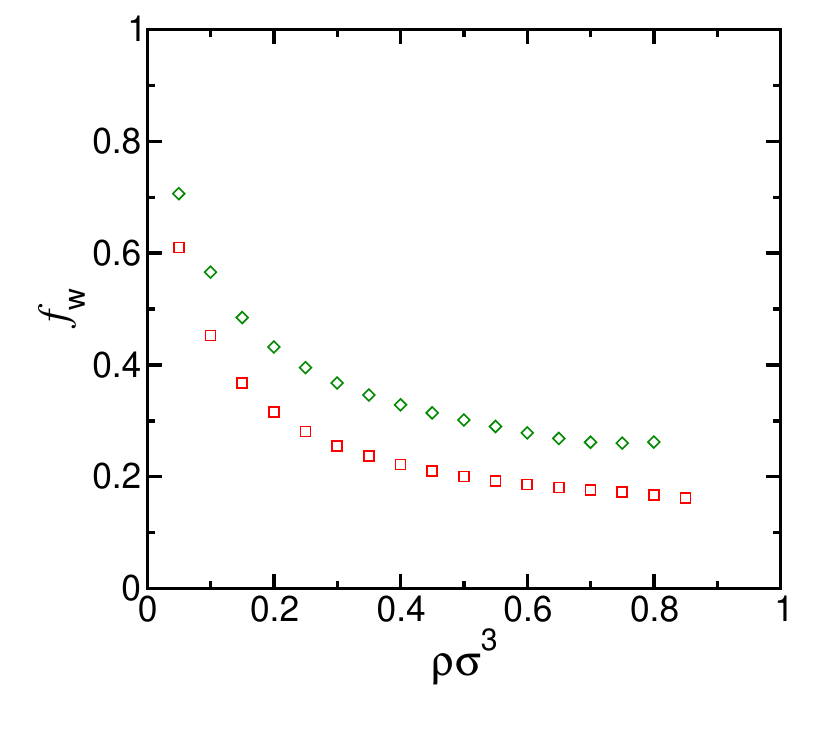}
    }
    \subfloat[][$H/\sigma=7.5$]{
      \includegraphics[width=0.2\textwidth]{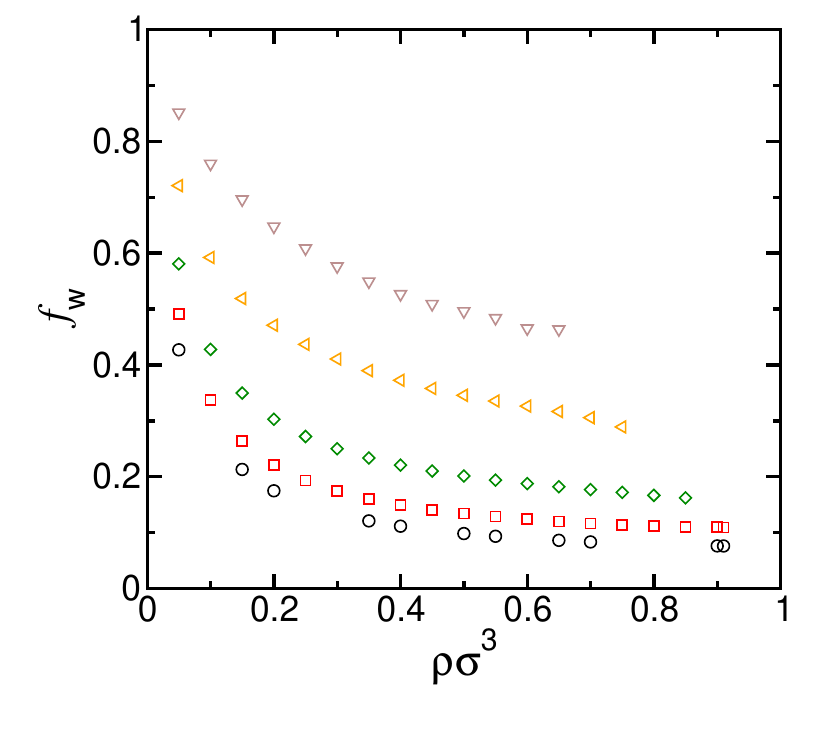}
    }
    \subfloat[][$H/\sigma=10.0$]{
      \includegraphics[width=0.2\textwidth]{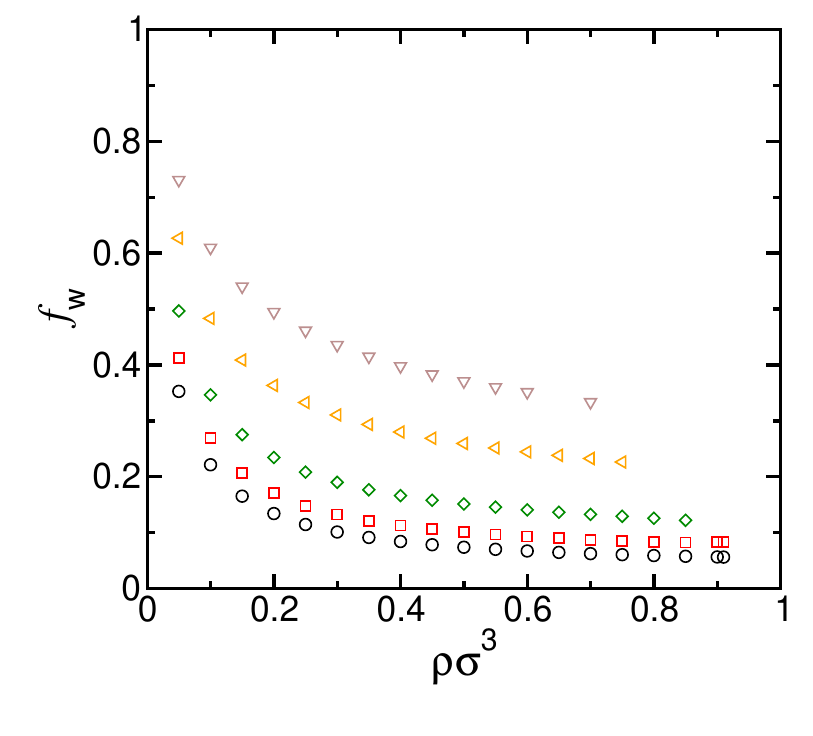}
    }
    \subfloat[][$H/\sigma=15.0$]{
      \includegraphics[width=0.2\textwidth]{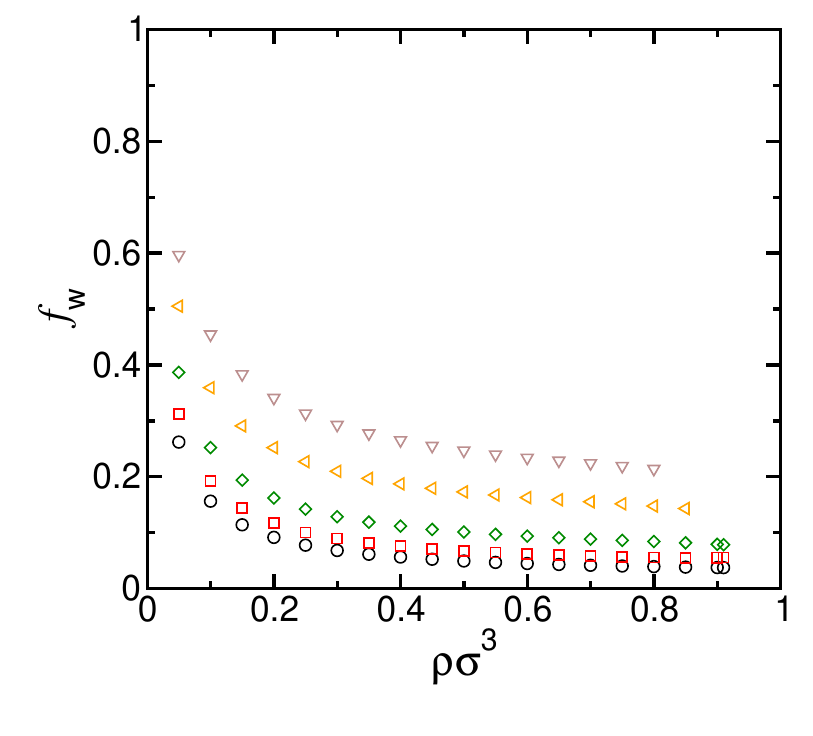}
    }
  }

  %dy/dx Hinv
    \subfloat[][$\lambda/\sigma=2.0$]{
      \includegraphics[width=0.2\textwidth]{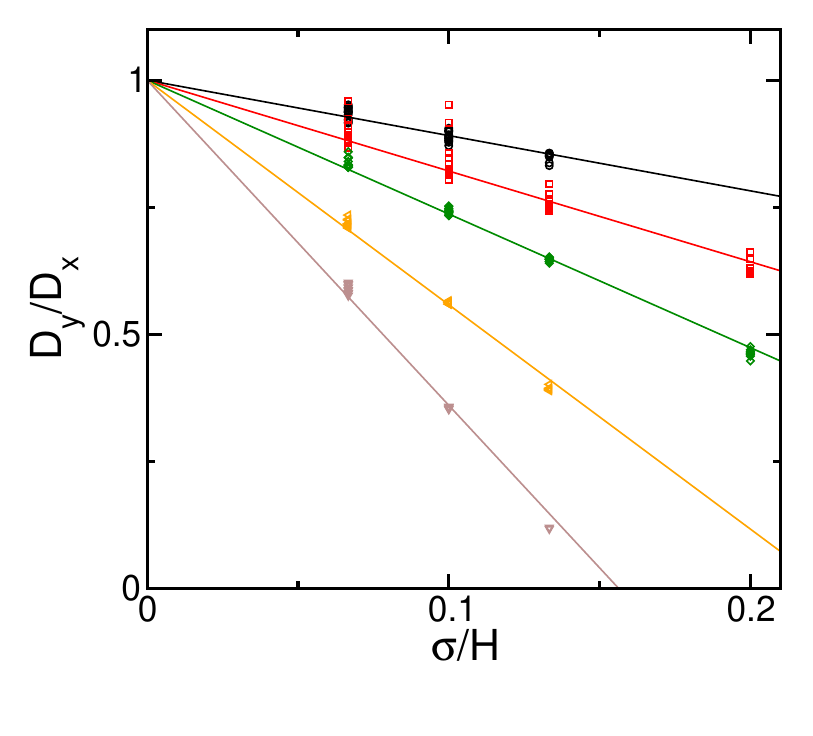}
    }

  \caption{Same as Fig.~\ref{fig:more_roughness_WL_1.5} with $\lambda/\sigma=2.0$.}
  \label{fig:more_roughness_WL_2.0}
\end{figure*}

%3.0
\begin{figure*}[t]
  \centering
  %DX
  \mbox{
    \subfloat[][$H/\sigma=5.0$]{
      \includegraphics[width=0.2\textwidth]{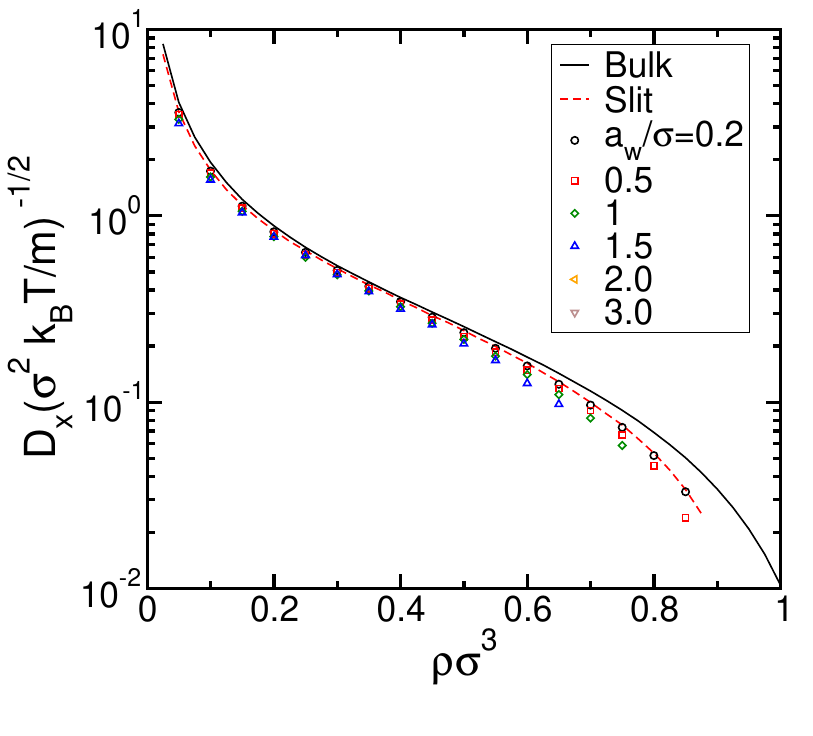}
      \label{fig:wavy_DX_vs_dens_h_5_WL_3.0}
    }
    \subfloat[][$H/\sigma=7.5$]{
      \includegraphics[width=0.2\textwidth]{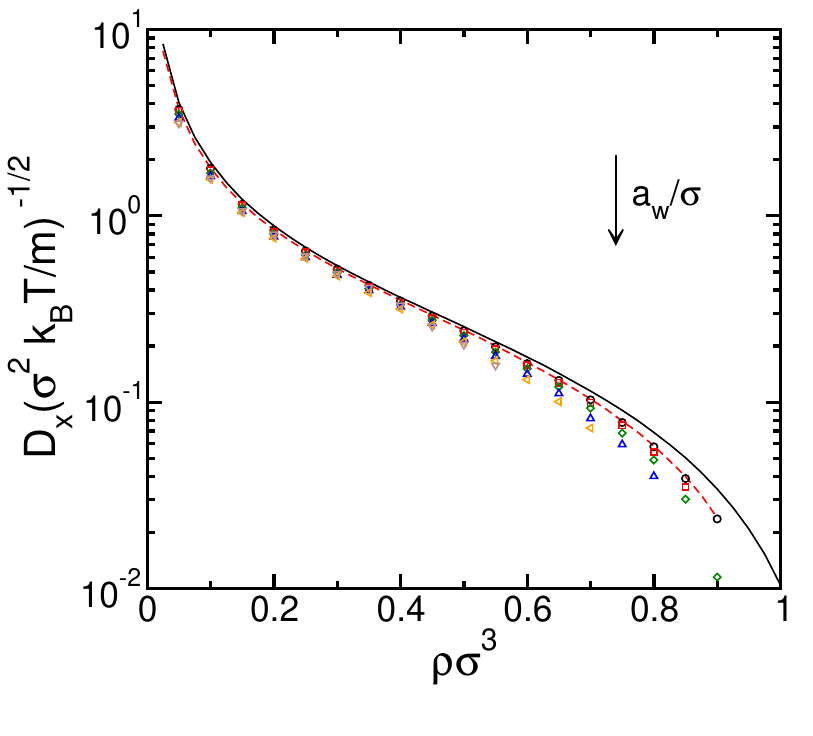}
      \label{fig:wavy_DX_vs_dens_h_7.5_WL_3.0}
    }
    \subfloat[][$H/\sigma=10.0$]{
      \includegraphics[width=0.2\textwidth]{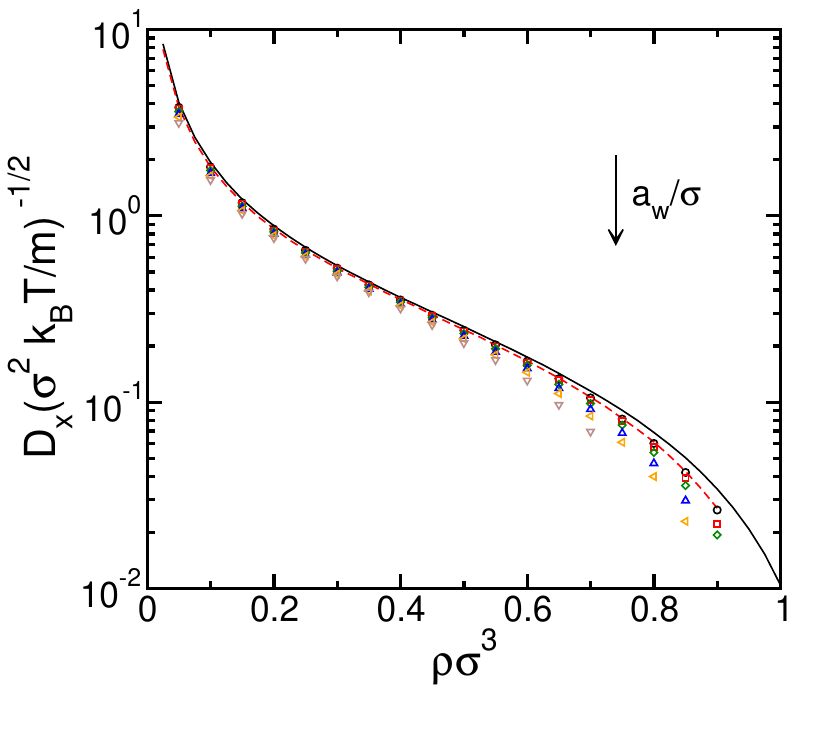}
      \label{fig:wavy_DX_vs_dens_h_10_WL_3.0}
    }
    \subfloat[][$H/\sigma=15.0$]{
      \includegraphics[width=0.2\textwidth]{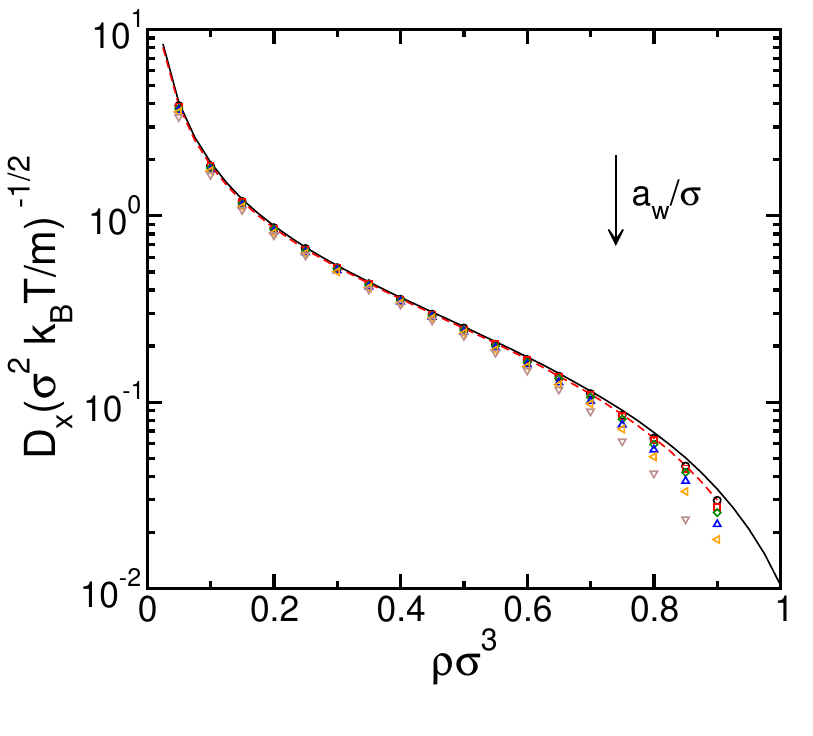}
      \label{fig:wavy_DX_vs_dens_h_15_WL_3.0}
    }

  }
  %DY
  \mbox{
    \subfloat[][$H/\sigma=5.0$]{
      \includegraphics[width=0.2\textwidth]{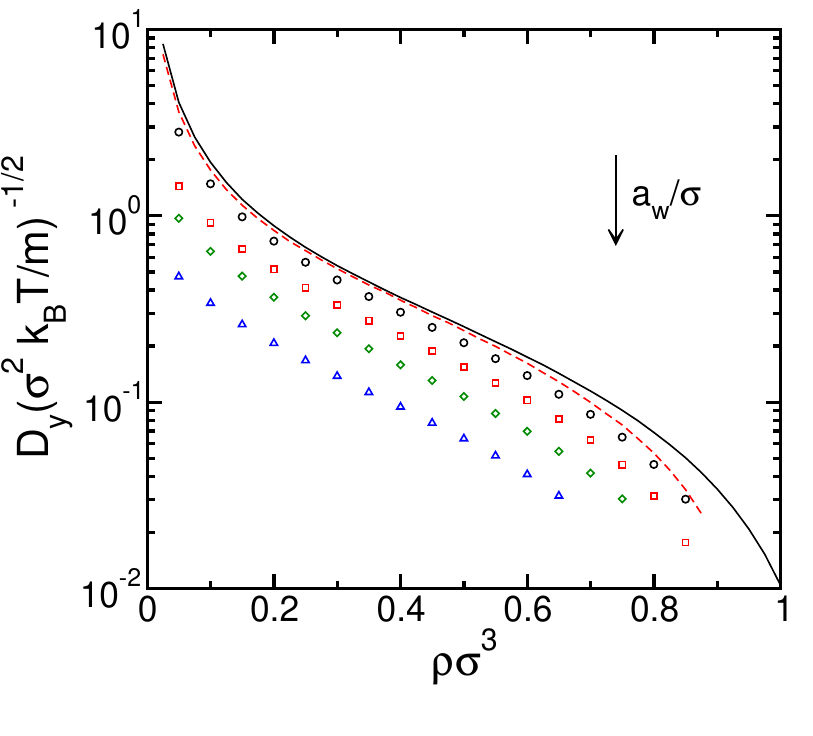}
      \label{fig:wavy_DY_vs_dens_h_5_WL_3.0}
    }
    \subfloat[][$H/\sigma=7.5$]{
      \includegraphics[width=0.2\textwidth]{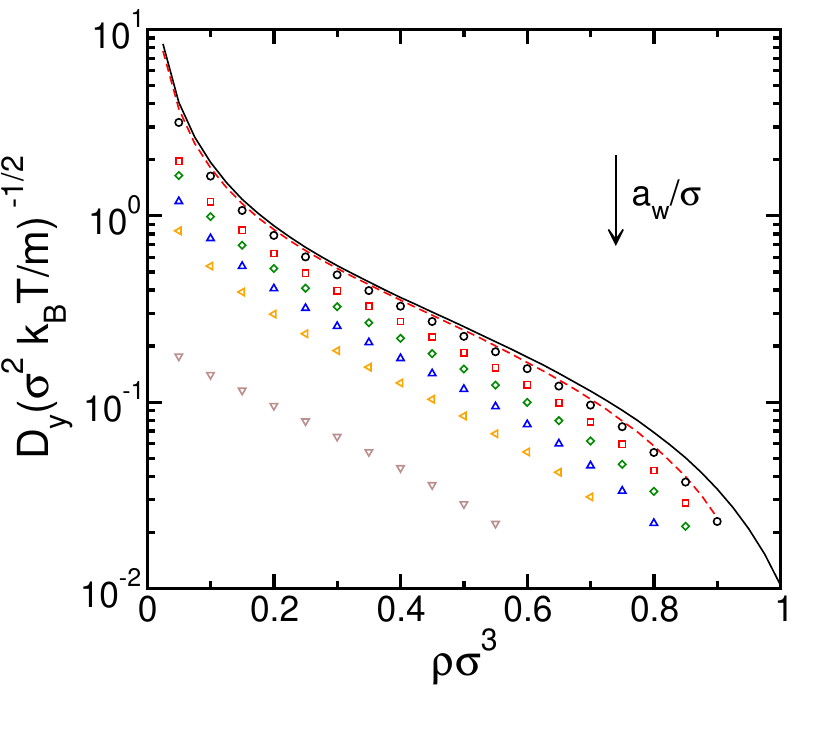}
      \label{fig:wavy_DY_vs_dens_h_7.5_WL_3.0}
    }
    \subfloat[][$H/\sigma=10.0$]{
      \includegraphics[width=0.2\textwidth]{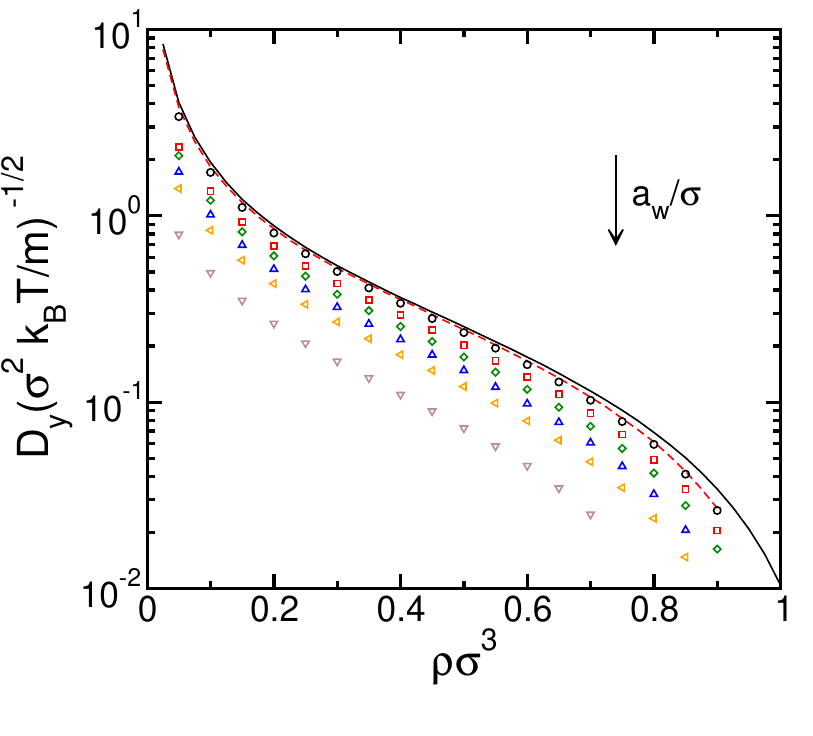}
      \label{fig:wavy_DY_vs_dens_h_10_WL_3.0}
    }
    \subfloat[][$H/\sigma=15.0$]{
      \includegraphics[width=0.2\textwidth]{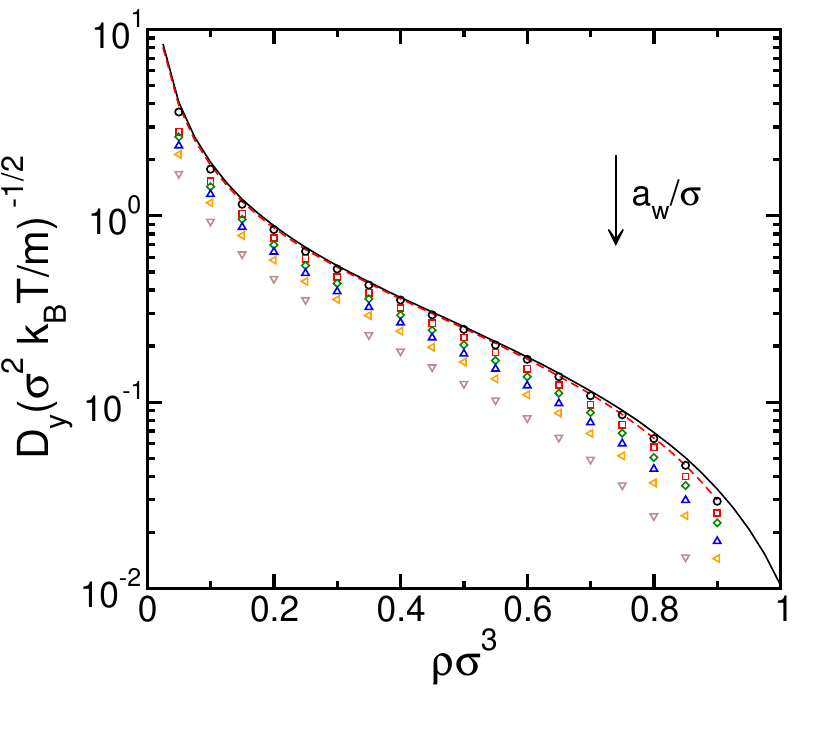}
      \label{fig:wavy_DY_vs_dens_h_15_WL_3.0}
    }
  }

  %DY/DX
  \mbox{
    \subfloat[][$H/\sigma=5.0$]{
      \includegraphics[width=0.2\textwidth]{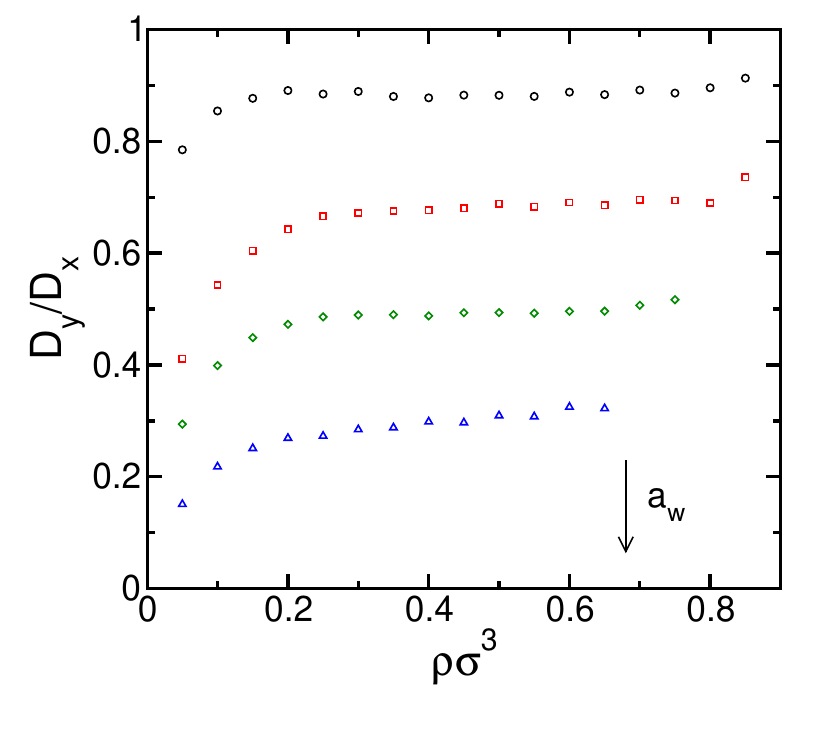}
      \label{fig:wavy_DY_by_DX_vs_dens_h_5_WL_3.0}
    }
    \subfloat[][$H/\sigma=7.5$]{
      \includegraphics[width=0.2\textwidth]{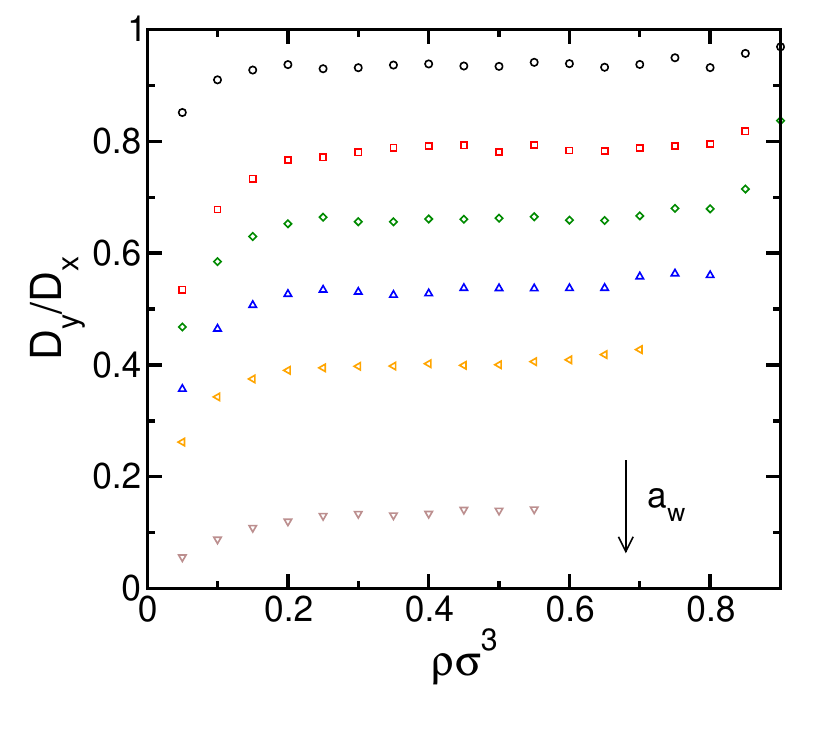}
      \label{fig:wavy_DY_by_DX_vs_dens_h_7p5_WL_3.0}
    }
    \subfloat[][$H/\sigma=10.0$]{
      \includegraphics[width=0.2\textwidth]{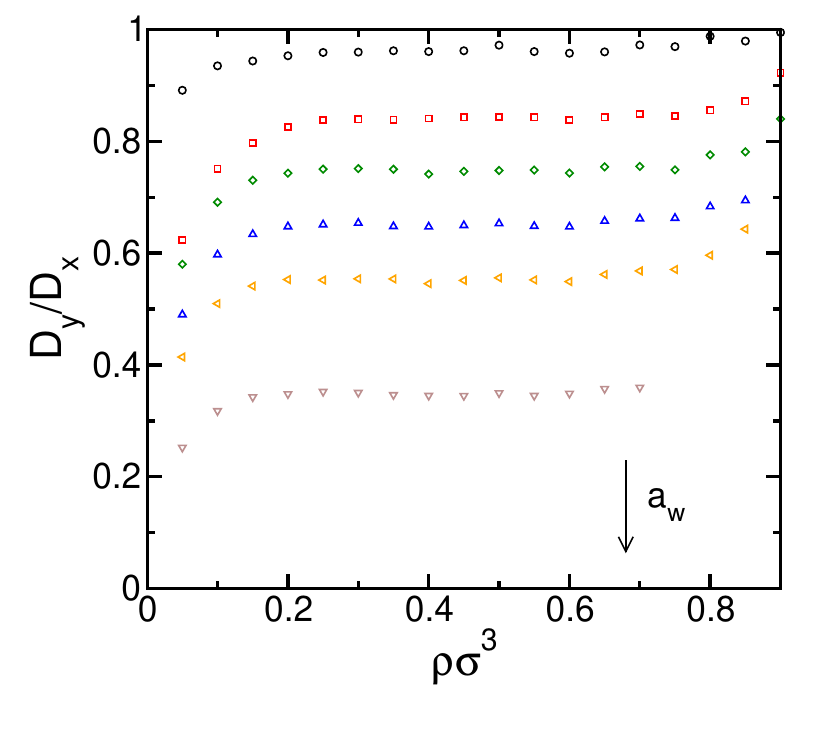}
      \label{fig:wavy_DY_by_DX_vs_dens_h_10_WL_3.0}
    }
    \subfloat[][$H/\sigma=15.0$]{
      \includegraphics[width=0.2\textwidth]{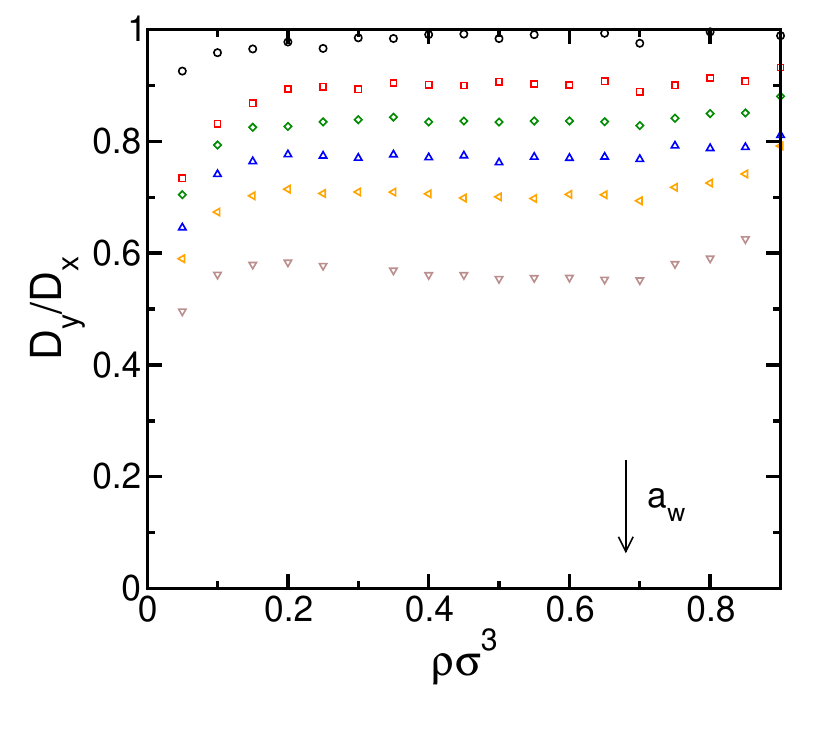}
      \label{fig:wavy_DY_by_DX_vs_dens_h_15_WL_3.0}
    }
  }

  \mbox{
    \subfloat[][$H/\sigma=5.0$]{
      \includegraphics[width=0.2\textwidth]{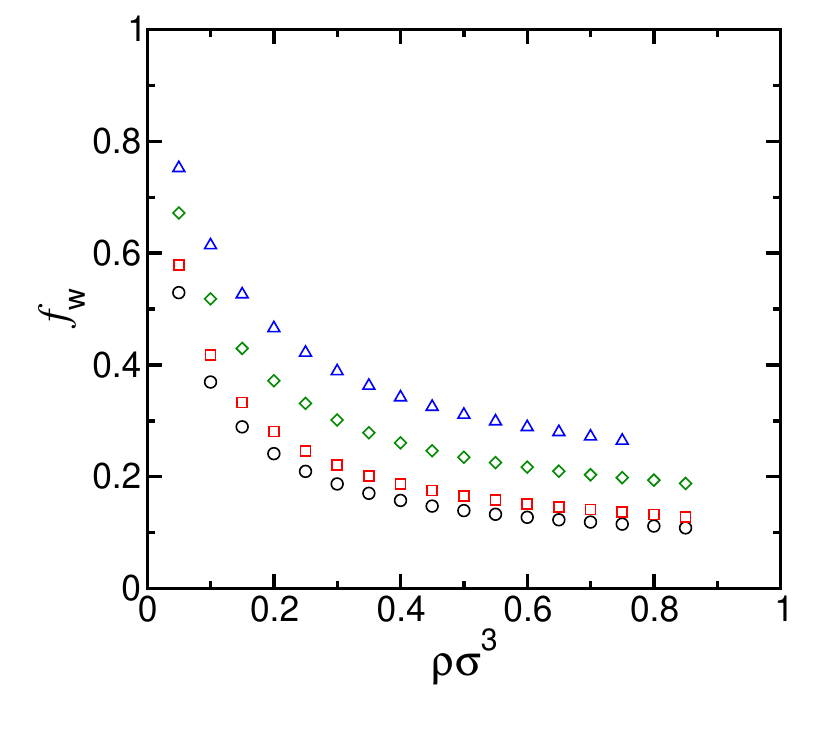}
    }
    \subfloat[][$H/\sigma=7.5$]{
      \includegraphics[width=0.2\textwidth]{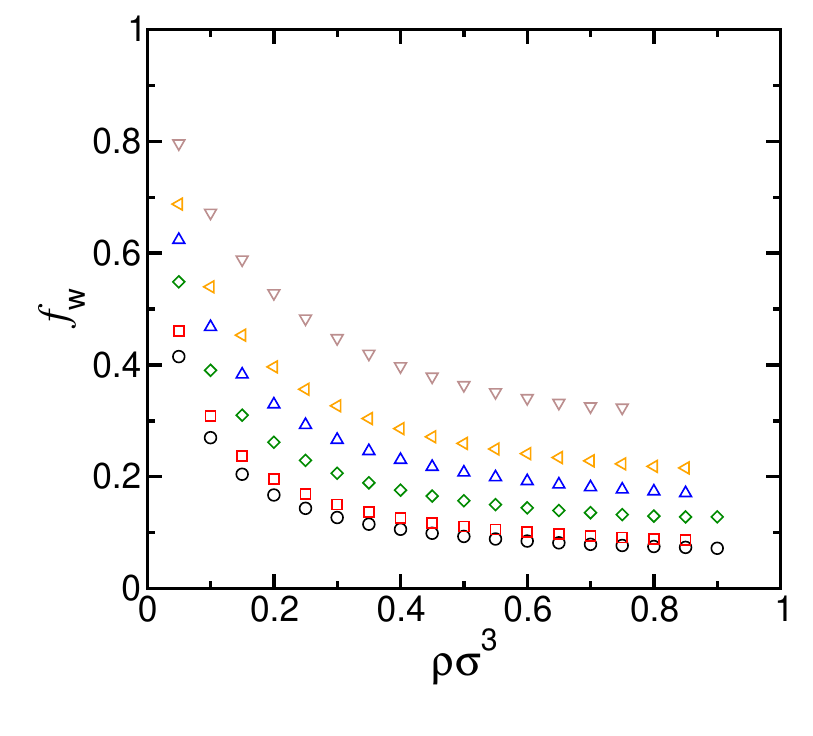}
    }
    \subfloat[][$H/\sigma=10.0$]{
      \includegraphics[width=0.2\textwidth]{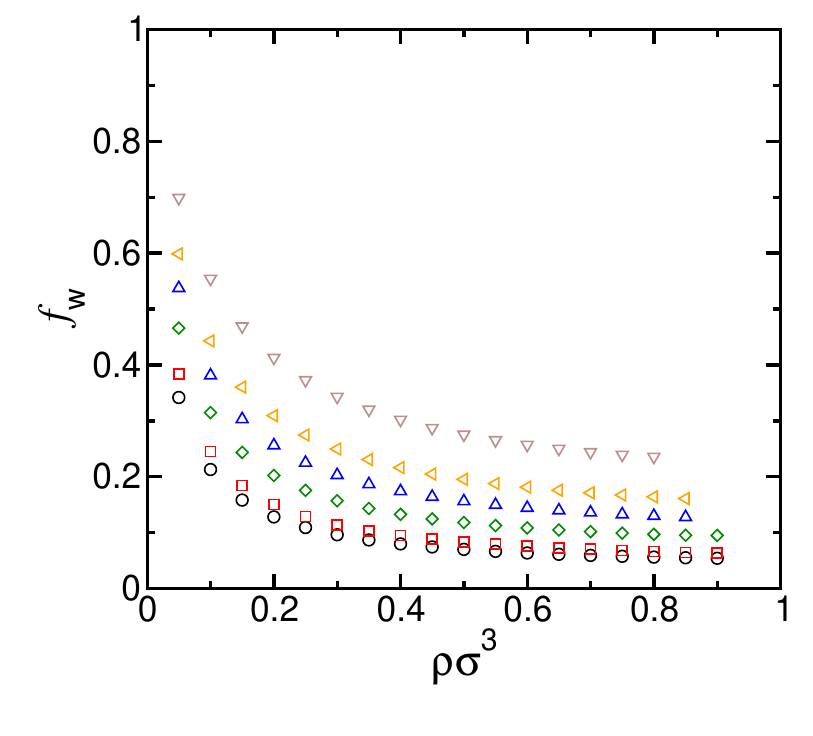}
    }
    \subfloat[][$H/\sigma=15.0$]{
      \includegraphics[width=0.2\textwidth]{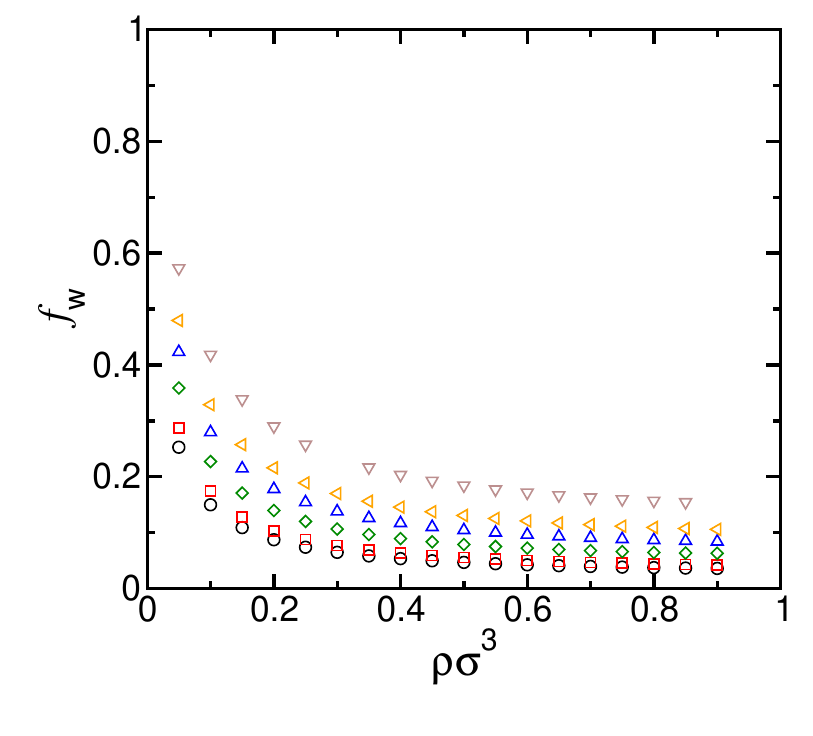}
    }
  }

  %dy/dx Hinv
  \subfloat[][$\lambda/\sigma=3.0$]{
    \includegraphics[width=0.2\textwidth]{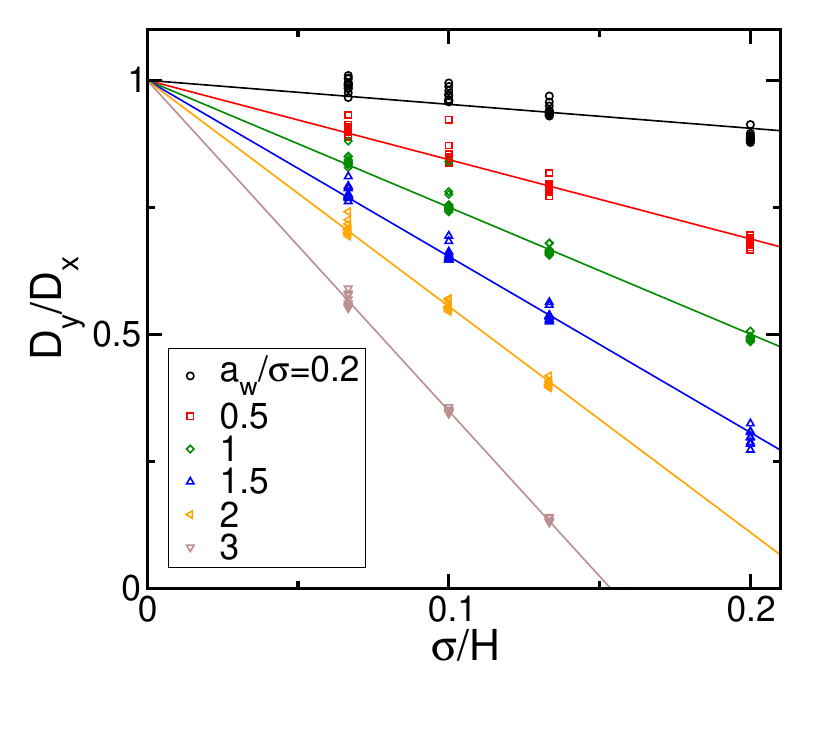}
  }

  \caption{Same as Fig.~\ref{fig:more_roughness_WL_1.5} with  $\lambda/\sigma=3.0$.}
  \label{fig:more_roughness_WL_3.0}
\end{figure*}

%6.0
\begin{figure*}[t]
  \centering
  %DX
  \mbox{
    \subfloat[][$H/\sigma=5.0$]{
      \includegraphics[width=0.2\textwidth]{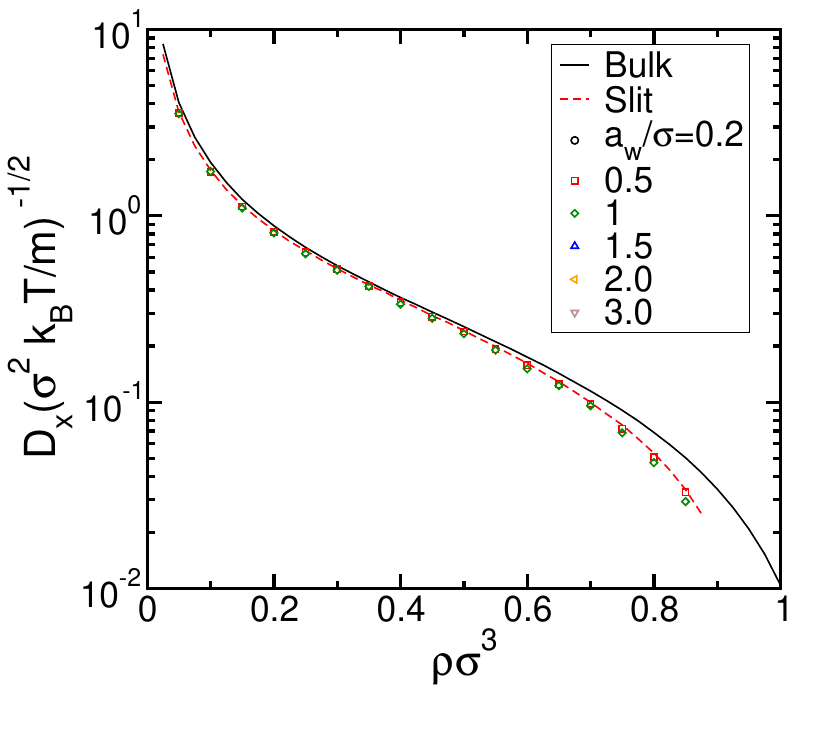}
      \label{fig:wavy_DX_vs_dens_h_5_WL_6.0}
    }
    \subfloat[][$H/\sigma=7.5$]{
      \includegraphics[width=0.2\textwidth]{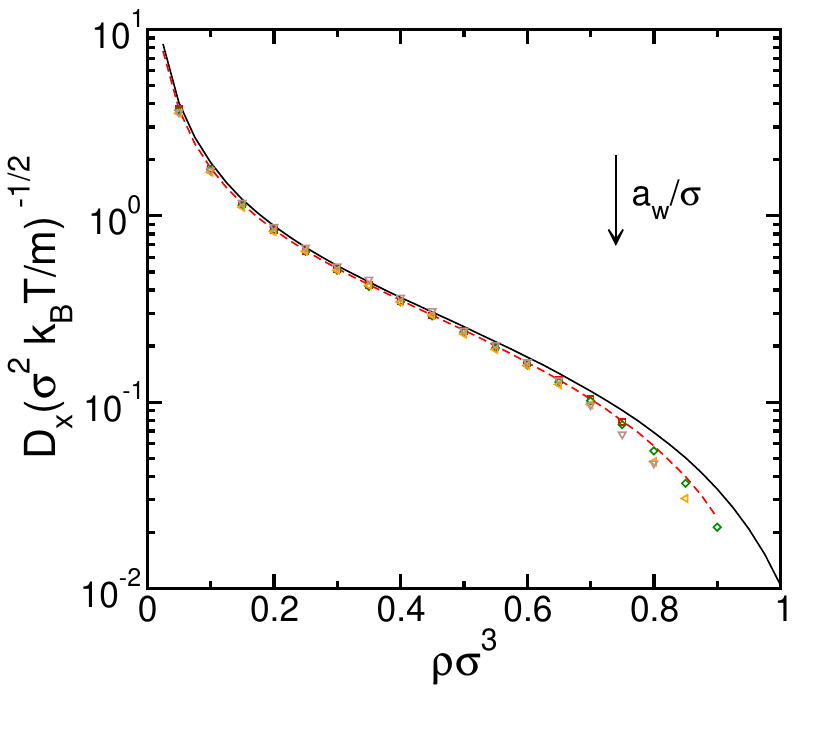}
      \label{fig:wavy_DX_vs_dens_h_7.5_WL_6.0}
    }
    \subfloat[][$H/\sigma=10.0$]{
      \includegraphics[width=0.2\textwidth]{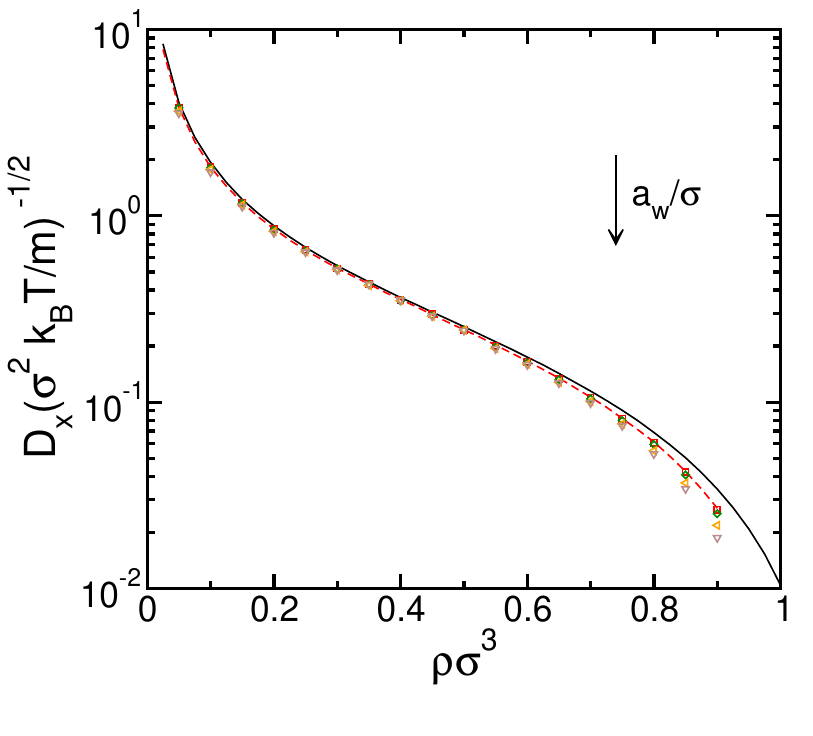}
      \label{fig:wavy_DX_vs_dens_h_10_WL_6.0}
    }
    \subfloat[][$H/\sigma=15.0$]{
      \includegraphics[width=0.2\textwidth]{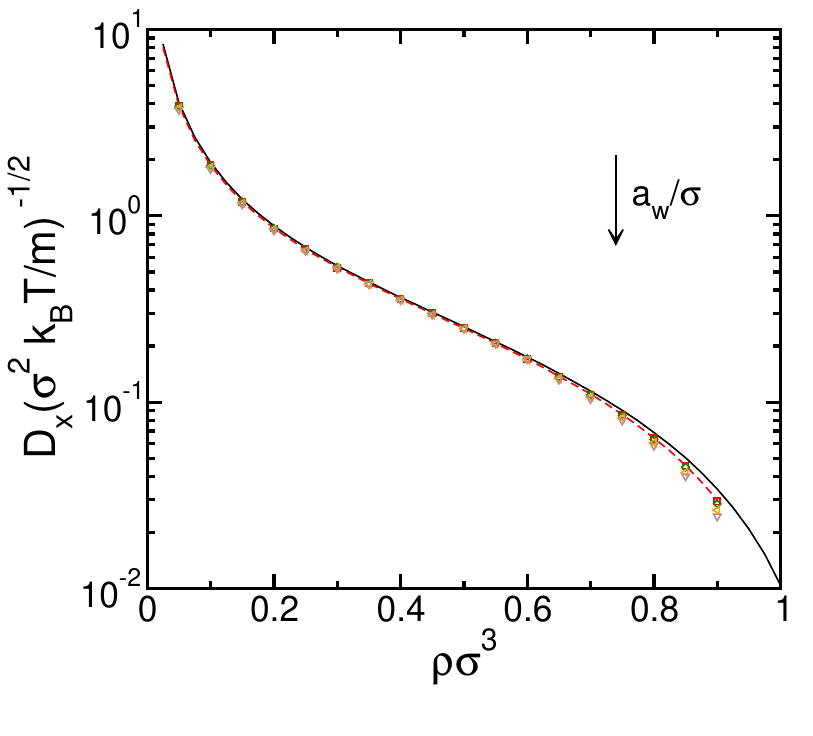}
      \label{fig:wavy_DX_vs_dens_h_15_WL_6.0}
    }

  }
  %DY
  \mbox{
    \subfloat[][$H/\sigma=5.0$]{
      \includegraphics[width=0.2\textwidth]{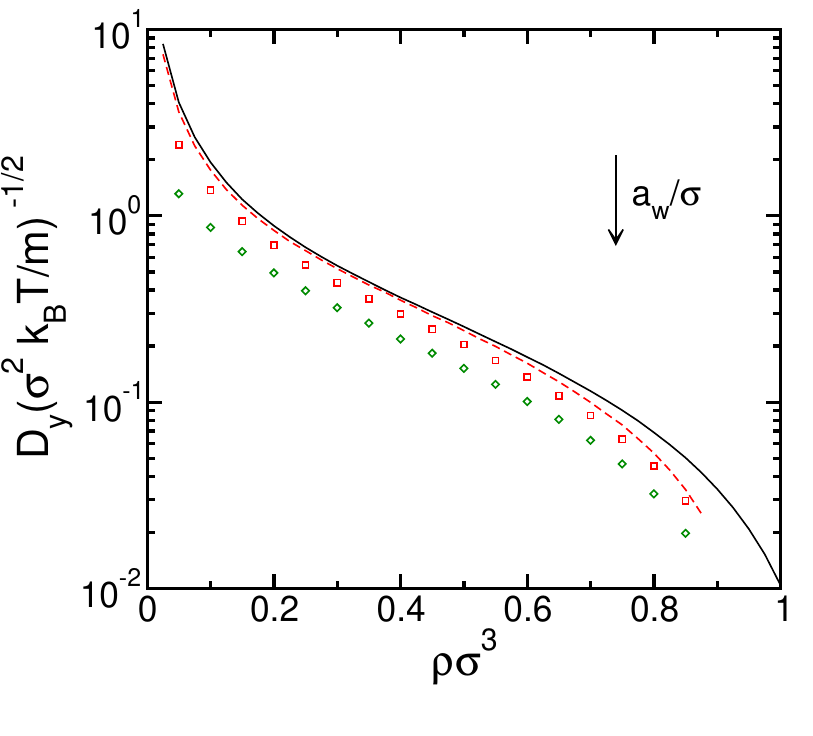}
      \label{fig:wavy_DY_vs_dens_h_5_WL_6.0}
    }
    \subfloat[][$H/\sigma=7.5$]{
      \includegraphics[width=0.2\textwidth]{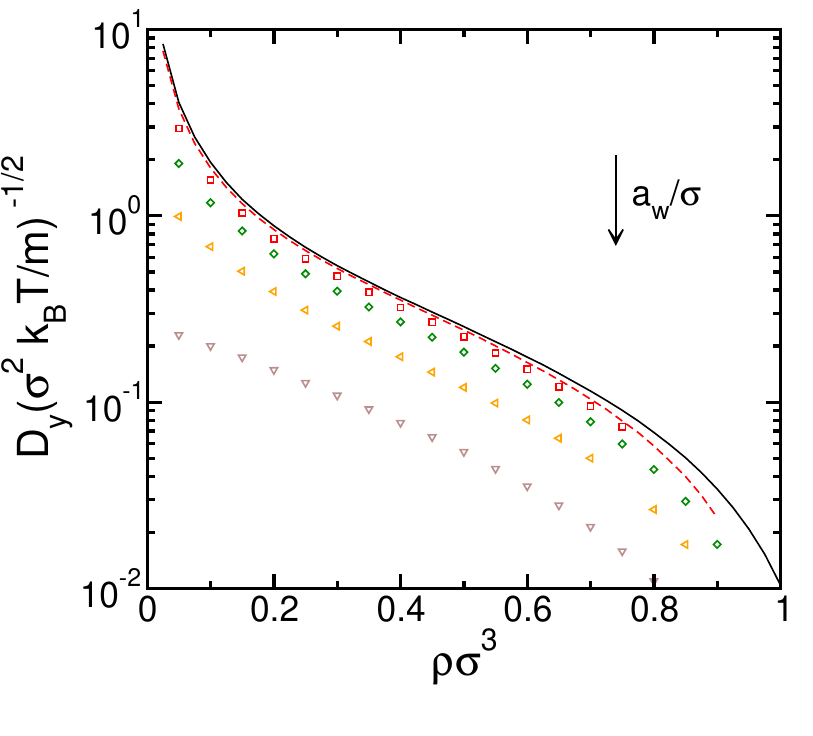}
      \label{fig:wavy_DY_vs_dens_h_7.5_WL_6.0}
    }
    \subfloat[][$H/\sigma=10.0$]{
      \includegraphics[width=0.2\textwidth]{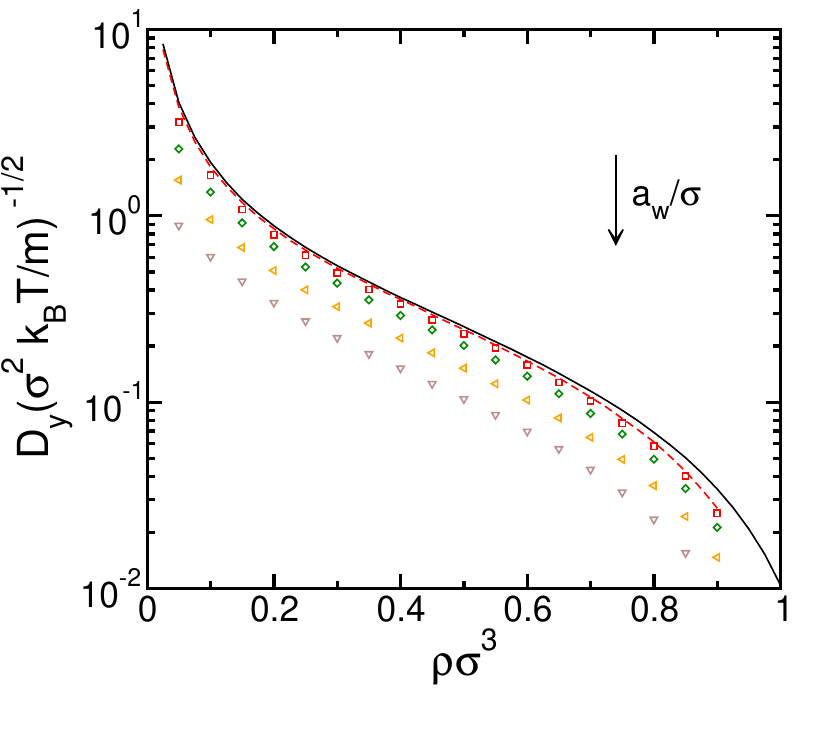}
      \label{fig:wavy_DY_vs_dens_h_10_WL_6.0}
    }
    \subfloat[][$H/\sigma=15.0$]{
      \includegraphics[width=0.2\textwidth]{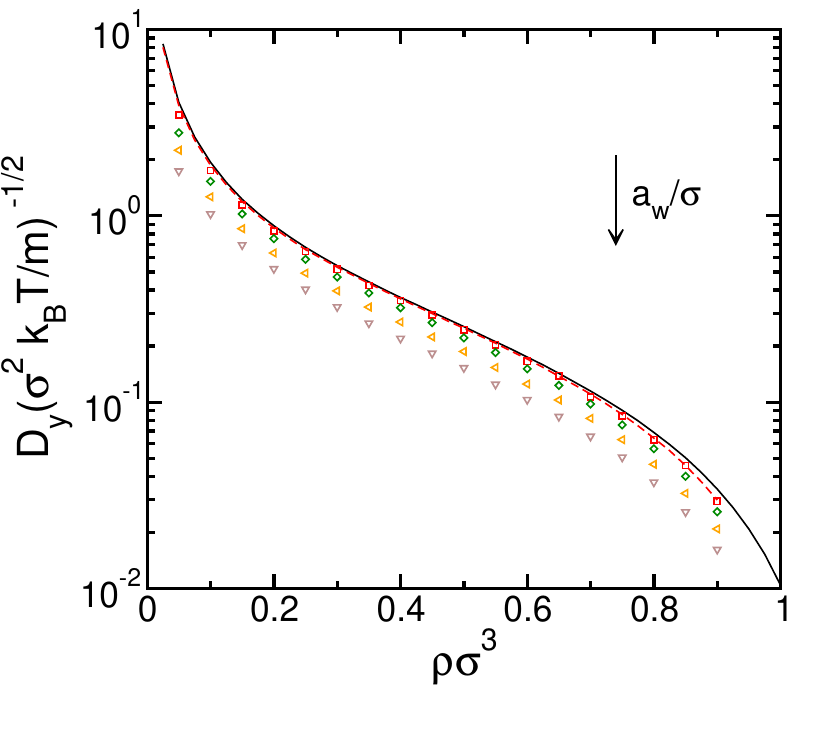}
      \label{fig:wavy_DY_vs_dens_h_15_WL_6.0}
    }
  }

  %DY/DX
  \mbox{
    \subfloat[][$H/\sigma=5.0$]{
      \includegraphics[width=0.2\textwidth]{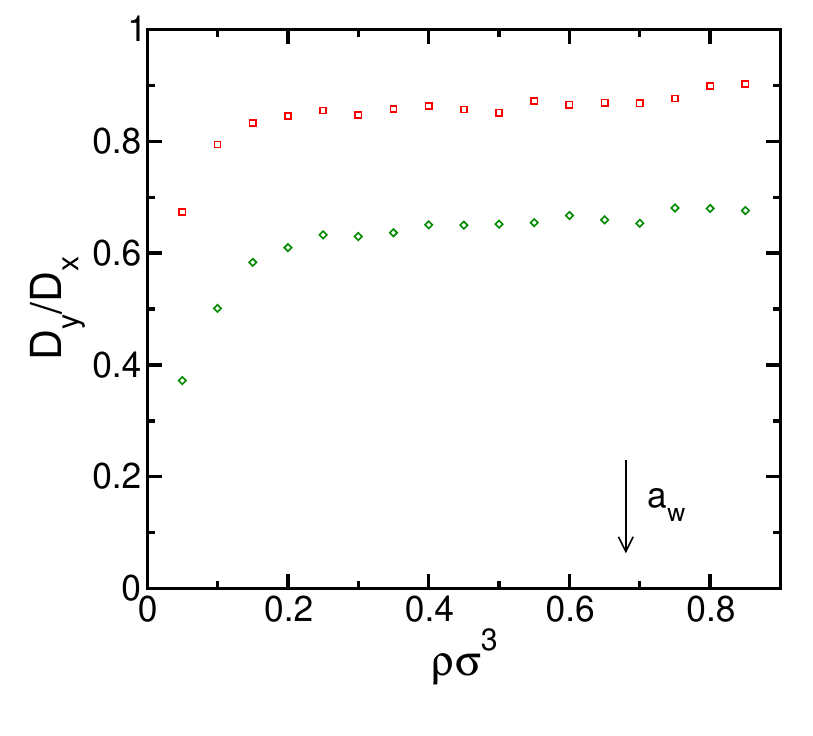}
      \label{fig:wavy_DY_by_DX_vs_dens_h_5_WL_6.0}
    }
    \subfloat[][$H/\sigma=7.5$]{
      \includegraphics[width=0.2\textwidth]{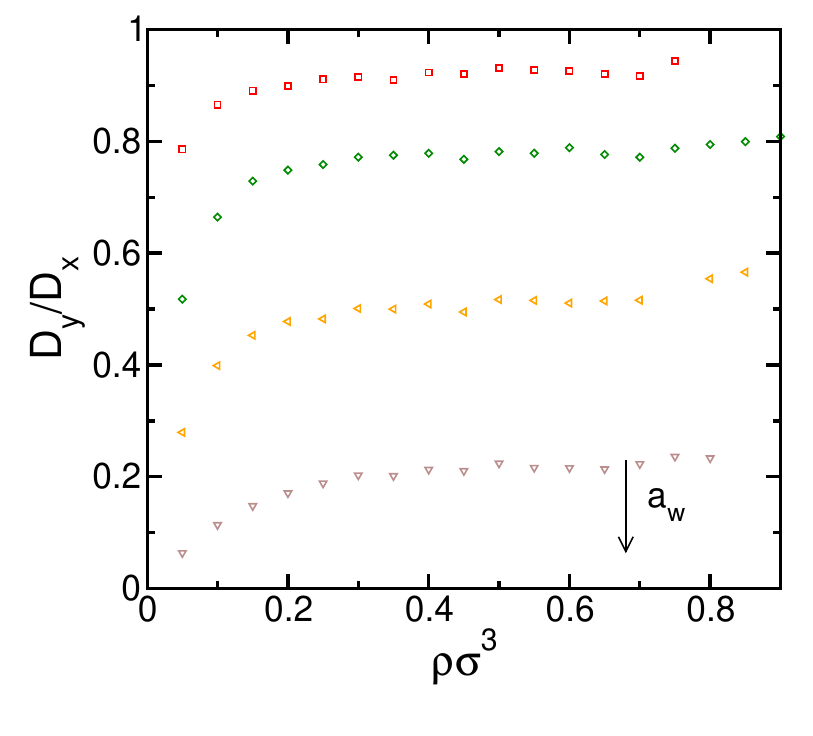}
      \label{fig:wavy_DY_by_DX_vs_dens_h_7p5_WL_6.0}
    }
    \subfloat[][$H/\sigma=10.0$]{
      \includegraphics[width=0.2\textwidth]{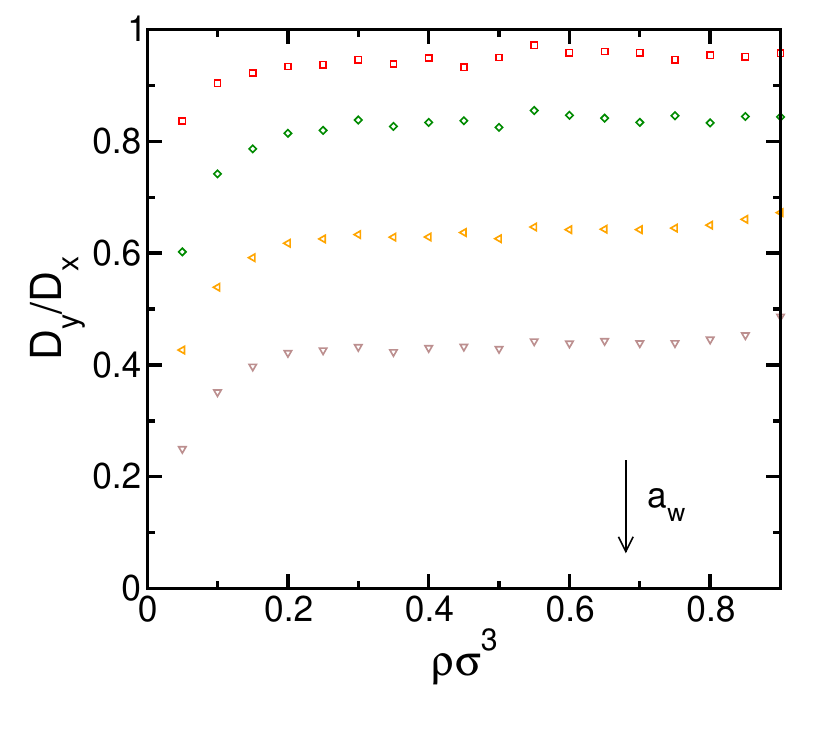}
      \label{fig:wavy_DY_by_DX_vs_dens_h_10_WL_6.0}
    }
    \subfloat[][$H/\sigma=15.0$]{
      \includegraphics[width=0.2\textwidth]{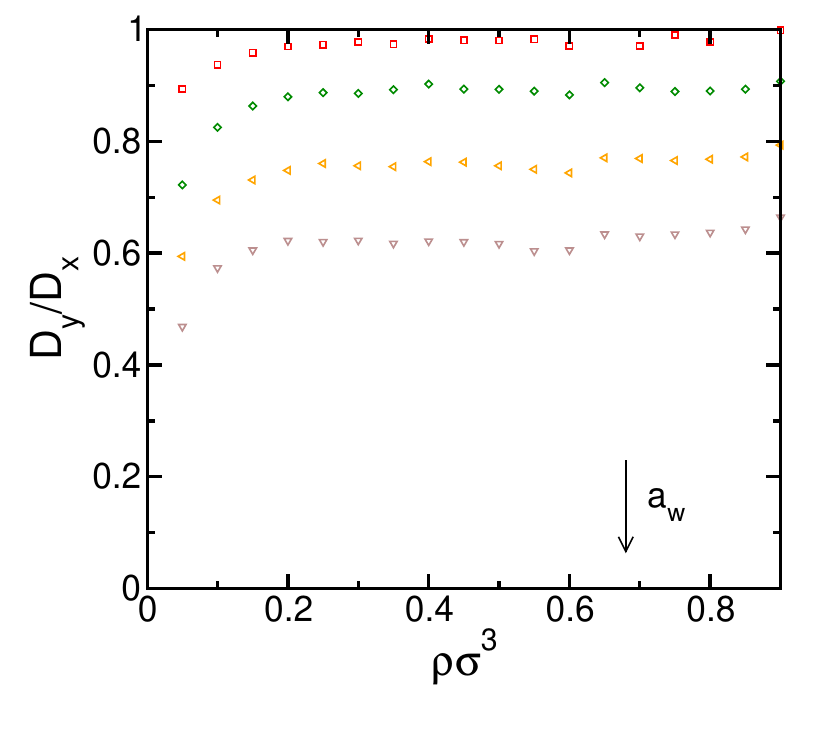}
      \label{fig:wavy_DY_by_DX_vs_dens_h_15_WL_6.0}
    }
  }

  \mbox{
    \subfloat[][$H/\sigma=5.0$]{
      \includegraphics[width=0.2\textwidth]{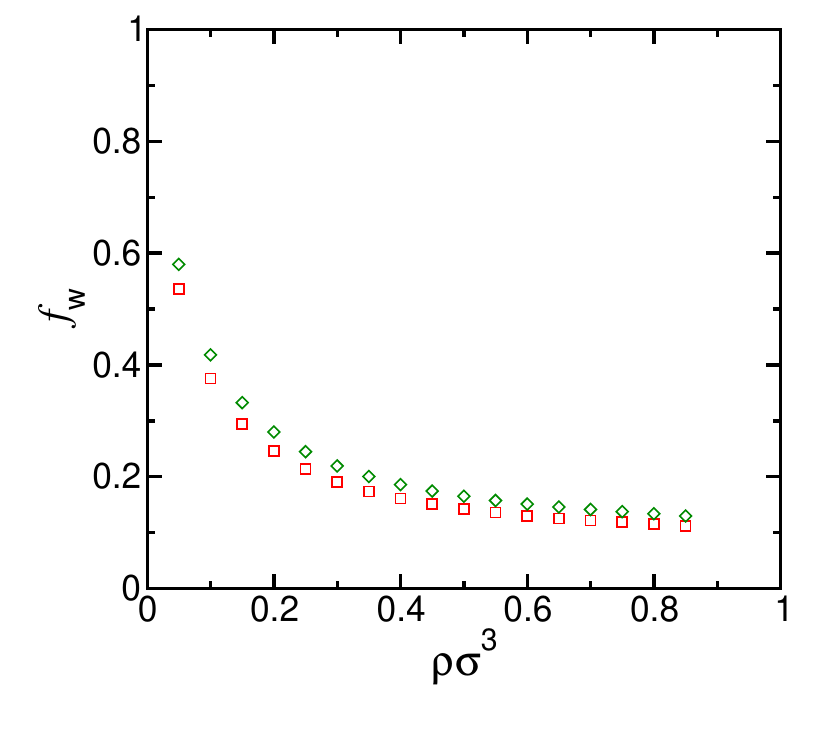}
    }
    \subfloat[][$H/\sigma=7.5$]{
      \includegraphics[width=0.2\textwidth]{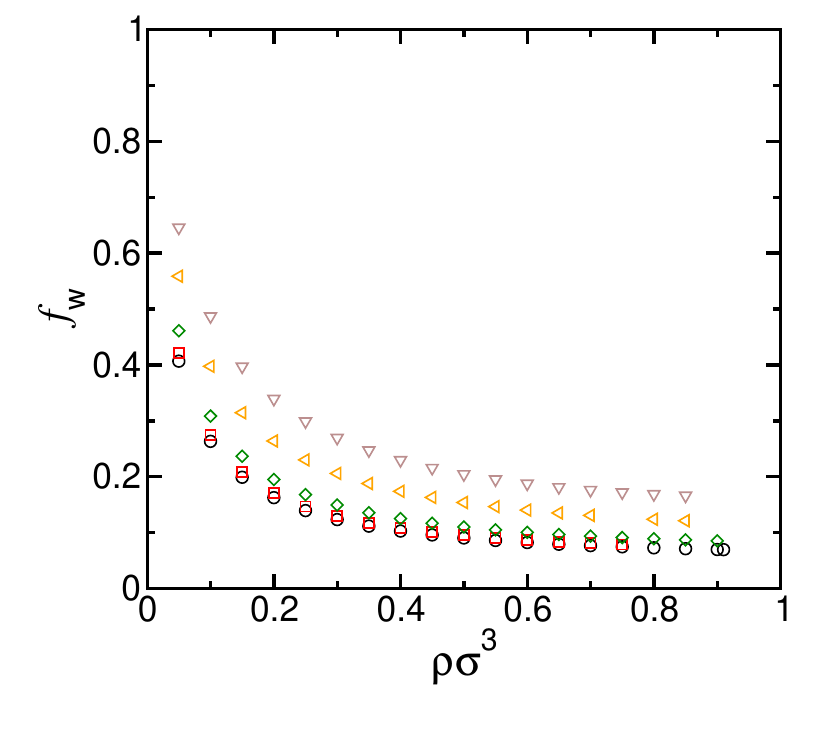}
    }
    \subfloat[][$H/\sigma=10.0$]{
      \includegraphics[width=0.2\textwidth]{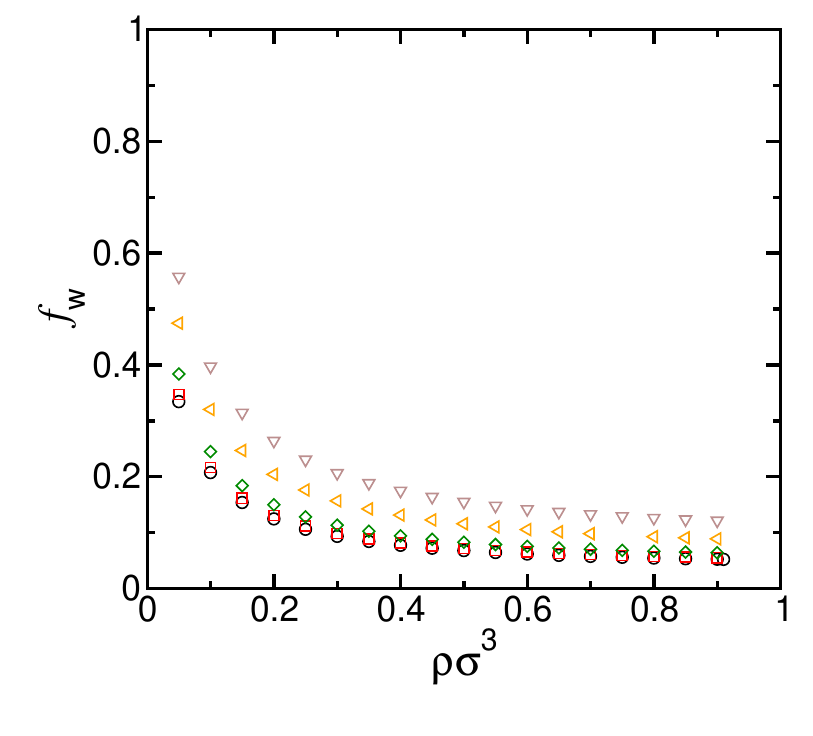}
    }
    \subfloat[][$H/\sigma=15.0$]{
      \includegraphics[width=0.2\textwidth]{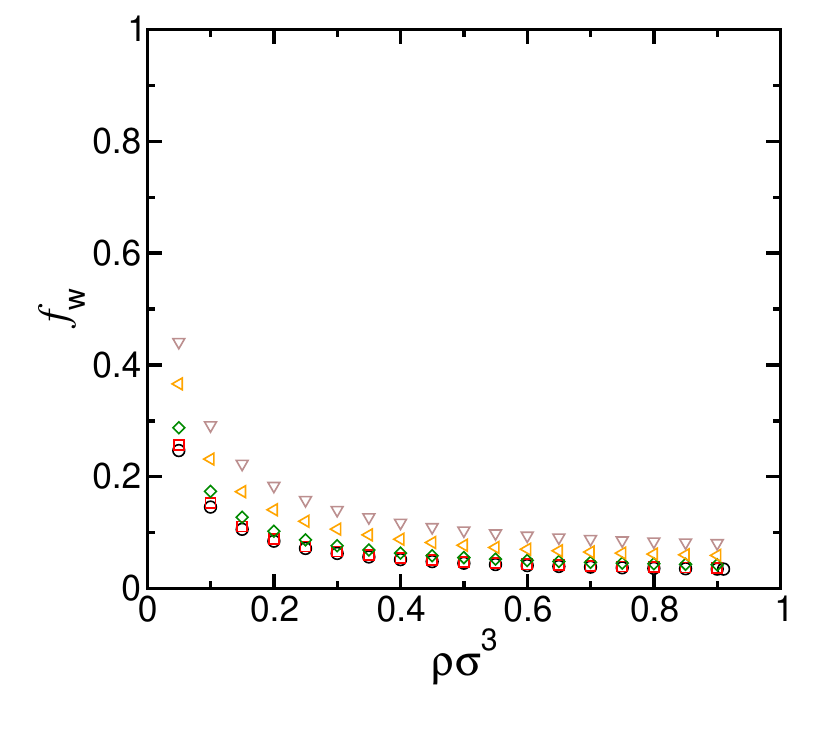}
    }
  }

  % dy/dx Hinv
  \subfloat[][$\lambda/\sigma=6.0$]{
    \includegraphics[width=0.2\textwidth]{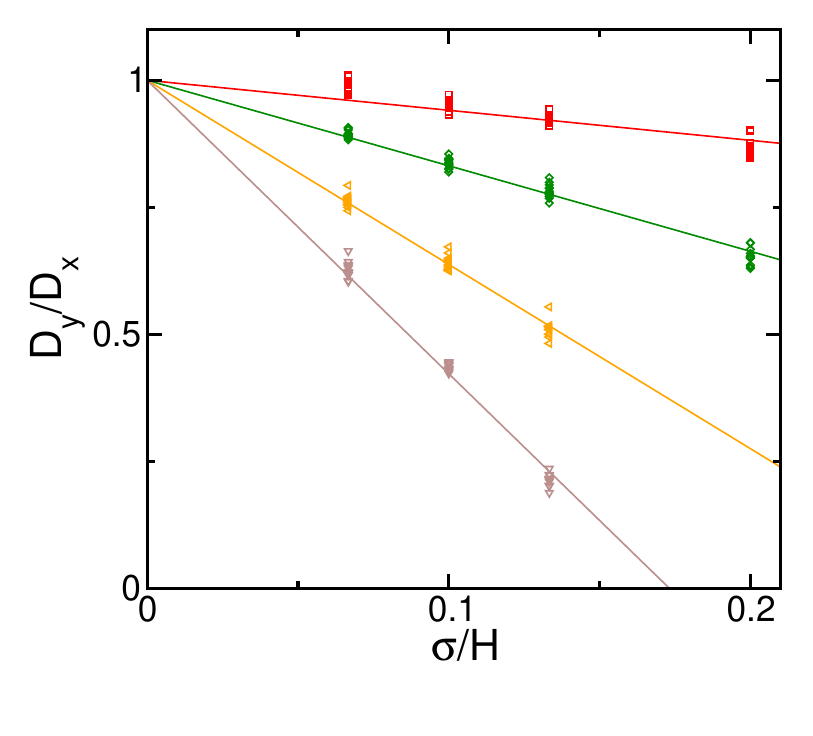}
  }

  \caption{Same as Fig.~\ref{fig:more_roughness_WL_1.5} with $\lambda/\sigma=6.0$.}
  \label{fig:more_roughness_WL_6.0}
\end{figure*}

\begin{figure*}[h]
  \centering
  \subfloat[][]{
    \includegraphics[width=6in]{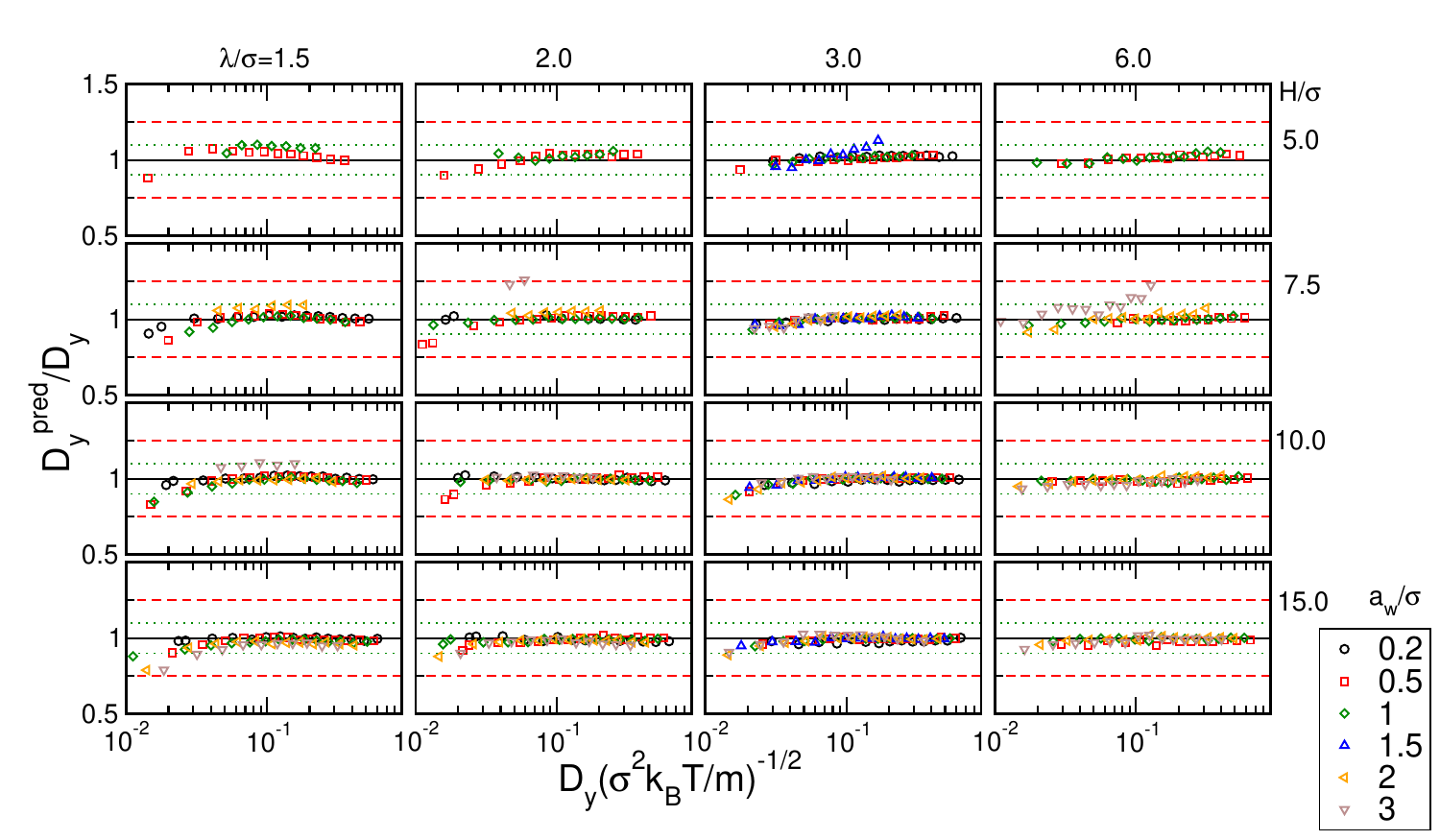}
    \label{fig:DY_prediction_exact}
  }\\
  \subfloat[][]{
    \includegraphics[width=6in]{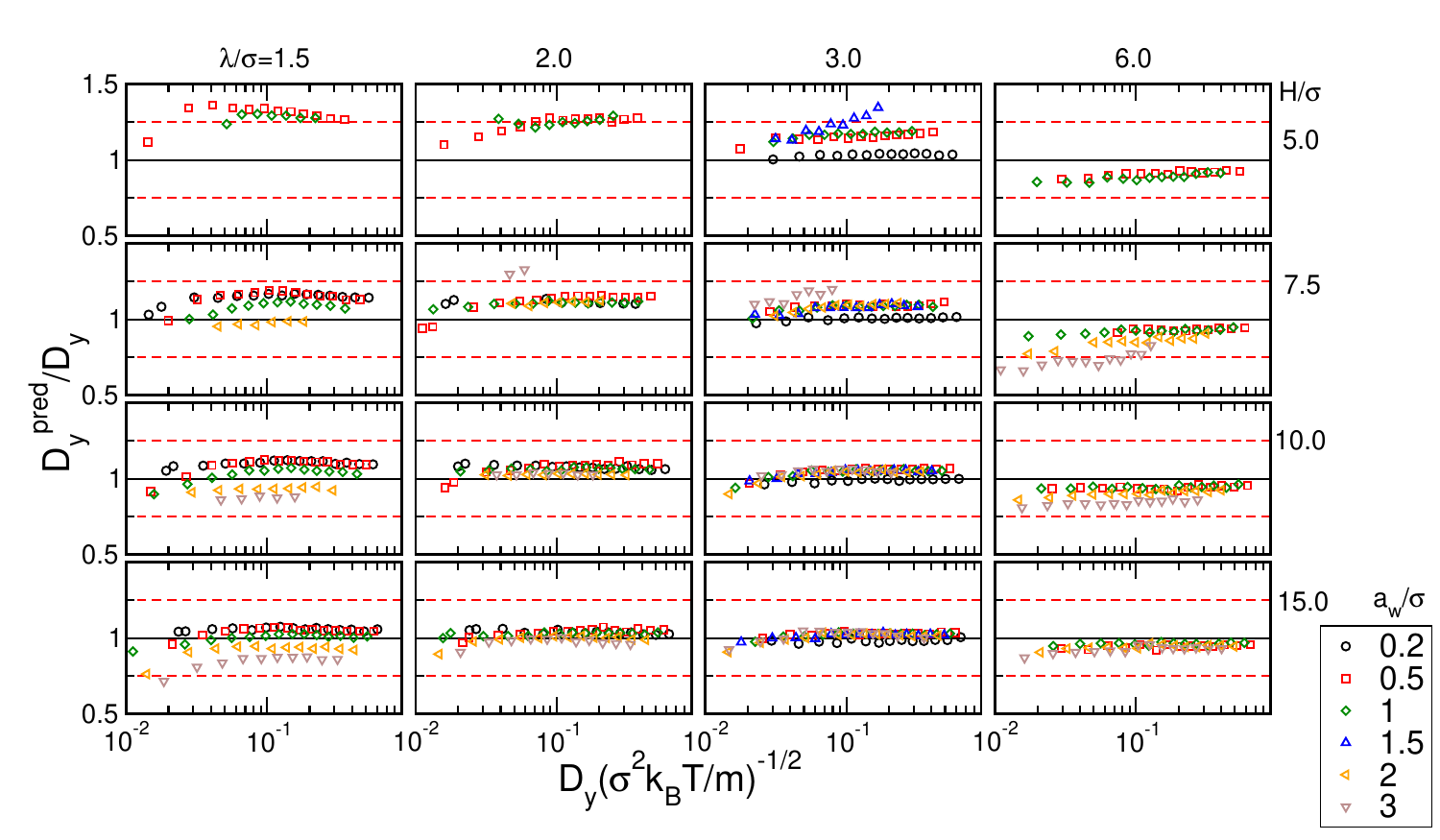}
    \label{fig:DY_prediction_approx}
  }
  \caption{Ratio of predicted value of $D_y$ from
     $D_y/D_x=1-C[H/\sigma]^{-1}$ to actual value of $D_y$
    versus $D_y$ for states with $\rho\sigma^3>0.2$. In (a), values of $C$
    used in prediction are taken from the full linear fits shown in
    subfigure (q) of the above figures.  In (b), the value of $C$ is
    taken from the linear fit $C=2.12a_w/\sigma$ (see main text).  The dotted green
    and dashed red lines correspond to $10\%$ and $25\%$ error bounds,
    respectively.}
  \label{fig:DY_predictions}
\end{figure*}

\end{document}